%% file: SI01rep.tex
\documentstyle[a4,12pt,epsf]{article}

\newcommand{\nn}{\nonumber}
\newcommand{\be}{\begin{equation}}
\newcommand{\ee}{\end{equation}}
\newcommand{\bea}{\begin{eqnarray}}
\newcommand{\eea}{\end{eqnarray}}
\newcommand{\bean}{\begin{eqnarray*}}
\newcommand{\eean}{\end{eqnarray*}}
\newcommand{\th}{\theta^2}
\newcommand{\thb}{\bar{\theta}^2}

\begin{document}
\noindent
\renewcommand{\thefootnote}{\alph{footnote}}
\begin{center}
{\large
Renormalization Group for Soft SUSY Breaking Parameters \\
and MSSM Coupled with Superconformal Field Theories
\footnote{
Talk at Summer Institute 2001, Yamanashi, Japan, 2001. 
}
}
\vspace*{10mm}\\
Haruhiko Terao
\footnote{E-mail: terao@hep.s.kanazawa-u.ac.jp}
\vspace*{5mm}\\
Institute for Theoretical Physics, Kanazawa University\\
Kanazawa 920-1192, Japan
\end{center}
\begin{abstract}
First we give a review of the spurion formalism and the exact 
renormalization group equations for soft supersymmetry breaking 
parameters in general gauge theories.
Next we discuss the minimal supersymmetric standard model coupled to 
superconformal theories leading to hierarchical Yukawa couplings by
large anomalous dimensions of quarks and leptons.
The soft scalar masses are found to satisfy the noble sum rule
in IR regime.
It is possible construct the models such that the degenerate 
squark/slepton masses are realized thanks to this IR sum rule.
However, it is found also that this degeneracy is slightly broken mainly by 
the radiative corrections of the SM gaugino mass insertion.
We show that the degeneracy is sufficient for the squark sector,
but weak for the slepton sector to avoid the supersymmetric flavor problems.
\end{abstract}

\renewcommand{\thefootnote}{\arabic{footnote}}
\setcounter{footnote}{0}

\section{Exact RG for soft SUSY breakings}
\subsection{Soft SUSY breakings and spurion superfields}
SUSY has been expected to realize the high energy theories in solving
the gauge hierarchy problem.
This is based on the fact that UV sensitivity is reduced to logarithmic,
i.e. absence of quadratic divergence, ensured by supersymmetry and renormalizability.
While SUSY must be broken at TeV scale.
The SUSY breaking terms should be introduced so as not to generate power
divergences. Such SUSY breakings are called soft in supersymmetric theories.
The type of soft SUSY breaking parameters have been clarified for
renormalizable theories by using spurion method \cite{gg}.
We note that all terms with dimension less than 4 are not allowed.

There have been proposed various SUSY breaking and mediation mechanisms.
On the other hand Minimal Supersymmetric Standard Model (MSSM) must be
compatible with severe constraints from precision experiments.
In order to see a given high energy model to satisfy the constraints or 
not, we need the RG framework for soft SUSY breaking parameters.

It has been known that the RG equations for the soft SUSY breaking parameters
can be derived from the beta functions in the rigid theories.
This correspondence is exact in the sense of all order perturbation
theory. Therefore non-perturbative analysis is totally beyond our present scope.
In this formalism also the spurions are found to be  quite
useful \cite{gg,yamada}.
However it is rather recent that the formalism for the ``exact" RG for 
soft parameters have been completed. 
In this section we review the formalism of the exact RG for soft SUSY
breaking parameters and try to sum up the findings spreading in the 
literatures \cite[-,14]{yamada}.

Let us start the review to see what kinds of the soft breaking parameters
are allowed.
Actually the proof relies heavily on the renormalizability of the theory.
So we first discuss how renormalizability remove quadratic divergences
in supersymmetric theories.
There we use the superfield Feynman rules in perturbation theory \cite{feynman}.

We consider the general renormalizable Lagrangian given by
\be
{\cal L} = \int d^4 \theta
\phi^{\dagger}_i e^V \phi^i + \int d^2 \theta W(\phi) + \mbox{h.c}
+ \int d^2 \theta \frac{1}{2g^2} \mbox{Tr} W^{\alpha} W_{\alpha} + \mbox{h.c.}
\ee
First we see $\theta$ integration in the $N=1$ super-Feynman rules.
The superfield propagators are all local in the $\theta$ coordinate
and are given explicitly by
\bea
\langle \phi^{\dagger}_i \phi^j \rangle &=&
\frac{1}{p^2 + m^2} \delta_i^j \delta^4(\theta - \theta') \\
\langle \phi^i \phi^j \rangle &=&
\frac{m D^2}{4 p^2 (p^2 + m^2)} \delta^{ij} \delta^4(\theta - \theta') \\
\langle V^A V^B \rangle &=&
-\frac{1}{p^2} \delta^{AB} \delta^4 (\theta - \theta')
\eea
Let us count degrees of divergence for Feynman diagrams.
For each vertex $\int d^4\theta$ integration and four covariant derivatives
$D$ or $\bar{D}$ are attached. If there are $E$ external legs of chiral
superfields, then the number of covariant derivatives is reduced by $2E$.
For each loop with n vertices, the $\theta$ integrations are reduced by partial
integration and by using the formula, $D^2 \bar{D}^2 D^2= 16 \Box D^2$, to
\be
\Gamma \sim \int d^4\theta_1 \int d^4\theta_n 
\delta^4(\theta_1 - \theta_n) (D^2_n)^l (\bar{D}^2_n)^k 
\delta^4(\theta_n - \theta_1),
\ee
where $k, l$ are 0 or 1. Therefore the $\theta$ integration does not vanish,
only if $k=l=1$. 
Thus the n-point Green functions are always given by the form of
\be
\Gamma = \sum_n \int d^4 x_1 \cdots \int d^4 x_n
\int d^4 \theta G(x_1, \cdots, x_n) f(\phi, V, D^{\alpha}\phi, \cdots)
\ee
Here the famous non-renormalization of the superpotential follows,
\footnote{
The non-renormalization is also shown by a simple argument based on holomorphic
property of the superpotential.
For the proofs in this line, see Ref.~\cite{seiberg1,weinberg}.
} 
since the radiative corrections appears only with $\int d^4\theta$ integration.

Now it is seen that the degree of divergence $D$ is given by
\be
D = 4 L - 2 P - C + 2 V - E - 2 L = 2 - C - E,
\label{degree}
\ee
where $L, P, C, V$ and $E$ denote the number of loops, propagators,
chiral propagators, $\langle \phi \phi \rangle$
and $\langle \phi^{\dagger} \phi^{\dagger} \rangle$,
vertices, and chiral external lines respectively.
By taking into account of the gauge invariance,
\footnote{
In abelian gauge theory, the formula (\ref{degree}) gives $D=2$ 
for the Fayet-Iliopoulos D-term, $\int d^4 \theta \xi V$. 
However this D-term does not receive
radiative corrections, since Tr$Y=0$ must be guaranteed to avoid 
gravitational and chiral anomalies \cite{fiterm}.
}
we find that the divergence is at most logarithmic.
Thus the naturalness of supersymmetric theories relies on
the renormalizability. For example, if the superpotential contains
$\phi^4$ term, which is non-renormalizable, then the quadratic 
divergence appears \cite{fl}.

The SUSY breaking terms can be incorporated into the superspace
perturbation by using spurion superfields.
For example, the scalar mass term $m^2 \phi^{\dagger}\phi$ is
rewritten as $\int d^4 \theta U \phi^{\dagger}\phi$
by introducing a spurion superfield $U=m^2 \th \thb$.
Thus, what ensures  absence of quadratic divergence in softly
broken theories is again renormalizability.
If all the SUSY breaking terms represented by means of spurions
do not destroy renormalizability, then divergences are at most
logarithmic. Therefore the vertex containing a spurion superfield,
which is treated as an external field in perturbation, should
not carry more than four covariant derivatives.
Thus the soft terms are restricted to the followings,
\bea
{\cal L}_{\mbox{soft}} &=& 
-\int d^4 \theta ({m^2}_i^j \th \thb) \phi_j^{\dagger} \phi^i
- \int d^2 \theta (M \th) \frac{1}{g^2} \mbox{Tr} W^{\alpha}W_{\alpha} \nn\\
& &- \int d^2 \theta \frac{1}{2}(\mu_{ij} \th) \phi^i \phi^j
- \int d^2 \theta \frac{1}{6}(h_{ijk} \th) \phi^i \phi^j \phi^k
- \mbox{h. c.},
\eea
which represent the soft scalar masses, B-terms, tri-linear scalar couplings
and the gaugino masses.
Note that, for example, the fermion mass term given by 
$\int d^4 \theta (m \th \thb) D^{\alpha}\phi D_{\alpha} \phi$ 
gives a hard breaking, though the dimension of this term is less 
than four.

\subsection{Softly broken Wess-Zumino models}
In this sub-section we derive the RG equations for softly broken
Wess-Zumino model, whose Lagrangian is given by
\be
{\cal L} = \int d^4 \theta (1 - m_{0i}^2 \th \thb) \phi_i^{\dagger}
\phi^i + \int d^2 \theta \frac{1}{6}(y_{0ijk} - h_{0ijk} \th)
\phi^i \phi^j \phi^k + \mbox{h.c.},
\ee
where the superscript $0$ shows that the coupling is bare.
Owing to the non-renormalization of the super potential, the effective
Lagrangian is given by
\footnote{The Z-factor introduced here may be inverse of the conventional
one.}
\be
{\cal L}_{\mbox{eff}} = \int d^4 \theta \tilde{Z}_i (\theta, \bar{\theta})
\phi_i^{\dagger} \phi^i + \int d^2 \theta \frac{1}{6}(y_{0ijk} - h_{0ijk} \th)
\phi^i \phi^j \phi^k + \mbox{h.c.}.
\ee
Here we have ignored mixings among the chiral superfields.
Since the mixing effects does not change the argument significantly,
let us assume the wave function renormalization to be diagonal for
a moment. The RG formulae with mixing is shown in Appendix.

It should be noted that the renormalization of (anti-)chiral
superfields must be also (anti-)chiral. Therefore 
we need to extract the (anti-)chiral wave function renormalization 
factors as \cite{kazakov}
\be
\tilde{Z}_i(\theta, \bar{\theta})  =
\tilde{Z}_{\phi^i}^{\dagger}(\bar{\theta})(1 - m^2_i \th \thb) 
\tilde{Z}_{\phi^i}(\theta),
\ee
where $m^2$ is nothing but the soft scalar mass.
By expanding $\tilde{Z}$ into
$\tilde{Z}= Z + Z^{(1)}\th + \bar{Z}^{(1)} \thb + Z^{(2)} \th \thb$,
we may represent the chiral wave function factors explicitly also 
in terms of the components as
\be
\tilde{Z}_{\phi^i} (\theta) = 
Z_i^{1/2} \left(
1 + Z_i^{-1} Z_i^{(1)} \th
\right).
\ee
By using these chiral wave function superfields, we may define the renormalized
couplings $y_{ijk}$ and $h_{ijk}$ by
\be
Y_{ijk}(\theta) \equiv y_{ijk} - h_{ijk} \th
= \tilde{Z}^{-1}_{\phi^i}(\theta)\tilde{Z}^{-1}_{\phi^j}(\theta)
\tilde{Z}^{-1}_{\phi^k}(\theta)
(y_{0ijk} - h_{0ijk} \th).
\ee

It is found that the RG equations for the chiral superfield $Y_{ijk}$ are
given by
\be
\mu \frac{d Y_{ijk}}{d \mu} =
- \left(
\frac{d \ln \tilde{Z}_{\phi^i}}{d \ln \mu} +
\frac{d \ln \tilde{Z}_{\phi^j}}{d \ln \mu} +
\frac{d \ln \tilde{Z}_{\phi^k}}{d \ln \mu} 
\right) Y_{ijk}.
\label{RGforY}
\ee
Here let us introduce superfield generalization of the anomalous dimensions
by using $\tilde{Z}$ as
\be
\tilde{\gamma}_i(\theta, \bar{\theta}) \equiv
- \frac{d \ln \tilde{Z}_i}{d \ln \mu} =
\gamma_i + \gamma^{(1)} \th + \bar{\gamma}^{(1)} \thb +
\gamma^{(2)} \th \thb,
\ee
where $\gamma_i$ are nothing but the ordinary anomalous diemnsions.
Then the RG equations for $y_{ijk}$ and $h_{ijk}$ are immediately
derived from the superfield equation given by (\ref{RGforY}) and found
to be
\bea
\mu \frac{d y_{ijk}}{d \mu} &=& 
\frac{1}{2} \left( \gamma_i + \gamma_j + \gamma_k \right)y_{ijk}, \nn \\
\mu \frac{d h_{ijk}}{d \mu}&=& 
\frac{1}{2} \left( \gamma_i + \gamma_j + \gamma_k \right)h_{ijk}
- \left( \gamma_i^{(1)} +\gamma_j^{(1)} +\gamma_k^{(1)} \right) y_{ijk}.
\eea
In this stage we do not know the explicit form of $\gamma^{(1)}$ or $\bar{\gamma}^{(1)}$,
therefore, the beta functions for the trilinear couplings $h_{ijk}$ either.

Fortunately the singular part of $\tilde{Z}_i$, {\it i.e.} $\tilde{\gamma}_i$
are found to be given, once the wave function renormalization
$Z_i$ for the rigid theories are known.
Explicitly the wave function superfields $\tilde{Z}_i$ are related with their 
rigid ones through \cite{kazakov}
\be
\tilde{Z}_i(\theta, \bar{\theta}) =
Z_i (\tilde{y}_{ijk}, \tilde{y}_{ijk}^{\dagger}),
\ee
where we introduced a new coupling superfield given by
\be
\tilde{y}_{ijk} = Y_{ijk} + \frac{1}{2} (m^2_i + m^2_j + m^2_k)y_{ijk} \th \thb.
\label{ytilde}
\ee
There have been known two kinds of arguments explaining this.
The first one is based on the superfield Feynman rules \cite{jj1,kazakov}.
It is noted that the superfield propagators in the softly broken theories are
modified from the rigid ones;
\be
\langle \phi^i \phi_j^{\dagger} \rangle_{\mbox{soft}}
= \left(1 + \frac{1}{2} m^2_i \th \thb \right)
\langle \phi^i \phi_j^{\dagger} \rangle_{\mbox{rigid}}
\left(1 + \frac{1}{2} m^2_j \th \thb \right).
\ee
The factors with the soft scalar masses may be absorbed by
redefining the Yukawa coupling superfields appearing in the superfield
Feynman rules to $\tilde{y}_{ijk}$ given by (\ref{ytilde})
as far as the singular parts of the diagrams are concerned.

The coupling superfields given by (\ref{ytilde}) are supported also by
the symmetry argument as follows \cite{gr,aglr}.
Once we suppose that the coupling superfields are not spurions but dynamical,
then the softly broken theories enjoy a global $U(1)^c_{\phi^i}$ symmetry
corresponding to each chiral superfield $\phi^i$.
Here the subscript $c$ denotes the complex extension.
Explicitly the transformation laws are given with chiral superfields $\tilde{T}_i$
as parameters by
\bea
\phi^i &\rightarrow& e^{\tilde{T}_i} \phi^i, ~~~~~
\phi_i^{\dagger} \rightarrow  \phi_i^{\dagger} e^{\bar{\tilde{T}}_i}, 
\nn \\
Y_{0ijk} &\rightarrow& e^{-\tilde{T}_i}Y_{0ijk}, \nn \\
\tilde{Z}_i &\rightarrow& e^{-\bar{\tilde{T}}_i} \tilde{Z}_i e^{-\tilde{T}_i}.
\eea
The physical quantities should be invariant under each $U(1)^c_{\phi^i}$
transformation.
The obvious invariant is the soft scalar mass 
$m^2_i = - \ln \tilde{Z}_i|_{\th \thb}$.
Also the combination $Y_{0ijk} \tilde{Z}_i^{-1} Y_{0ij'k'}^{\dagger}$
gives another invariant.
Therefore each contraction between the Yukawa couplings should be 
accompanied with the factor $(1 + m^2_i \th \thb)$,
which is realized with the extension given by (\ref{ytilde}).

Thus the components in the superfields of anomalous dimensions 
$\tilde{\gamma}$ are given explicitly from the rigid anomalous dimensions
$\gamma$ as functions of Yukawa couplings.
By introducing the following differential operators,
\bea
D_1 &=& - h_{ijk} \frac{\partial}{\partial y_{ijk}}, \nn \\
D_2 &=& D_1 \bar{D}_1  + \frac{1}{2}\left(m^2_i + m^2_j + m^2_k \right)
\left(y_{ijk}\frac{\partial}{\partial y_{ijk}} + 
\bar{y}^{ijk}\frac{\partial}{\partial \bar{y}^{ijk}}
\right),
\eea
those are given explicitly by
$\gamma^{(1)}_i = D_1 \gamma_i, \bar{\gamma}^{(1)}_i = \bar{D}_1 \gamma_i$ and
$\gamma^{(2)}_i = D_2 \gamma_i$.
Resultantly we arrive at the ``exact" beta functions for the soft SUSY breaking
parameters as well as Yukawa couplings, which are found to be
\bea
\mu \frac{d y_{ijk}}{d \mu} &=&
\frac{1}{2} \left( \gamma_i +\gamma_j +\gamma_k \right) y_{ijk}, \nn \\
\mu \frac{d h_{ijk}}{d \mu} &=&
\frac{1}{2} \left( \gamma_i +\gamma_j +\gamma_k \right) h_{ijk}
-\left( D_1 \gamma_i + D_1 \gamma_j + D_1 \gamma_k
\right)y_{ijk}, \nn \\
\mu \frac{d m^2_i}{d \mu} &=&
D_2 \gamma_i.
\eea

\subsection{Softly broken supersymmetric gauge theories}
In this subsection we extend the arguments to  the gauge theories, whose Lagrangian is
given by
\be
{\cal L} = \int d^4 \theta
\phi_i^{\dagger} e^{V} \phi^i + \int d^2 \theta S_0 \mbox{Tr} W^{\alpha}W_{\alpha}
+ \mbox{h.c.},
\ee
where $S_0 = 1/2g_{h0}^2 - i \Theta/16\pi^2$ is a holomorphic combination
of the bare gauge coupling and the vacuum angle.
By the holomorphy argument, it is shown that the holomorphic gauge coupling
$g_h$ is renormalized only at 1-loop \cite{seiberg1,weinberg}.
Explicitly the RG equation for the holomorphic coupling is given by
\be
\mu \frac{d S}{d \mu} = \frac{1}{16\pi^2}
\left( 3 T_G - \sum_i T_i \right),
\label{holomorphic}
\ee
where $T_G= C_2(G)$ and $T_i = T(R_i)$ for the gauge representation $R_i$
of the chiral superfield $\phi^i$.

It has been known for some time that this holomorphic gauge coupling
is not the physical coupling defined as the 1PI vertex function \cite{nsvz,am}.
The relation between these couplings are found to be
\bea
8\pi^2(S + S^{\dagger}) - \sum_i T_i \ln Z_i &=& F(\alpha) \nn \\
&=& \frac{1}{\alpha} + T_G \ln \alpha + \sum_{n > 0} a_n \alpha^n,
\label{gaugerel}
\eea
where $\alpha = g^2/8\pi^2$ and $a_n$ are scheme dependent constants.
The so-called Novikov-Shifman-Veinstein- Zaharov (NSVZ) scheme corresponds
to the case of $a_n=0$ for all $n$.
Explicit relations between the NSVZ scheme \cite{nsvz} 
and dimensional reduction (DRED)
scheme have been also found up to four-loop order in perturbation \cite{jjp3}.
It should be noted that the right hand side of eq.~(\ref{gaugerel})
is an invariant under the global $U(1)^c_{\phi}$ symmetry \cite{gr,aglr}.
The transformations with a complex parameter $T$ are given by
\bea
\phi^i &\rightarrow& \phi^i e^T, \nn \\
Z_i &\rightarrow& e^{-\bar{T}} Z_i e^{-T}, \nn \\
S &\rightarrow& S - \frac{T}{8\pi^2} \sum_i T_i.
\eea
The shift in the transformation of the gauge coupling follows
from the Konishi anomaly \cite{konishi}, which is supersymmetric extension of
the chiral anomaly.

The ``exact" beta function for the gauge coupling is derived from the relation
given by (\ref{gaugerel}).  By noting that the holomorphic gauge coupling
satisfies the one loop RG equation (\ref{holomorphic}), the beta function is
found to be
\be
\beta_{\alpha} = \mu \frac{d \alpha}{d \mu} = 
\frac{1}{F'(\alpha)} \left[
3 T_G - \sum_i T_i (1- \gamma_i) \right],
\ee
where $\gamma_i$ denotes the anomalous dimension for $\phi^i$ and
is defined by 
\be
\gamma_i = - d \ln Z_i / d \ln \mu. 
\ee

In the NSVZ scheme, the beta function is given by
\be
\beta_{\alpha} = - \frac{\alpha^2}{1 - T_G \alpha}
\left[ 3 T_G - \sum_i T_i (1 - \gamma_i) \right].
\ee

Now we discuss the softly broken gauge theories. 
The soft scalar masses and the gaugino mass are incorporated
in the effective Lagrangian by using spurion superfields as
\be
{\cal L}_{\mbox{eff}} = \int d^4 \theta
\tilde{Z}_i \phi_i^{\dagger} e^V \phi^i +
\int d^2 \theta \tilde{S} \mbox{Tr} W^{\alpha} W_{\alpha}
+ \mbox{h.c.}.
\ee
Here the spurions are given explicitly in terms of the gaugino mass $M$
and the soft scalar masses $m_i^2$ as
\bea
\tilde{S} &=& \frac{1}{g_h^2}(1 - 2M\th), \nn \\
\tilde{Z}_i &=& \tilde{Z}_{\phi^i} (1 - m_i^2 \th \thb ) \tilde{\bar{Z}}_{\phi^i}.
\eea

The spurion superfield corresponding to the physical gauge coupling
may be obtained by extending the relation between the holomorphic
and the physical couplings to superfields;
\be
8\pi^2(\tilde{S} + \tilde{S}^{\dagger}) - \sum_i T_i \ln \tilde{Z}_i 
= F(\tilde{\alpha}).
\label{gaugerelext}
\ee
Then it is found immediately that the spurion for the gauge coupling
should be introduced as
\be
\frac{1}{\tilde{g}^2} = \frac{1}{g^2}
\left(
1 - M \th - \bar{M} \thb - \Delta_g \th \thb
\right)
\ee
or for $\alpha = g^2/8\pi^2$,
\be
\tilde{\alpha} = \alpha \left(
1 + M \th \bar{M} \thb + (2 M \bar{M} + \Delta_g) \th \thb
\right).
\ee
Here $\Delta_g$ is also found to be
\be
\Delta_g = \frac{1}{\alpha F'(\alpha)} \left[
\sum_i T_i m^2_i - (\alpha^2 F'(\alpha))' M \bar{M}
\right].
\ee
In the NSVZ scheme, it is reduced to 
\be
\Delta_g = -\frac{\alpha}{1 - T_G \alpha} \left[
\sum_i T_i m^2_i - T_G M \bar{M}
\right].
\label{NSVZdelta}
\ee
which coincides with the known result found through
different arguments \cite{kkz,jjp2}.

At a glance $\Delta_g$ may seem to be irrelevant, since it does
not remain in the Lagrangian after $\theta$ integration.
In practice, however, the RG equations for soft scalar masses
depend on this factor, as is seen later on.
Especially $\Delta_g$ plays a significant role to realize
the IR sum rules among soft scalar masses in superconformal theories,
which will be discussed in the next section.
Historically, necessity of such a factor was recognized 
by comparing with the perturbative results in DRED scheme.
There  the 2-loop beta function for the soft scalar
mass depends on the ``$\epsilon$-scalar" mass, which is induced
by radiative corrections. 
However it is possible to remove this ``$\epsilon$-scalar" mass
dependence in a modified scheme, DRED' \cite{jjmvy}.
In this scheme, in turn, $\Delta_g$ must be introduced \cite{jjp1}.
\footnote{
The relation between $\th \thb$ component of the gauge coupling
spurion superfield and the ``$\epsilon$-scalar" mass is 
discussed also in Ref.~\cite{aglr}
}
Afterwards the exact form given by (\ref{NSVZdelta}) has been found
on the RG invariant trajectory \cite{kkz}.
General validity of the formula has been also confirmed by
comparison with perturbative results obtained in DRED' scheme \cite{jjp2}.
It has been also claimed that $\Delta_g$ is necessary to be introduced
even when assuming supersymmetric regularization and is identified with
a soft SUSY breaking mass for the ghost superfield \cite{kv}.

Now it is rather simple task to derive the RG equations for all soft 
SUSY breaking parameters
in gauge theories with general Yukawa interactions by performing
Grasmannian expansion. 
It is important to note that the singular part of the wave function
renormalization $\tilde{Z}_i$ may be given from the rigid one as
\be
\tilde{Z}_i = Z_i ( \tilde{\alpha}, \tilde{y}_{ijk}, \tilde{\bar{y}}_{ijk}).
\ee
Then the differential operators $D_1$ and $D_2$ are extended to
\bea
D_1&=& M \alpha \frac{\partial}{\partial \alpha} 
- h_{ijk} \frac{\partial}{\partial y_{ijk}}, \\
D_2&=&\bar{D}_1 D_1 + (M \bar{M} + \Delta_g) \alpha \frac{\partial}{\partial \alpha}\nn \\
& & 
+\frac{1}{2}\left(m^2_i + m^2_j + m^2_k \right)
\left(y_{ijk}\frac{\partial}{\partial y_{ijk}} + 
\bar{y}_{ijk}\frac{\partial}{\partial \bar{y}_{ijk}}
\right).
\eea
Consequently the RG equations for the gaugino mass, the tri-linear couplings and 
the soft scalar masses are obtained as
\bea
\mu \frac{d M}{d \mu} &=& \left. \mu \frac{d \ln \tilde{\alpha}}{d \mu} \right|_{\th}
= D_1 \left( \frac{\beta_{\alpha}}{\alpha} \right), \\
\mu \frac{d h_{ijk}}{d \mu} &=&
\frac{1}{2} \left( \gamma_i +\gamma_j +\gamma_k \right) h_{ijk}
-\left( D_1 \gamma_i + D_1 \gamma_j + D_1 \gamma_k
\right)y_{ijk}, \\
\mu \frac{d m^2_i}{d \mu} &=& D_2 \gamma_i.
\eea
These equations are valid in all orders of perturbation theory.

\subsection{RG invariant relations}
It is remarkable that the spurion formalism enable us to find
the RG equations for all soft SUSY breaking parameters, once we
know the beta functions in the rigid theories.
However the spurion formalism tells us more than that.
Interestingly the fact that the spurion superfields satisfy the RG
equations, automatically ensures existence of RG invariant relations
for the soft parameters \cite{jj2}.

Suppose spurion superfields $X_i$ satisfy
the RG equations $\mu (dX_i/d\mu) = \beta_i(X) $.
After expanding the superfields into
$X_i = x_i + y_i \th + \bar{y}_i \thb + z_i \th \thb$,
the RG equations for the components, which are regarded as couplings,
are easily found to be
\bea
\mu \frac{d x_i}{d \mu} &=& \beta_i(x), \nn \\
\mu \frac{d y_i}{d \mu} &=& \sum_j y_j 
\frac{\partial \beta_i}{\partial x_j}(x), \nn \\
\mu \frac{d z_i}{d \mu} &=& \sum_j z_j 
\frac{\partial \beta_i}{\partial x_j}(x)
+ \frac{1}{2} \sum_{k, l} y_k \bar{y}_l
\frac{\partial^2 \beta_i}{\partial x_k \partial x_l} (x).
\label{rginv1}
\eea

On the other hand, if there are the RG invariant
relations among couplings given by
\bea
y_i(\mu) = f_i\left(x(\mu)\right), \nn \\
z_i(\mu) = g_i\left(x(\mu)\right),
\eea
for all $\mu$, then the couplings must satisfy also
the following equations,
\bea
\mu \frac{d y_i}{d \mu} &=& \sum_j \beta_j(x)
\frac{\partial f_i}{\partial x_j}(x), \nn \\
\mu \frac{d z_i}{d \mu} &=& \sum_j \beta_j(x) 
\frac{\partial g_i}{\partial x_j}(x).
\label{rginv2}
\eea
{}From eqs.~(\ref{rginv1}) and eqs.~(\ref{rginv2}), it is
seen that the functions $f_i$ and $g_i$ are subject to some
differential equations. By solving these equations, 
we may find the general RG invariant relations;
\bea
y_i &=& \kappa \beta_i(x), \nn \\
z_i &=& \frac{1}{2} |\kappa|^2 \sum_k \beta_k(x) 
\frac{\partial \beta_i}{\partial x_k}(x)
=\frac{1}{2} |\kappa|^2 \mu \frac{d \beta_i}{d \mu},
\eea
where $\kappa$ is an integration constant.

By applying the above argument to the spurion superfields appearing
in the supersymmetric gauge-Yukawa theories, we can find out
the RG invariant relations for the soft SUSY breaking parameters.
The results are as follows;
\bea
M &=& \kappa \frac{\beta_{\alpha}}{\alpha}, \label{hsrel} \\
h_{ijk} &=& - \kappa \beta_{y_{ijk}}, \\
B_{ij} &=& -\kappa \beta_{\mu_{ij}}, \\
(m^2)_i^j &=& \frac{1}{2} |\kappa|^2 \mu \frac{d \gamma_i^j}{d \mu}.
\eea
Eq.~(\ref{hsrel}) was firstly discovered by Hisano and Shifman \cite{hs}.
It should be noted also that these relations are nothing but the conditions
for the soft parameters obtained in anomaly mediated SUSY
breaking scenario \cite{anomaly}, where the parameter $\kappa$ is the garavitino
mass.

\section{IR behavior of sfermion masses in MSSM coupled to SCFTs}
\subsection{Yukawa hierarchy by large anomalous dimensions}
In this section, we discuss the interesting models proposed recently by
Nelson and Strassler \cite{ns1}, which may realize Yukawa hierarchy dynamically.
Specially we focus on degeneracy of low energy squark and slepton masses
in the general models of this type \cite{kt,ns2}. 
In this study, it is found that the
``exact"" RG equations for soft SUSY breaking parameters discussed
in the previous section are found to be quite useful.

Hierarchical structure of the fermion mass matrices has been one of the 
mysteries in particle physics. Recently this problem has been more
attractive stimulated by findings of neutrino oscillation. 
The most popular scenario
leading to fermion mass hierarchy is the Froggatt-Nielsen mechanism \cite{fn}, 
where the effective Yukawa couplings are suppressed accordingly
to flavor dependent charges $n_{q_i}$ of quarks/leptons under an
extra $U(1)_X$ symmetry.
With introducing a SM gauge singlet field $\chi$ with $U(1)_X$ charge $-1$,
the non-renormalizable interactions are allowed in the superpotential;
\be
W =
\tilde{y}_{ij} \left( \frac{\chi}{M_0} \right)^{n_{q_i}+n_{q_j}} q_i q_j H,
\ee
where $M_0$ is cutoff scale of the theory.
In this scenario, we suppose that only the $U(1)_X$ charges generate large 
flavor dependence, and take the bare couplings $\tilde{y}_{ij}$ to be
O(1).
If this singlet $\chi$ acquires vacuum expectation value of $M_c$, then
the non-renormalizable interactions are reduced to the Yukawa interactions 
with hierarchical couplings given by
\be
y_{ij} \sim \left( \frac{M_c}{M_0} \right)^{n_{q_i}+n_{q_j}}.
\ee
Indeed we may explain the fermion masses and mixings by using this type of 
Yukawa matrices with assigning the $U(1)_X$ charges properly.

Nelson and Strassler showed that such Yukawa matrices are realized also
by assuming additional interactions of quarks/leptons with unknown strongly
coupled sectors, which are (nearly) on the superconformal (SC) fixed points.
In order to see this mechanism, let us consider N=1 supersymmetric $SU(N_c)$ 
QCD with Lagrangian,
\be
{\cal L} = \int d^4\theta \left(Q_f^{\dagger} e^{-V} Q_f 
+\bar{Q}_f e^{-V} \bar{Q}_f^{\dagger}\right) +
\frac{1}{2g^2}\int d^2\theta \mbox{Tr} W^2 + \mbox{h.c.}.
\ee
It has been known for some time \cite{fpt} that there exists a IR fixed point
if the number of flavors $N_f$ lies in the so-called conformal window,
$3/2N_c < N_f < 3N_c$.
There the gauge beta function must vanish.
The exact beta function in NSVZ scheme given by
\be
\beta_g = \mu \frac{dg}{d\mu} =
-\frac{g^3}{16\pi^2}
\frac{3N_c-N_f(1-\gamma_Q)}{1-\frac{N_c}{8\pi^2}g^2}
\ee
tells us that the anomalous dimensions of $Q$ at the fixed point
$\gamma_Q^*$ is exactly determined and is found to be
\be 
 -1 < \gamma^*_Q = - \frac{3N_c-N_f}{N_f} < 0.
\ee
We note that this anomalous dimension is negative and order of one in general.

Now we introduce a singlet field $q$ with the superpotential
$W = \lambda q \bar{Q}_1 Q_1$. 
Then this interaction is relevant at the IR fixed point of the QCD
due to the negative anomalous dimension of $Q_1$ and $\bar{Q}$.
Therefore the Yukawa coupling $\lambda$, hereafter we call it SC-Yukawa
coupling, grows rapidly.
The RG flows towards IR are shown in Fig.~1, in which A stands for 
the QCD fixed point. The flow line starting near the fixed point approaches
to a new fixed point B.
\begin{figure}[htb]
\begin{center}
\epsfxsize=0.6\textwidth
\leavevmode
\epsffile{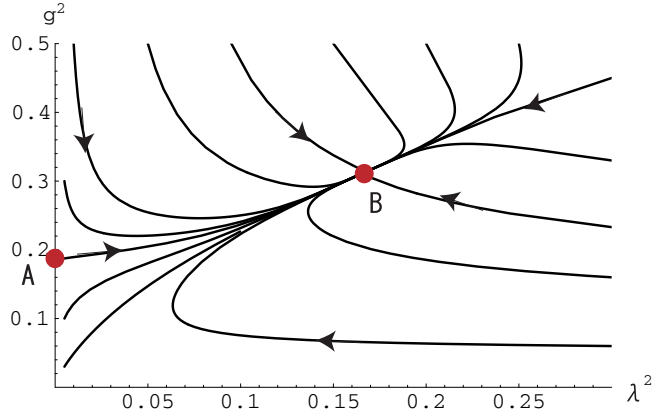}
\caption{
RG flows of N=1 SQCD with Yukawa interaction
}
\end{center}
\end{figure}
At this IR attractive fixed point, the beta function 
for the SC-Yukawa coupling,
$\beta_{\lambda} = (1/2)(\gamma_q + 2 \gamma_{Q_1}) \lambda$,
as well as the gauge beta function should vanish.
Therefore the anomalous dimension of $q$ is determined to be $
\gamma^*_q = - 2\gamma^*_{Q_1}$, which is positive and order of one.

Now we suppose that quark/lepton superfields $q_i$ couple to 
$\Phi_i$ and  $\bar{\Phi}_i$ in the SC-sector through the superpotential
\be
W=\lambda_i q_i\bar{\Phi}_i \Phi_i + y_{ij}q_iq_jH,
\ee
where $H$ is the SM Higgs and $y_{ij}$ are the Yukawa couplings
to be hierarchical in the end.
In the Nelson-Strassler model the flavor dependence
is generated by the SC dynamics giving different anomalous dimensions for
the SC-sector field such as $\gamma^*_{\Phi_1} < \gamma^*_{\Phi_2} < \cdots < 0$.
Then the anomalous dimensions of the quarks/leptons at the IR attractive
fixed point become $\gamma^*_{q_1} > \gamma^*_{q_2} > \cdots > 0$.
As results, the Yukawa couplings $y_{ij}$ decrease rapidly with power of
$\gamma^*_{q_i} + \gamma^*_{q_j}$, which give rise to the hierarchical
structure at low energy.
Here note also that mixed terms like $\lambda'_2 q_1 \bar{\Phi}_2 \Phi_2$, 
which are allowed by the symmetry, are irrelevant at the fixed point.
Namely the structure of generation in the quarks/leptons are determined  
by the interaction with the SC-sector.

However, the SC-Yukawa interaction should terminate at a certain
scale to produce realistic Yukawa couplings.
Therefore we also assume that all the SC-fields decouple from the
SM-sector at scale $M_c$, for example, by mass terms 
$M_c \bar{\Phi}_i \Phi_i$.
Then the Yukawa couplings in the SM model turn out to be
\be
y_{ij} \sim 
\left(\frac{M_c}{M_0}\right)^{\gamma^*_{q_i}+\gamma^*_{q_j}}
y_{ij}(M_0) 
\ee
at low energy. Thus the hierarchical Yukawa matrix of the Froggatt-Nielsen
type can be realized. 
In this scenario we may consider the model so that the third generation,
at least top quark, does not couple to the SC-sector for the order one Yukawa
coupling. It should be noted also that the tri-linear couplings
also become hierarchical with the same mechanism.

\subsection{IR sum rules for soft scalar masses in SCFTs}
It is very non-trivial for supersymmetric models to be compatible
with constraints from precision experiments of FCNC, LFV, EDM and so on \cite{fcnc}.
In order to avoid the so-called supersymmetric flavor problem, there have been
proposed three types of solutions for the soft scalar mass matrices of 
squarks/sleptons; 1. degenerate masses \cite{degenerate}, 
2. alignment to Yukawa matrices \cite{alignment} and
3. decoupling of the first and the second generations \cite{decoupling}.
As for the degenerate solution, it is rather non-trivial to make masses of
quarks/leptons hierarchical with keeping the squark/slepton masses to be
flavor universal.
Related to the Froggatt-Nielsen mechanism, therefore, possibilities of the 
alignment \cite{alignment} and the decoupling solutions \cite{anomalousU1}
have been investigated. 

However totally successful models have not been known.
For example, the decoupling solution may be naturally realized in the
anomalous $U(1)_X$ SUSY breaking scenario due to the flavor dependent 
D-term contributions \cite{anomalousU1}.
On the other hand, however, it has been also claimed that
the heavy squarks in the first and the second
generations make the radiative EW breaking unstable \cite{decoupling} 
and even bring about color breaking by tachyonic s-top \cite{decoupling2}.
Hence, mechanisms leading to Yukawa hierarchy with naturally 
avoiding the supersymmetric flavor problems are highly desired.

The remarkable property of the Nelson-Strassler models is not only to 
generate flavor hierarchy. 
Interestingly enough, these models have possibility to realize degenerate
squark/slepton masses in low energy irrespectively of the SUSY breaking 
dynamics.
The key observation is the soft scalar masses can be suppressed owing to
strong interaction with the SC-sector \cite{kkkz,lr}.
After the SC-sector is decoupled, the soft masses grow slowly by
radiative corrections from the SM gaugino masses, which are flavor
blind. Therefore it may be expected that masses of the first and the
second generation sfermions become degenerate and the supersymmetric 
flavor problem is also avoided.\footnote{
Here we concern only the block diagonal part of
squark/slepton mass matrices, {\it i.e.} $(m^2)_{LL}$ and $(m^2)_{RR}$,
which are given by the soft scalar masses. The off-diagonal parts,
$(m^2)_{LR}$ induced from the tri-linear couplings are less problematic
due to alignment. For studies of experimental constraints for the
off-diagonal parts, see Ref.~\cite{ns2}.
}
A sketch of running sfermion masses in the Nelson-Strassler model
is shown in Fig.~2. 
\begin{figure}[htb]
\begin{center}
\epsfxsize=0.6\textwidth
\leavevmode
\epsffile{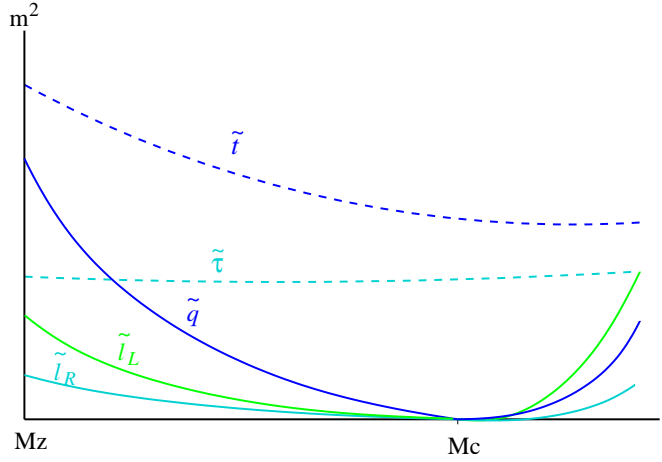}
\caption{A sketch of running sfermion masses.
}
\end{center}
\end{figure}

In this subsection, we examine the behavior of soft scalar masses
added to general superconformal gauge theories by using the exact RG
equations for them. As results, it is found that the soft scalar masses
enjoy sum rules at IR \cite{kt,ns2}. 
Actually we can construct models coupled to SCFTs such that the 
suppression of squark/slepton masses are guaranteed thanks to 
these sum rules. 

Now we suppose a SCFT on a IR attractive fixed point. Then, by definition,
any infinitesimal variation of the gauge coupling $g$ and the 
Yukawa coupling $y^{ijk}$
from their fixed point values should decrease towards IR.
This is represented explicitly as follows.
The linear perturbation of the coupled RG equations
for $\alpha=g^2/8\pi^2$ and $\alpha_{y}^{ijk}=|y^{ijk}|^2/8\pi^2$ around
the fixed point gives the equations given by
\be
\mu \frac{d}{d\mu} \left(
\begin{array}{c}
\delta \alpha \\
\delta \alpha_y^{lmn}
\end{array}
\right)
= \left(
\begin{array}{cc}
\left. \frac{\partial \beta_{\alpha}}{\partial \alpha}
\right|_*&
\left. \frac{\partial \beta_{\alpha}}{\partial \alpha_y^{ijk}}
\right|_* \\
\left. \frac{\partial \beta^{lmn}_{\alpha_y}}{\partial \alpha}
\right|_* &
\left. \frac{\partial \beta^{lmn}_{\alpha_y}}{\partial \alpha^{ijk}_y}
\right|_*
\end{array}
\right)
\left(
\begin{array}{c}
\delta \alpha \\
\delta \alpha_y^{ijk}
\end{array}
\right),
\label{linear}
\ee
where $|_*$ stands for evaluation at the fixed point.
The IR attractive nature implies that the matrix appearing in this equation is
positive definite.

The spurion formalism discussed in the previous section enables us
to see behavior of soft SUSY breaking parameters added to this SCFT
quite easily. We note that the spurion superfields corresponding
to $\alpha$ and $\alpha_y^{ijk}$ satisfy the same RG equations as
the rigid ones,
\bea
& &\mu \frac{d \tilde{\alpha}}{d \mu} = 
\beta_{\alpha}(\tilde{\alpha}, \tilde{\alpha}_y), \nn \\
& &\mu \frac{d \tilde{\alpha}_y^{ijk}}{d \mu} = 
\beta_{\alpha_y}^{ijk}(\tilde{\alpha}, \tilde{\alpha}_y).
\eea
The Grassmanian expansion with respect to $\th$ and $\thb$ of
these equations gives us the RG equations for the soft SUSY breaking
parameters. Since the RG equations for the gauge coupling and the 
Yukawa coupling are not modified, we fix them to be the fixed
point values. Then the Grassmanian expansion corresponds to the
linear perturbation of the RG equations around the fixed point.
In other words, the $\th$ and the $\th \thb$ parts satisfy the
exactly same equation given by (\ref{linear}) respectively.

{}From the $\th$ terms of the spurions, the infinitesimal variations
in eqn.~(\ref{linear}) may be replaced as
$\delta{\alpha} = \alpha_* M \theta^2$ and
$\delta{\alpha^{ijk}_y} =  -(1/8\pi^2)
\bar{y}^{ijk}_*h^{ijk} \theta^2$.
Therefore the gaugino mass $M$ and the tri-linear couplings $h^{ijk}$
are found to be suppressed with a certain power of scale.
In the NSVZ scheme, the $\th \thb$ terms are given by
\bea
\delta{\alpha} &=& -\frac{\alpha^2_*}{1-T_G \alpha_*} \sum_i T_i m^2_i
\theta^2  \bar{\theta}^2  \nn \\
\delta{\alpha^{ijk}_y} &=& \alpha^{ijk}_{y*}
(m^2_i+m^2_j+m^2_k)
\theta^2  \bar{\theta}^2,
\eea
where we have omitted $M$ and $h^{ijk}$.
Thus it is found that the soft scalar masses satisfy the 
IR sum rules,\footnote{
It is speculated that similar sum rules hold in the presence of
non-renormalizable interactions in the superpotential of SCFT \cite{ns1,kt}.
}
\bea
& &\sum_i T_i m^2_i \rightarrow 0,\nn \\
& &m^2_i + m^2_j + m^2_k \rightarrow 0. 
\eea
Note that the second  condition holds for each Yukawa coupling 
$y_{ijk}\phi^i \phi^j \phi^k$.

So far we have assumed the soft scalar masses to be flavor diagonal.
In taking the mixing effects into considerations, 
these sum rules are found to be valid among the diagonal components. 
It can be also shown that each off-diagonal component of the soft 
scalar mass matrix is suppressed by power law. For details, see 
Ref.~\cite{ns2}.

\subsection{Convergence of the soft scalar masses and flavor dependence}
In the Nelson-Strassler models it is necessary for squark/slepton masses
of the first and second generations vanish at the decoupling scale $M_c$
to realize degeneracy at low energy.
Indeed there exist some types of models in which the sum rules
enforce the squark/slepton masses to be suppressed.
The condition for these models is that the anomalous dimension
of quark/lepton superfield is fixed uniquely by the
relations among anomalous dimensions, which follows from the 
vanishing beta functions.\footnote{
This condition may be restated that the R-charge of quark/lepton 
superfield is uniquely fixed.
} 

Here let us present a simple toy model with one generation.
We consider $SU(3)_{SC}$ as the
gauge group of the SCFT and $SU(3)_C$ as the SM gauge group.
The field contents and their representation under the group
are as follows;
\begin{center}
\begin{tabular}{c|ccccccc}
 & $Q$ & $\bar{Q}$ & $P$ & $\bar{P}$ & $q_i$ & $\bar{q}_i$ & $H$ \\ \hline
SU(3)$_{SC}$ & {\bf 3} & {\bf \=3} &  {\bf 3} & {\bf \=3} & 
{\bf 1}& {\bf 1}& {\bf 1} \\
SU(3)$_{C}$ & {\bf \=3} & {\bf 3} &  {\bf 3} & {\bf \=3} & 
{\bf 3}& {\bf \=3}& {\bf 1} 
\end{tabular}
\end{center}
We also assume the superpotential given by
\be
W = \lambda(q_1\bar{Q}P+ \bar{q}_1 \bar{P}Q) + y_i\bar{q}_iq_iH
\ee
Then the IR sum rules are found to be
\bea
& &m^2_Q + m^2_P + m^2_{\bar{Q}} + m^2_{\bar{P}}  \rightarrow  0, \nn \\
& &m^2_{q_1}+ m^2_{\bar{Q}} + m^2_P  \rightarrow 0, \nn \\
& &m^2_{\bar{q}_1}+ m^2_Q + m^2_{\bar{P}}  \rightarrow 0.
\eea
{}From these equations, it follows that $m^2_{q_1} + m^2_{\bar{q}_1}$
vanishes at IR. Therefore the squark mass is suppressed by imposing 
left-right symmetry for their initial values.
Examples of RG flows for the squark (mass)$^2$ with the SM gaugino 
(mass)$^2$ in this model are shown with various initial values in Fig.~3.
There $t$ denotes $\ln(\mu/M_0)$.
Thus the information of the bare squark masses are
completely lost at the decoupling scale ($t=-3$ in Fig.~3). 
\begin{figure}[htb]
\begin{center}
\epsfxsize=0.6\textwidth
\leavevmode
\epsffile{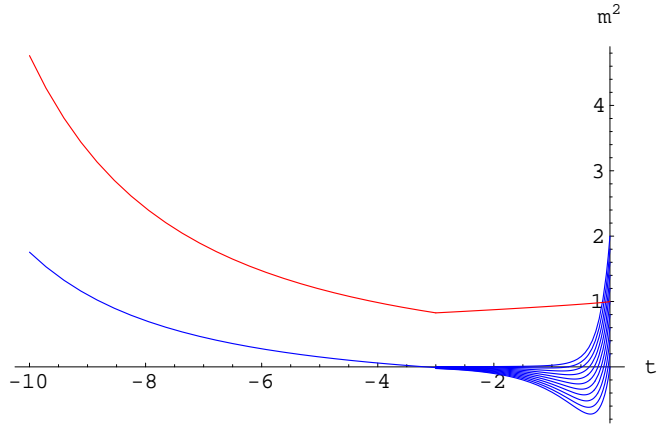}
\caption{
RG flows of SM-sector gaugino (mass)$^2$ and squark (mass)$^2$.
}
\end{center}
\end{figure}

However, the Nelson-Strassler models are not exactly superconformal,
since the SM-sector interactions act as weak perturbation.
The significant effects appear through the SM gaugino mass insertions.
By taking into account of this corrections at one loop, the RG equations
for the soft scalar masses are changed to be
\be
\mu \frac{dm^2_i}{d \mu} = {\cal M}_{ij} m^2_j
- \sum_{a=1,2,3} C_{ia}\alpha_a M_a^2,
\ee
where ${\cal M}_{ij}$ stands for the positive definite matrix
appearing in eq.~(\ref{linear}). 
Also $\alpha_a$ and  $M_a$ denote the gauge couplings and the
gaugino masses of the SM gauge groups of 
$SU(3)\times SU(2) \times U(1)$, and $C_{ia}=4C_2(R_i)$.
The eigenvalues of the matrix ${\cal M}$ are supposed to be
order of one in general.
It should be noted that this matrix is necessarily flavor dependent
in the Nelson-Strassler models.

With this additional term, the soft scalar masses for squark/slepton
do not disappear, but converge a non-vanishing values at the 
decoupling scale.
Compared with the rapid suppression of the soft scalar masses, 
the evolution of the gauge couplings and the gaugino masses in the
SM-sector is very slow.
Therefore we may evaluate the convergence as
\be
m^2_{q^i} \rightarrow \frac{C_{ia}}{\Gamma_i}\alpha_a(M_c)M^2_a(M_c),
\ee
where $\Gamma_i$ is the smallest eigenvalue of matrix {\cal M}.
This factor is of order one and flavor dependent.
Thus, though the soft scalar masses are suppressed,
small flavor dependence remains at the decoupling scale in practice.
This non-degeneracy appears as a sizable effect to squark/slepton 
spectra at the weak scale in considering the supersymmetric flavor
problems.

\subsection{Degeneracy of squark/slepton masses in low energy}
Before going into the low energy degeneracy of squark/slepton masses,
let us see the gross aspect of their mass spectra.
After the SC-sector is decoupled, the squark/slepton masses are
generated by radiative correction with SM gaugino mass insertion.
This correction is flavor blind and may be evaluated at one-loop 
perturbation.
For the doublet squark, the mass squared at the weak scale is given by
\be
m_{Q^i}^2(M_Z) = m_{Q^i}^2(M_C) + 
{8 \over 9}[{\alpha^2_3(M_Z) \over \alpha^2_3(M_C) } -1 ] M^2_3(M_c)
+ \cdots.
\ee
If we use the GUT relation for the gaugino masses, 
$M_a(\mu) = \alpha_a(\mu) M$, then this relation may be represented also as
\be
m_{Q^i}^2(M_Z) - m_{Q^i}^2(M_c) =
{8 \over 9}\left[\alpha^2_3(M_Z) - \alpha^2_3(M_c)\right] 
M^2 + \cdots.
\ee

Since the soft masses for the first two generations at the decoupling
scale is suppressed, the low energy masses are determined by the radiative
corrections. Then ratio of squark/slepton masses and the gluino mass $M_3$ 
is fixed, once the decoupling scale is given.
In Fig.~4 these ratio are plotted for some range of the decoupling scale.
The dotted lines represent the ratio of $M_2$ and $M_1$ to $M_3$.
It is seen that the right-handed slepton appears as the lightest
supersymmetric particle (LSP).
This is because the right-handed slepton carries only $U(1)$ charge,
and, therefore, the radiative correction is small. 
However the LSP must be either the lightest neutralino or the sneutrino,
unless R-parity is violated.
\begin{center}
\input s-spec.tex
\newline
Figure 4: Ratio of sfermion masses (bold line), $M_2$, $M_1$ to $M_3$.
\end{center}

In the Nelson-Strassler model the LSP problem may be ameliorated by
considering the hypercharge D-term contribution.
Actually the hypercharge D-term contribution cannot be ignored for 
slepton masses.
Now the interaction with the SC-sector does not constrain
soft masses of the third generation sfermion or higgs particles.
Therefore the D-term contribution given by $S=$Tr$Y m_i^2(M_c)$ 
is not determined.
In Fig.~5, the low energy right-handed slepton mass is shown 
in ratio to $M_3$ in the case of $S=0$ and $S=-0.5 M^2_3$.
It is seen that the LSP problem can be avoided, as long as
the decoupling scale is larger than some intermediate scale, say, 
$10^{10}$GeV.
\begin{center}
\input spec-es.tex
\newline
Figure 5: $U(1)_Y$ D-term contribution to right handed slepton mass.
\end{center}

Finally we examine how good degeneracy between the low energy 
sfermion masses of the first two generations can be achieved
in the Nelson-Strassler models.
The important quantities in the view point of flavor changing processes
are the off-diagonal elements of the soft scalar masses in the
base that quark/lepton masses are diagonalized.
These are usually represented as $(\delta^q_{12})_{LL}$ or
$(\delta^q_{12})_{RR}$ in the literatures.
In the original base, the off-diagonal elements are suppressed enough
by the strong interaction with the SC-sector. 
Therefore, what we should evaluate here is the degree of splitting 
among the diagonal components of the soft scalar mass matrices;
\be
\delta^q_{12} = \frac{m^2_{q^2}-m^2_{q^1}}{\bar{m}^2_q},
\ee
where $\bar{m}^2_q$ denotes average of the diagonal masses
of squarks/sleptons $q$.
Then the off-diagonal elements $(\delta^q_{12})_{LL}$ or 
$(\delta^q_{12})_{RR}$ may be obtained by $\delta^q_{12}$ 
times the mixing angle of the quarks/leptons.

The splitting obtained in the Nelson-Strassler model is now 
evaluated as 
\bea
\delta^q_{12} &=& \Delta_{\tilde{q}}
\left(\frac{1}{\Gamma_2}-\frac{1}{\Gamma_1}\right), \\
\Delta_{\tilde{q}} &=&
C_{qa}\alpha_a(M_c)\frac{M_a^2(M_c)}{\bar{m}^2_q(M_Z)}.
\eea
Here the factor $\Gamma_i$ are model dependent but $\Delta_{\tilde{q}}$
is determined without any ambiguities except for the decoupling 
scale $M_c$.
It should be noted here that the quantity $\Delta_{\tilde{q}}$
are independent of the gaugino mass $M$, since the average mass 
$\bar{m}_q$ is also proportional to $M$.

In Fig.~6 and 7, $\Delta_{\tilde{q}}$ are explicitly shown for 
$q = Q, d, L$ and $e$ for some range of the decoupling scale.
The result for $q = u$ is similar to those for $q=d$, therefore
has been omitted.
As results, degeneracy of squark masses is found to be less than 1\%.
This value is good enough in avoiding the supersymmetric flavor
problem. For example, $(\delta^d_{12})_{LL}$ is reduced to be
less than 0.2\% due to the Cabbibo angle, which satisfies
the experimental bound from K$^0$-\={K}$^0$ mixing \cite{fcnc}.

However the degeneracy in the slepton sector is not strong as 
the squarks. Especially for the right-handed sleptons $\tilde{e}$,
the expected degeneracy is at most a few \%.
Therefore the constraints from lepton flavor violation (LFV) processes,
{\it e.g.} $\mu \rightarrow e + \gamma$, seem severe to be 
satisfied in general.
For models with the factor $\Gamma_i \gg 1$, this restriction is
ameliorated.
There would be other ways of improving this situation.
For example, if the Nelson-Strassler mechanism would work 
within the GUT framework, the slepton masses will be significantly
changed.
Indeed degeneracy of the right-handed sleptons is found to be mach
improved in the flipped $SU(5)$ GUT \cite{knnt}.

\begin{center}
\input Delta-q.tex
\newline
Figure~6: $\Delta_{\tilde Q}$ and $\Delta_{\tilde d}$ against $M_c$.
\input Delta-l.tex
\newline
Figure~7: $\Delta_{\tilde Q}$ and $\Delta_{\tilde d}$ against $M_c$.
\end{center}

\section{Summary and discussions}
In the first section recent development on the exact RG equations 
for the soft SUSY breaking parameters was reviewed.
There we saw that the spurion formalism is a quite powerful method
in discussing the RG.
The simple derivation of the gauge coupling spurion superfield
has been given also.

In the second section we have discussed the Nelson-Strassler type of 
models: SSMs coupled with SCFTs.
First it was shown that the hierarchical texture of the Yukawa couplings
can be realized by large anomalous dimensions for quarks and leptons.
The interaction with superconformal sector brings about the large
anomalous dimensions.
Then behavior of the soft SUSY breaking parameters in these models
are investigated in details.
We have found the IR sum rules among soft scalar masses in the general
superconformal field theories.
In the proof of the sum rules also, the exact RG equations and the
spurion formalism discussed in the first section play an essential
role.
With these IR sum rule, we found the conditions for the models
to realize degenerate sfermion masses at low energy.

It is remarkable that the Nelson-Strassler models offer us 
possibilities to realize Yukawa hierarchy with avoiding the supersymmetric
flavor problem.
Moreover the low energy sfermion masses for the first two generations
are determined irrespectively of the SUSY breaking mechanism, or
the initial values of SUSY breaking parameters at cutoff scale.
However, it has been pointed out also that the flavor dependence in the
sfermion masses are not completely washed out.
In a realistic case with non-vanishing gauge couplings 
of the SM sector, the radiative corrections of SM gaugino mass insertion 
in RG equations of sfermion masses play a significant role.
These corrections are small but flavor dependent. 
Therefore the sfermion masses lose complete universality at low energy.

We have shown explicitly how much degeneracy we obtain
between sfermion masses in the MSSM.
For squarks we can have suppression strong enough to avoid the FCNC problem.
On the other hand, for sleptons such suppression was found to be weak
in general.

Lastly let us mention also the possibilities
to achieve completely universal sfermion masses.
The origin of non-degeneracy of sfermion masses in the Nelson-Strassler
models lies in that the couplings to the SC-sector are necessarily
flavor dependent.
If these couplings are also flavor universal, then we will obtain
completely degenerate masses.
Then what about Yukawa hierarchy?
Interestingly the hierarchy of Yukawa couplings in SM-sector can be
induced by changing the bare SC-Yukawa couplings to be flavor
dependent. We call such scenarios ``Yukawa hierarchy transfer" \cite{knt}.
There flavor universality in squark/slepton masses is compatible with
the hierarchical quark/lepton masses.
For details, see the report by Nakano in this proceeding of 
the Summer Institute.

\section*{Acknowledgements}
The author is grateful to the organizers for the stimulating 
Summer Institute with nice atmosphere.
It is also a pleasure to thank Tatsuo Kobayashi, Hiroaki Nakano and Tatsuya
Noguchi for fruitful collaborations. Most of the work reviewed in this paper
was done in collaboration with Tatsuo Kobayashi.

\appendix

\section{Exact RG equations with mixings}
In the section~1. we have been ignoring the field mixing in the effective
Lagrangian. In the Appendix we consider the exact RG equations for the
soft SUSY breaking parameters with taking mixing effect into considerations in
the Wess-Zumino models.
Actually it is not difficult to incorporate the field mixing by improving
the argument given in section~1 \cite{ns2}.

The effective Lagrangian is given by
\be
{\cal L}_{\mbox{eff}} = 
\int d^4 \theta \tilde{Z}(\theta, \bar{\theta})_i^j \phi_j^{\dagger} \phi^i
+ \int d^4 \theta \frac{1}{6}Y_{0ijk}\phi^i \phi^j \phi^k + \mbox{h.c.},
\ee
where $Y_{0ijk} = (y_{0ijk} - h_{0ijk} \th \thb)$ is the chiral
spurion superfield of the bare couplings.
What to be considered newly is that the wave function renormalization
factors are matrices.
We extract the (anti-)chiral part from $\tilde{Z}$ by
\be
\tilde{Z}(\theta, \bar{\theta})_i^j =
\tilde{Z}_{\phi}^{\dagger}(\bar{\theta})_k^j
(\delta_l^k - (m^2)_l^k \th \thb) \tilde{Z}_{\phi} (\theta) _i^l.
\ee
Note that the soft scalar masses also appear as a matrix.
By using the chiral superfield matrix $\tilde{Z}_{\phi}$, the
renormalized couplings are defined by
\be
Y_{ijk}(\theta) = y_{ijk} - h_{ijk} \th =
Y_{0i'j'k'}(\theta)
\tilde{Z}^{-1}_{\phi}(\theta)_i^{i'}\tilde{Z}^{-1}_{\phi}(\theta)_j^{j'}
\tilde{Z}^{-1}_{\phi}(\theta)_k^{k'}.
\ee

The effective Lagrangian turns out to be invariant under global 
$U(N)^c_{\phi}$ transformation in taking spurions as dynamical
superfields.
The transformations are extended to
\bea
\phi &\rightarrow& e^{\tilde{T}} \phi, ~~~~
\phi^{\dagger} \rightarrow \phi^{\dagger} e^{\bar{\tilde{T}}}, \nn \\
\tilde{Z} &\rightarrow& e^{-\bar{\tilde{T}}} \tilde{Z} e^{-\tilde{T}}, \nn \\
Y_0 &\rightarrow& Y_0 \left(e^{-\tilde{T}} \otimes
e^{-\tilde{T}} \otimes e^{-\tilde{T}} \right),
\eea
where $\tilde{T} = \tilde{T}(\theta)$ is a matrix of chiral superfield.
We also define a matrix superfield
\be
\tilde{X}(\theta, \bar{\theta})_i^j =
\left(\delta_k^j - \frac{1}{2} (m^2)_k^j \th \thb \right)
\tilde{Z}_{\phi}(\theta)_i^k,
\ee
which satisfies $\tilde{X}^{\dagger} \tilde{X} = \tilde{Z}$ and the
transformation $\tilde{X} \rightarrow \tilde{X}\exp (-\tilde{T})$
under $U(N)^c_{\phi}$.

The physical quantities must be invariant under this global symmetry.
The soft scalar masses are again invariant, since
$(m^2)_i^j = - (\ln \tilde{Z})_i^j |_{\th \thb}$.
The invariant combination made of the Yukawa coupling superfields
is given by
\bea
\tilde{y}_{ijk} &=& Y_{0i'j'k'} (\tilde{X}^{-1})_i^{i'}
(\tilde{X}^{-1})_j^{j'}(\tilde{X}^{-1})_k^{k'}, \nn \\
&=& y_{ijk} - h_{ijk} \th
+\frac{1}{2} \left[
y_{i'jk}(m^2)_i^{i'} +y_{ij'k}(m^2)_j^{j'} +y_{ijk'}(m^2)_k^{k'} 
\right] \th \thb.
\label{ytilder}
\eea
The key point is again that the wave function factor $\tilde{Z}$ is
given by replacing the Yukawa couplings with the invariant combination
$\tilde{y}$ in the rigid one; 
$\tilde{Z}(\theta, \bar{\theta})_i^j = Z(\tilde{y}, \bar{\tilde{y}})_i^j$.

Next we introduce the matrix of anomalous dimensions extended to
superfield. By taking account that the anomalous dimensions also must be 
invariant under $U(N)^c_{\phi}$, we may define them as
\be
\tilde{\gamma}(\theta, \bar{\theta})_i^j = 
\left(
\tilde{X} \frac{d \tilde{X}^{-1}}{d \ln \mu} +
 \frac{d (\tilde{X}^{\dagger})^{-1}}{d \ln \mu}\tilde{X}^{\dagger}
\right)_i^j.
\ee

If we expand the anomalous dimension matrix superfield with respect to
$\th, \thb$ as
$\tilde{\gamma}(\theta, \bar{\theta}) =
\gamma + \gamma^{(1)}\th + \bar{\gamma}^{(1)} \thb + \gamma^{(2)} \th \thb$,
then the each component enjoys the following relations;
\be
\gamma = Z_{\phi} \frac{d Z^{-1}_{\phi}}{d \ln \mu} + 
\frac{d (Z^{\dagger}_{\phi})^{-1}}{d \ln \mu} Z_{\phi}^{\dagger},
\ee 
which gives the hermitian matrix of anomalous dimensions,
\bea
\gamma^{(1)} &=& \left. \tilde{X} \frac{d \tilde{X}^{-1}}{d \ln \mu}\right|_{\th}
= D_1 \gamma,\\
\gamma^{(2)} &=& \frac{d m^2}{d\ln \mu} = D_2 \gamma.
\eea
Here the differential operators are given explicitly by
\bea
D_1 &=& - h_{ijk} \frac{\partial}{\partial y_{ijk}}, \\
D_2 &=& D_1 \bar{D}_1 
+ \frac{1}{2}(m^2)_b^a \left(
y_{ajk} \frac{\partial}{\partial y_{bjk}} + 
y_{iak} \frac{\partial}{\partial y_{ibk}} + 
y_{ija} \frac{\partial}{\partial y_{ijb}} 
\right) \nn \\
& &+ \frac{1}{2}(m^2)_a^b \left(
\bar{y}^{ajk} \frac{\partial}{\partial \bar{y}^{bjk}} + 
\bar{y}^{iak} \frac{\partial}{\partial \bar{y}^{ibk}} + 
\bar{y}^{ija} \frac{\partial}{\partial \bar{y}^{ijb}} 
\right)
\eea

Now let us consider the RG equations for the $U(N)^c_{\phi}$ invariant
couplings $\tilde{y}$.
By taking derivative with respect to renormalization scale $\mu$ in
eq.~(\ref{ytilder}), we obtain
\be
\mu \frac{d \tilde{y}}{d \mu}=
\tilde{y} \left(
\tilde{X} \frac{d \tilde{X}^{-1}}{d \ln \mu} \otimes 1 \otimes 1 
+ 1 \otimes \tilde{X} \frac{d \tilde{X}^{-1}}{d \ln \mu} \otimes 1 
+ 1 \otimes 1 \otimes \tilde{X} \frac{d \tilde{X}^{-1}}{d \ln \mu} 
\right).
\ee
The RG equations for the soft SUSY breaking parameters and the Yukawa couplings
may be immediately obtained by expanding this equation with respect to $\th$ and $\thb$.
The results are written down neatly by using the differential operators as
\bea
\mu \frac{d y_{ijk}}{d \mu} &=&
\frac{1}{2} \left(
y_{i'jk} \gamma_i^{i'} +y_{ij'k} \gamma_j^{j'} +y_{ijk'} \gamma_k^{k'}
\right), \\
\mu \frac{d h_{ijk}}{d \mu} &=&
\frac{1}{2} \left(
h_{i'jk} \gamma_i^{i'} +h_{ij'k} \gamma_j^{j'} +h_{ijk'} \gamma_k^{k'}
\right) \nn \\
& & - \left(
y_{i'jk}D_1 \gamma_i^{i'} +y_{ij'k} D_1 \gamma_j^{j'} +y_{ijk'} D_1 \gamma_k^{k'}
\right), \\
\mu \frac{d (m^2)_i^j}{d \mu} &=&
D_2 \gamma_i^j.
\eea

\end{document}

%% file: s-spec.tex
\setlength{\unitlength}{0.240900pt}
\ifx\plotpoint\undefined\newsavebox{\plotpoint}\fi
\sbox{\plotpoint}{\rule[-0.200pt]{0.400pt}{0.400pt}}%
\begin{picture}(1500,900)(0,0)
\font\gnuplot=cmr10 at 10pt
\gnuplot
\sbox{\plotpoint}{\rule[-0.200pt]{0.400pt}{0.400pt}}%
\put(161.0,123.0){\rule[-0.200pt]{4.818pt}{0.400pt}}
\put(141,123){\makebox(0,0)[r]{0}}
\put(1419.0,123.0){\rule[-0.200pt]{4.818pt}{0.400pt}}
\put(161.0,197.0){\rule[-0.200pt]{4.818pt}{0.400pt}}
\put(141,197){\makebox(0,0)[r]{0.1}}
\put(1419.0,197.0){\rule[-0.200pt]{4.818pt}{0.400pt}}
\put(161.0,270.0){\rule[-0.200pt]{4.818pt}{0.400pt}}
\put(141,270){\makebox(0,0)[r]{0.2}}
\put(1419.0,270.0){\rule[-0.200pt]{4.818pt}{0.400pt}}
\put(161.0,344.0){\rule[-0.200pt]{4.818pt}{0.400pt}}
\put(141,344){\makebox(0,0)[r]{0.3}}
\put(1419.0,344.0){\rule[-0.200pt]{4.818pt}{0.400pt}}
\put(161.0,418.0){\rule[-0.200pt]{4.818pt}{0.400pt}}
\put(141,418){\makebox(0,0)[r]{0.4}}
\put(1419.0,418.0){\rule[-0.200pt]{4.818pt}{0.400pt}}
\put(161.0,492.0){\rule[-0.200pt]{4.818pt}{0.400pt}}
\put(141,492){\makebox(0,0)[r]{0.5}}
\put(1419.0,492.0){\rule[-0.200pt]{4.818pt}{0.400pt}}
\put(161.0,565.0){\rule[-0.200pt]{4.818pt}{0.400pt}}
\put(141,565){\makebox(0,0)[r]{0.6}}
\put(1419.0,565.0){\rule[-0.200pt]{4.818pt}{0.400pt}}
\put(161.0,639.0){\rule[-0.200pt]{4.818pt}{0.400pt}}
\put(141,639){\makebox(0,0)[r]{0.7}}
\put(1419.0,639.0){\rule[-0.200pt]{4.818pt}{0.400pt}}
\put(161.0,713.0){\rule[-0.200pt]{4.818pt}{0.400pt}}
\put(141,713){\makebox(0,0)[r]{0.8}}
\put(1419.0,713.0){\rule[-0.200pt]{4.818pt}{0.400pt}}
\put(161.0,786.0){\rule[-0.200pt]{4.818pt}{0.400pt}}
\put(141,786){\makebox(0,0)[r]{0.9}}
\put(1419.0,786.0){\rule[-0.200pt]{4.818pt}{0.400pt}}
\put(161.0,860.0){\rule[-0.200pt]{4.818pt}{0.400pt}}
\put(141,860){\makebox(0,0)[r]{1}}
\put(1419.0,860.0){\rule[-0.200pt]{4.818pt}{0.400pt}}
\put(161.0,123.0){\rule[-0.200pt]{0.400pt}{4.818pt}}
\put(161,82){\makebox(0,0){10}}
\put(161.0,840.0){\rule[-0.200pt]{0.400pt}{4.818pt}}
\put(374.0,123.0){\rule[-0.200pt]{0.400pt}{4.818pt}}
\put(374,82){\makebox(0,0){11}}
\put(374.0,840.0){\rule[-0.200pt]{0.400pt}{4.818pt}}
\put(587.0,123.0){\rule[-0.200pt]{0.400pt}{4.818pt}}
\put(587,82){\makebox(0,0){12}}
\put(587.0,840.0){\rule[-0.200pt]{0.400pt}{4.818pt}}
\put(800.0,123.0){\rule[-0.200pt]{0.400pt}{4.818pt}}
\put(800,82){\makebox(0,0){13}}
\put(800.0,840.0){\rule[-0.200pt]{0.400pt}{4.818pt}}
\put(1013.0,123.0){\rule[-0.200pt]{0.400pt}{4.818pt}}
\put(1013,82){\makebox(0,0){14}}
\put(1013.0,840.0){\rule[-0.200pt]{0.400pt}{4.818pt}}
\put(1226.0,123.0){\rule[-0.200pt]{0.400pt}{4.818pt}}
\put(1226,82){\makebox(0,0){15}}
\put(1226.0,840.0){\rule[-0.200pt]{0.400pt}{4.818pt}}
\put(1439.0,123.0){\rule[-0.200pt]{0.400pt}{4.818pt}}
\put(1439,82){\makebox(0,0){16}}
\put(1439.0,840.0){\rule[-0.200pt]{0.400pt}{4.818pt}}
\put(161.0,123.0){\rule[-0.200pt]{307.870pt}{0.400pt}}
\put(1439.0,123.0){\rule[-0.200pt]{0.400pt}{177.543pt}}
\put(161.0,860.0){\rule[-0.200pt]{307.870pt}{0.400pt}}
\put(5,451){\makebox(0,0){$m_{\tilde f}/M_3$}}
\put(800,21){\makebox(0,0){$\log_{10} (M_C)$ [GeV]}}
\put(1226,713){\makebox(0,0)[r]{$\tilde Q$}}
\put(1226,160){\makebox(0,0)[r]{$\tilde e$}}
\put(1226,418){\makebox(0,0)[r]{$\tilde L$}}
\put(161.0,123.0){\rule[-0.200pt]{0.400pt}{177.543pt}}
\sbox{\plotpoint}{\rule[-0.400pt]{0.800pt}{0.800pt}}%
\put(161,742){\usebox{\plotpoint}}
\put(161,740.84){\rule{3.132pt}{0.800pt}}
\multiput(161.00,740.34)(6.500,1.000){2}{\rule{1.566pt}{0.800pt}}
\put(174,741.84){\rule{3.132pt}{0.800pt}}
\multiput(174.00,741.34)(6.500,1.000){2}{\rule{1.566pt}{0.800pt}}
\put(200,742.84){\rule{3.132pt}{0.800pt}}
\multiput(200.00,742.34)(6.500,1.000){2}{\rule{1.566pt}{0.800pt}}
\put(213,743.84){\rule{3.132pt}{0.800pt}}
\multiput(213.00,743.34)(6.500,1.000){2}{\rule{1.566pt}{0.800pt}}
\put(226,744.84){\rule{2.891pt}{0.800pt}}
\multiput(226.00,744.34)(6.000,1.000){2}{\rule{1.445pt}{0.800pt}}
\put(238,745.84){\rule{3.132pt}{0.800pt}}
\multiput(238.00,745.34)(6.500,1.000){2}{\rule{1.566pt}{0.800pt}}
\put(187.0,744.0){\rule[-0.400pt]{3.132pt}{0.800pt}}
\put(264,746.84){\rule{3.132pt}{0.800pt}}
\multiput(264.00,746.34)(6.500,1.000){2}{\rule{1.566pt}{0.800pt}}
\put(277,747.84){\rule{3.132pt}{0.800pt}}
\multiput(277.00,747.34)(6.500,1.000){2}{\rule{1.566pt}{0.800pt}}
\put(290,748.84){\rule{3.132pt}{0.800pt}}
\multiput(290.00,748.34)(6.500,1.000){2}{\rule{1.566pt}{0.800pt}}
\put(251.0,748.0){\rule[-0.400pt]{3.132pt}{0.800pt}}
\put(316,749.84){\rule{3.132pt}{0.800pt}}
\multiput(316.00,749.34)(6.500,1.000){2}{\rule{1.566pt}{0.800pt}}
\put(329,750.84){\rule{3.132pt}{0.800pt}}
\multiput(329.00,750.34)(6.500,1.000){2}{\rule{1.566pt}{0.800pt}}
\put(342,751.84){\rule{3.132pt}{0.800pt}}
\multiput(342.00,751.34)(6.500,1.000){2}{\rule{1.566pt}{0.800pt}}
\put(303.0,751.0){\rule[-0.400pt]{3.132pt}{0.800pt}}
\put(368,752.84){\rule{2.891pt}{0.800pt}}
\multiput(368.00,752.34)(6.000,1.000){2}{\rule{1.445pt}{0.800pt}}
\put(380,753.84){\rule{3.132pt}{0.800pt}}
\multiput(380.00,753.34)(6.500,1.000){2}{\rule{1.566pt}{0.800pt}}
\put(355.0,754.0){\rule[-0.400pt]{3.132pt}{0.800pt}}
\put(406,754.84){\rule{3.132pt}{0.800pt}}
\multiput(406.00,754.34)(6.500,1.000){2}{\rule{1.566pt}{0.800pt}}
\put(419,755.84){\rule{3.132pt}{0.800pt}}
\multiput(419.00,755.34)(6.500,1.000){2}{\rule{1.566pt}{0.800pt}}
\put(393.0,756.0){\rule[-0.400pt]{3.132pt}{0.800pt}}
\put(445,756.84){\rule{3.132pt}{0.800pt}}
\multiput(445.00,756.34)(6.500,1.000){2}{\rule{1.566pt}{0.800pt}}
\put(458,757.84){\rule{3.132pt}{0.800pt}}
\multiput(458.00,757.34)(6.500,1.000){2}{\rule{1.566pt}{0.800pt}}
\put(432.0,758.0){\rule[-0.400pt]{3.132pt}{0.800pt}}
\put(484,758.84){\rule{3.132pt}{0.800pt}}
\multiput(484.00,758.34)(6.500,1.000){2}{\rule{1.566pt}{0.800pt}}
\put(497,759.84){\rule{3.132pt}{0.800pt}}
\multiput(497.00,759.34)(6.500,1.000){2}{\rule{1.566pt}{0.800pt}}
\put(471.0,760.0){\rule[-0.400pt]{3.132pt}{0.800pt}}
\put(522,760.84){\rule{3.132pt}{0.800pt}}
\multiput(522.00,760.34)(6.500,1.000){2}{\rule{1.566pt}{0.800pt}}
\put(510.0,762.0){\rule[-0.400pt]{2.891pt}{0.800pt}}
\put(548,761.84){\rule{3.132pt}{0.800pt}}
\multiput(548.00,761.34)(6.500,1.000){2}{\rule{1.566pt}{0.800pt}}
\put(561,762.84){\rule{3.132pt}{0.800pt}}
\multiput(561.00,762.34)(6.500,1.000){2}{\rule{1.566pt}{0.800pt}}
\put(535.0,763.0){\rule[-0.400pt]{3.132pt}{0.800pt}}
\put(587,763.84){\rule{3.132pt}{0.800pt}}
\multiput(587.00,763.34)(6.500,1.000){2}{\rule{1.566pt}{0.800pt}}
\put(574.0,765.0){\rule[-0.400pt]{3.132pt}{0.800pt}}
\put(613,764.84){\rule{3.132pt}{0.800pt}}
\multiput(613.00,764.34)(6.500,1.000){2}{\rule{1.566pt}{0.800pt}}
\put(626,765.84){\rule{3.132pt}{0.800pt}}
\multiput(626.00,765.34)(6.500,1.000){2}{\rule{1.566pt}{0.800pt}}
\put(600.0,766.0){\rule[-0.400pt]{3.132pt}{0.800pt}}
\put(652,766.84){\rule{2.891pt}{0.800pt}}
\multiput(652.00,766.34)(6.000,1.000){2}{\rule{1.445pt}{0.800pt}}
\put(639.0,768.0){\rule[-0.400pt]{3.132pt}{0.800pt}}
\put(677,767.84){\rule{3.132pt}{0.800pt}}
\multiput(677.00,767.34)(6.500,1.000){2}{\rule{1.566pt}{0.800pt}}
\put(664.0,769.0){\rule[-0.400pt]{3.132pt}{0.800pt}}
\put(703,768.84){\rule{3.132pt}{0.800pt}}
\multiput(703.00,768.34)(6.500,1.000){2}{\rule{1.566pt}{0.800pt}}
\put(716,769.84){\rule{3.132pt}{0.800pt}}
\multiput(716.00,769.34)(6.500,1.000){2}{\rule{1.566pt}{0.800pt}}
\put(690.0,770.0){\rule[-0.400pt]{3.132pt}{0.800pt}}
\put(742,770.84){\rule{3.132pt}{0.800pt}}
\multiput(742.00,770.34)(6.500,1.000){2}{\rule{1.566pt}{0.800pt}}
\put(729.0,772.0){\rule[-0.400pt]{3.132pt}{0.800pt}}
\put(768,771.84){\rule{3.132pt}{0.800pt}}
\multiput(768.00,771.34)(6.500,1.000){2}{\rule{1.566pt}{0.800pt}}
\put(755.0,773.0){\rule[-0.400pt]{3.132pt}{0.800pt}}
\put(794,772.84){\rule{2.891pt}{0.800pt}}
\multiput(794.00,772.34)(6.000,1.000){2}{\rule{1.445pt}{0.800pt}}
\put(781.0,774.0){\rule[-0.400pt]{3.132pt}{0.800pt}}
\put(819,773.84){\rule{3.132pt}{0.800pt}}
\multiput(819.00,773.34)(6.500,1.000){2}{\rule{1.566pt}{0.800pt}}
\put(806.0,775.0){\rule[-0.400pt]{3.132pt}{0.800pt}}
\put(845,774.84){\rule{3.132pt}{0.800pt}}
\multiput(845.00,774.34)(6.500,1.000){2}{\rule{1.566pt}{0.800pt}}
\put(832.0,776.0){\rule[-0.400pt]{3.132pt}{0.800pt}}
\put(871,775.84){\rule{3.132pt}{0.800pt}}
\multiput(871.00,775.34)(6.500,1.000){2}{\rule{1.566pt}{0.800pt}}
\put(858.0,777.0){\rule[-0.400pt]{3.132pt}{0.800pt}}
\put(897,776.84){\rule{3.132pt}{0.800pt}}
\multiput(897.00,776.34)(6.500,1.000){2}{\rule{1.566pt}{0.800pt}}
\put(884.0,778.0){\rule[-0.400pt]{3.132pt}{0.800pt}}
\put(923,777.84){\rule{3.132pt}{0.800pt}}
\multiput(923.00,777.34)(6.500,1.000){2}{\rule{1.566pt}{0.800pt}}
\put(910.0,779.0){\rule[-0.400pt]{3.132pt}{0.800pt}}
\put(948,778.84){\rule{3.132pt}{0.800pt}}
\multiput(948.00,778.34)(6.500,1.000){2}{\rule{1.566pt}{0.800pt}}
\put(936.0,780.0){\rule[-0.400pt]{2.891pt}{0.800pt}}
\put(974,779.84){\rule{3.132pt}{0.800pt}}
\multiput(974.00,779.34)(6.500,1.000){2}{\rule{1.566pt}{0.800pt}}
\put(961.0,781.0){\rule[-0.400pt]{3.132pt}{0.800pt}}
\put(1000,780.84){\rule{3.132pt}{0.800pt}}
\multiput(1000.00,780.34)(6.500,1.000){2}{\rule{1.566pt}{0.800pt}}
\put(987.0,782.0){\rule[-0.400pt]{3.132pt}{0.800pt}}
\put(1026,781.84){\rule{3.132pt}{0.800pt}}
\multiput(1026.00,781.34)(6.500,1.000){2}{\rule{1.566pt}{0.800pt}}
\put(1013.0,783.0){\rule[-0.400pt]{3.132pt}{0.800pt}}
\put(1065,782.84){\rule{3.132pt}{0.800pt}}
\multiput(1065.00,782.34)(6.500,1.000){2}{\rule{1.566pt}{0.800pt}}
\put(1039.0,784.0){\rule[-0.400pt]{6.263pt}{0.800pt}}
\put(1090,783.84){\rule{3.132pt}{0.800pt}}
\multiput(1090.00,783.34)(6.500,1.000){2}{\rule{1.566pt}{0.800pt}}
\put(1078.0,785.0){\rule[-0.400pt]{2.891pt}{0.800pt}}
\put(1116,784.84){\rule{3.132pt}{0.800pt}}
\multiput(1116.00,784.34)(6.500,1.000){2}{\rule{1.566pt}{0.800pt}}
\put(1103.0,786.0){\rule[-0.400pt]{3.132pt}{0.800pt}}
\put(1142,785.84){\rule{3.132pt}{0.800pt}}
\multiput(1142.00,785.34)(6.500,1.000){2}{\rule{1.566pt}{0.800pt}}
\put(1129.0,787.0){\rule[-0.400pt]{3.132pt}{0.800pt}}
\put(1181,786.84){\rule{3.132pt}{0.800pt}}
\multiput(1181.00,786.34)(6.500,1.000){2}{\rule{1.566pt}{0.800pt}}
\put(1155.0,788.0){\rule[-0.400pt]{6.263pt}{0.800pt}}
\put(1207,787.84){\rule{3.132pt}{0.800pt}}
\multiput(1207.00,787.34)(6.500,1.000){2}{\rule{1.566pt}{0.800pt}}
\put(1194.0,789.0){\rule[-0.400pt]{3.132pt}{0.800pt}}
\put(1245,788.84){\rule{3.132pt}{0.800pt}}
\multiput(1245.00,788.34)(6.500,1.000){2}{\rule{1.566pt}{0.800pt}}
\put(1220.0,790.0){\rule[-0.400pt]{6.022pt}{0.800pt}}
\put(1271,789.84){\rule{3.132pt}{0.800pt}}
\multiput(1271.00,789.34)(6.500,1.000){2}{\rule{1.566pt}{0.800pt}}
\put(1258.0,791.0){\rule[-0.400pt]{3.132pt}{0.800pt}}
\put(1297,790.84){\rule{3.132pt}{0.800pt}}
\multiput(1297.00,790.34)(6.500,1.000){2}{\rule{1.566pt}{0.800pt}}
\put(1284.0,792.0){\rule[-0.400pt]{3.132pt}{0.800pt}}
\put(1336,791.84){\rule{3.132pt}{0.800pt}}
\multiput(1336.00,791.34)(6.500,1.000){2}{\rule{1.566pt}{0.800pt}}
\put(1310.0,793.0){\rule[-0.400pt]{6.263pt}{0.800pt}}
\put(1362,792.84){\rule{2.891pt}{0.800pt}}
\multiput(1362.00,792.34)(6.000,1.000){2}{\rule{1.445pt}{0.800pt}}
\put(1349.0,794.0){\rule[-0.400pt]{3.132pt}{0.800pt}}
\put(1400,793.84){\rule{3.132pt}{0.800pt}}
\multiput(1400.00,793.34)(6.500,1.000){2}{\rule{1.566pt}{0.800pt}}
\put(1374.0,795.0){\rule[-0.400pt]{6.263pt}{0.800pt}}
\put(1413.0,796.0){\rule[-0.400pt]{6.263pt}{0.800pt}}
\put(161,250){\usebox{\plotpoint}}
\put(174,248.84){\rule{3.132pt}{0.800pt}}
\multiput(174.00,248.34)(6.500,1.000){2}{\rule{1.566pt}{0.800pt}}
\put(161.0,250.0){\rule[-0.400pt]{3.132pt}{0.800pt}}
\put(200,249.84){\rule{3.132pt}{0.800pt}}
\multiput(200.00,249.34)(6.500,1.000){2}{\rule{1.566pt}{0.800pt}}
\put(213,250.84){\rule{3.132pt}{0.800pt}}
\multiput(213.00,250.34)(6.500,1.000){2}{\rule{1.566pt}{0.800pt}}
\put(187.0,251.0){\rule[-0.400pt]{3.132pt}{0.800pt}}
\put(238,251.84){\rule{3.132pt}{0.800pt}}
\multiput(238.00,251.34)(6.500,1.000){2}{\rule{1.566pt}{0.800pt}}
\put(226.0,253.0){\rule[-0.400pt]{2.891pt}{0.800pt}}
\put(264,252.84){\rule{3.132pt}{0.800pt}}
\multiput(264.00,252.34)(6.500,1.000){2}{\rule{1.566pt}{0.800pt}}
\put(251.0,254.0){\rule[-0.400pt]{3.132pt}{0.800pt}}
\put(290,253.84){\rule{3.132pt}{0.800pt}}
\multiput(290.00,253.34)(6.500,1.000){2}{\rule{1.566pt}{0.800pt}}
\put(303,254.84){\rule{3.132pt}{0.800pt}}
\multiput(303.00,254.34)(6.500,1.000){2}{\rule{1.566pt}{0.800pt}}
\put(277.0,255.0){\rule[-0.400pt]{3.132pt}{0.800pt}}
\put(329,255.84){\rule{3.132pt}{0.800pt}}
\multiput(329.00,255.34)(6.500,1.000){2}{\rule{1.566pt}{0.800pt}}
\put(316.0,257.0){\rule[-0.400pt]{3.132pt}{0.800pt}}
\put(355,256.84){\rule{3.132pt}{0.800pt}}
\multiput(355.00,256.34)(6.500,1.000){2}{\rule{1.566pt}{0.800pt}}
\put(342.0,258.0){\rule[-0.400pt]{3.132pt}{0.800pt}}
\put(380,257.84){\rule{3.132pt}{0.800pt}}
\multiput(380.00,257.34)(6.500,1.000){2}{\rule{1.566pt}{0.800pt}}
\put(368.0,259.0){\rule[-0.400pt]{2.891pt}{0.800pt}}
\put(406,258.84){\rule{3.132pt}{0.800pt}}
\multiput(406.00,258.34)(6.500,1.000){2}{\rule{1.566pt}{0.800pt}}
\put(419,259.84){\rule{3.132pt}{0.800pt}}
\multiput(419.00,259.34)(6.500,1.000){2}{\rule{1.566pt}{0.800pt}}
\put(393.0,260.0){\rule[-0.400pt]{3.132pt}{0.800pt}}
\put(445,260.84){\rule{3.132pt}{0.800pt}}
\multiput(445.00,260.34)(6.500,1.000){2}{\rule{1.566pt}{0.800pt}}
\put(432.0,262.0){\rule[-0.400pt]{3.132pt}{0.800pt}}
\put(471,261.84){\rule{3.132pt}{0.800pt}}
\multiput(471.00,261.34)(6.500,1.000){2}{\rule{1.566pt}{0.800pt}}
\put(458.0,263.0){\rule[-0.400pt]{3.132pt}{0.800pt}}
\put(497,262.84){\rule{3.132pt}{0.800pt}}
\multiput(497.00,262.34)(6.500,1.000){2}{\rule{1.566pt}{0.800pt}}
\put(484.0,264.0){\rule[-0.400pt]{3.132pt}{0.800pt}}
\put(522,263.84){\rule{3.132pt}{0.800pt}}
\multiput(522.00,263.34)(6.500,1.000){2}{\rule{1.566pt}{0.800pt}}
\put(535,264.84){\rule{3.132pt}{0.800pt}}
\multiput(535.00,264.34)(6.500,1.000){2}{\rule{1.566pt}{0.800pt}}
\put(510.0,265.0){\rule[-0.400pt]{2.891pt}{0.800pt}}
\put(561,265.84){\rule{3.132pt}{0.800pt}}
\multiput(561.00,265.34)(6.500,1.000){2}{\rule{1.566pt}{0.800pt}}
\put(548.0,267.0){\rule[-0.400pt]{3.132pt}{0.800pt}}
\put(587,266.84){\rule{3.132pt}{0.800pt}}
\multiput(587.00,266.34)(6.500,1.000){2}{\rule{1.566pt}{0.800pt}}
\put(574.0,268.0){\rule[-0.400pt]{3.132pt}{0.800pt}}
\put(613,267.84){\rule{3.132pt}{0.800pt}}
\multiput(613.00,267.34)(6.500,1.000){2}{\rule{1.566pt}{0.800pt}}
\put(600.0,269.0){\rule[-0.400pt]{3.132pt}{0.800pt}}
\put(639,268.84){\rule{3.132pt}{0.800pt}}
\multiput(639.00,268.34)(6.500,1.000){2}{\rule{1.566pt}{0.800pt}}
\put(652,269.84){\rule{2.891pt}{0.800pt}}
\multiput(652.00,269.34)(6.000,1.000){2}{\rule{1.445pt}{0.800pt}}
\put(626.0,270.0){\rule[-0.400pt]{3.132pt}{0.800pt}}
\put(677,270.84){\rule{3.132pt}{0.800pt}}
\multiput(677.00,270.34)(6.500,1.000){2}{\rule{1.566pt}{0.800pt}}
\put(664.0,272.0){\rule[-0.400pt]{3.132pt}{0.800pt}}
\put(703,271.84){\rule{3.132pt}{0.800pt}}
\multiput(703.00,271.34)(6.500,1.000){2}{\rule{1.566pt}{0.800pt}}
\put(690.0,273.0){\rule[-0.400pt]{3.132pt}{0.800pt}}
\put(729,272.84){\rule{3.132pt}{0.800pt}}
\multiput(729.00,272.34)(6.500,1.000){2}{\rule{1.566pt}{0.800pt}}
\put(716.0,274.0){\rule[-0.400pt]{3.132pt}{0.800pt}}
\put(755,273.84){\rule{3.132pt}{0.800pt}}
\multiput(755.00,273.34)(6.500,1.000){2}{\rule{1.566pt}{0.800pt}}
\put(742.0,275.0){\rule[-0.400pt]{3.132pt}{0.800pt}}
\put(781,274.84){\rule{3.132pt}{0.800pt}}
\multiput(781.00,274.34)(6.500,1.000){2}{\rule{1.566pt}{0.800pt}}
\put(794,275.84){\rule{2.891pt}{0.800pt}}
\multiput(794.00,275.34)(6.000,1.000){2}{\rule{1.445pt}{0.800pt}}
\put(768.0,276.0){\rule[-0.400pt]{3.132pt}{0.800pt}}
\put(819,276.84){\rule{3.132pt}{0.800pt}}
\multiput(819.00,276.34)(6.500,1.000){2}{\rule{1.566pt}{0.800pt}}
\put(806.0,278.0){\rule[-0.400pt]{3.132pt}{0.800pt}}
\put(845,277.84){\rule{3.132pt}{0.800pt}}
\multiput(845.00,277.34)(6.500,1.000){2}{\rule{1.566pt}{0.800pt}}
\put(832.0,279.0){\rule[-0.400pt]{3.132pt}{0.800pt}}
\put(871,278.84){\rule{3.132pt}{0.800pt}}
\multiput(871.00,278.34)(6.500,1.000){2}{\rule{1.566pt}{0.800pt}}
\put(858.0,280.0){\rule[-0.400pt]{3.132pt}{0.800pt}}
\put(897,279.84){\rule{3.132pt}{0.800pt}}
\multiput(897.00,279.34)(6.500,1.000){2}{\rule{1.566pt}{0.800pt}}
\put(910,280.84){\rule{3.132pt}{0.800pt}}
\multiput(910.00,280.34)(6.500,1.000){2}{\rule{1.566pt}{0.800pt}}
\put(884.0,281.0){\rule[-0.400pt]{3.132pt}{0.800pt}}
\put(936,281.84){\rule{2.891pt}{0.800pt}}
\multiput(936.00,281.34)(6.000,1.000){2}{\rule{1.445pt}{0.800pt}}
\put(923.0,283.0){\rule[-0.400pt]{3.132pt}{0.800pt}}
\put(961,282.84){\rule{3.132pt}{0.800pt}}
\multiput(961.00,282.34)(6.500,1.000){2}{\rule{1.566pt}{0.800pt}}
\put(948.0,284.0){\rule[-0.400pt]{3.132pt}{0.800pt}}
\put(987,283.84){\rule{3.132pt}{0.800pt}}
\multiput(987.00,283.34)(6.500,1.000){2}{\rule{1.566pt}{0.800pt}}
\put(974.0,285.0){\rule[-0.400pt]{3.132pt}{0.800pt}}
\put(1013,284.84){\rule{3.132pt}{0.800pt}}
\multiput(1013.00,284.34)(6.500,1.000){2}{\rule{1.566pt}{0.800pt}}
\put(1026,285.84){\rule{3.132pt}{0.800pt}}
\multiput(1026.00,285.34)(6.500,1.000){2}{\rule{1.566pt}{0.800pt}}
\put(1000.0,286.0){\rule[-0.400pt]{3.132pt}{0.800pt}}
\put(1052,286.84){\rule{3.132pt}{0.800pt}}
\multiput(1052.00,286.34)(6.500,1.000){2}{\rule{1.566pt}{0.800pt}}
\put(1039.0,288.0){\rule[-0.400pt]{3.132pt}{0.800pt}}
\put(1078,287.84){\rule{2.891pt}{0.800pt}}
\multiput(1078.00,287.34)(6.000,1.000){2}{\rule{1.445pt}{0.800pt}}
\put(1065.0,289.0){\rule[-0.400pt]{3.132pt}{0.800pt}}
\put(1103,288.84){\rule{3.132pt}{0.800pt}}
\multiput(1103.00,288.34)(6.500,1.000){2}{\rule{1.566pt}{0.800pt}}
\put(1090.0,290.0){\rule[-0.400pt]{3.132pt}{0.800pt}}
\put(1129,289.84){\rule{3.132pt}{0.800pt}}
\multiput(1129.00,289.34)(6.500,1.000){2}{\rule{1.566pt}{0.800pt}}
\put(1142,290.84){\rule{3.132pt}{0.800pt}}
\multiput(1142.00,290.34)(6.500,1.000){2}{\rule{1.566pt}{0.800pt}}
\put(1116.0,291.0){\rule[-0.400pt]{3.132pt}{0.800pt}}
\put(1168,291.84){\rule{3.132pt}{0.800pt}}
\multiput(1168.00,291.34)(6.500,1.000){2}{\rule{1.566pt}{0.800pt}}
\put(1155.0,293.0){\rule[-0.400pt]{3.132pt}{0.800pt}}
\put(1194,292.84){\rule{3.132pt}{0.800pt}}
\multiput(1194.00,292.34)(6.500,1.000){2}{\rule{1.566pt}{0.800pt}}
\put(1181.0,294.0){\rule[-0.400pt]{3.132pt}{0.800pt}}
\put(1220,293.84){\rule{2.891pt}{0.800pt}}
\multiput(1220.00,293.34)(6.000,1.000){2}{\rule{1.445pt}{0.800pt}}
\put(1207.0,295.0){\rule[-0.400pt]{3.132pt}{0.800pt}}
\put(1245,294.84){\rule{3.132pt}{0.800pt}}
\multiput(1245.00,294.34)(6.500,1.000){2}{\rule{1.566pt}{0.800pt}}
\put(1258,295.84){\rule{3.132pt}{0.800pt}}
\multiput(1258.00,295.34)(6.500,1.000){2}{\rule{1.566pt}{0.800pt}}
\put(1232.0,296.0){\rule[-0.400pt]{3.132pt}{0.800pt}}
\put(1284,296.84){\rule{3.132pt}{0.800pt}}
\multiput(1284.00,296.34)(6.500,1.000){2}{\rule{1.566pt}{0.800pt}}
\put(1271.0,298.0){\rule[-0.400pt]{3.132pt}{0.800pt}}
\put(1310,297.84){\rule{3.132pt}{0.800pt}}
\multiput(1310.00,297.34)(6.500,1.000){2}{\rule{1.566pt}{0.800pt}}
\put(1297.0,299.0){\rule[-0.400pt]{3.132pt}{0.800pt}}
\put(1336,298.84){\rule{3.132pt}{0.800pt}}
\multiput(1336.00,298.34)(6.500,1.000){2}{\rule{1.566pt}{0.800pt}}
\put(1349,299.84){\rule{3.132pt}{0.800pt}}
\multiput(1349.00,299.34)(6.500,1.000){2}{\rule{1.566pt}{0.800pt}}
\put(1323.0,300.0){\rule[-0.400pt]{3.132pt}{0.800pt}}
\put(1374,300.84){\rule{3.132pt}{0.800pt}}
\multiput(1374.00,300.34)(6.500,1.000){2}{\rule{1.566pt}{0.800pt}}
\put(1362.0,302.0){\rule[-0.400pt]{2.891pt}{0.800pt}}
\put(1400,301.84){\rule{3.132pt}{0.800pt}}
\multiput(1400.00,301.34)(6.500,1.000){2}{\rule{1.566pt}{0.800pt}}
\put(1387.0,303.0){\rule[-0.400pt]{3.132pt}{0.800pt}}
\put(1426,302.84){\rule{3.132pt}{0.800pt}}
\multiput(1426.00,302.34)(6.500,1.000){2}{\rule{1.566pt}{0.800pt}}
\put(1413.0,304.0){\rule[-0.400pt]{3.132pt}{0.800pt}}
\put(161,173){\usebox{\plotpoint}}
\put(174,171.84){\rule{3.132pt}{0.800pt}}
\multiput(174.00,171.34)(6.500,1.000){2}{\rule{1.566pt}{0.800pt}}
\put(161.0,173.0){\rule[-0.400pt]{3.132pt}{0.800pt}}
\put(213,172.84){\rule{3.132pt}{0.800pt}}
\multiput(213.00,172.34)(6.500,1.000){2}{\rule{1.566pt}{0.800pt}}
\put(187.0,174.0){\rule[-0.400pt]{6.263pt}{0.800pt}}
\put(251,173.84){\rule{3.132pt}{0.800pt}}
\multiput(251.00,173.34)(6.500,1.000){2}{\rule{1.566pt}{0.800pt}}
\put(226.0,175.0){\rule[-0.400pt]{6.022pt}{0.800pt}}
\put(290,174.84){\rule{3.132pt}{0.800pt}}
\multiput(290.00,174.34)(6.500,1.000){2}{\rule{1.566pt}{0.800pt}}
\put(264.0,176.0){\rule[-0.400pt]{6.263pt}{0.800pt}}
\put(329,175.84){\rule{3.132pt}{0.800pt}}
\multiput(329.00,175.34)(6.500,1.000){2}{\rule{1.566pt}{0.800pt}}
\put(303.0,177.0){\rule[-0.400pt]{6.263pt}{0.800pt}}
\put(368,176.84){\rule{2.891pt}{0.800pt}}
\multiput(368.00,176.34)(6.000,1.000){2}{\rule{1.445pt}{0.800pt}}
\put(342.0,178.0){\rule[-0.400pt]{6.263pt}{0.800pt}}
\put(393,177.84){\rule{3.132pt}{0.800pt}}
\multiput(393.00,177.34)(6.500,1.000){2}{\rule{1.566pt}{0.800pt}}
\put(380.0,179.0){\rule[-0.400pt]{3.132pt}{0.800pt}}
\put(432,178.84){\rule{3.132pt}{0.800pt}}
\multiput(432.00,178.34)(6.500,1.000){2}{\rule{1.566pt}{0.800pt}}
\put(406.0,180.0){\rule[-0.400pt]{6.263pt}{0.800pt}}
\put(471,179.84){\rule{3.132pt}{0.800pt}}
\multiput(471.00,179.34)(6.500,1.000){2}{\rule{1.566pt}{0.800pt}}
\put(445.0,181.0){\rule[-0.400pt]{6.263pt}{0.800pt}}
\put(497,180.84){\rule{3.132pt}{0.800pt}}
\multiput(497.00,180.34)(6.500,1.000){2}{\rule{1.566pt}{0.800pt}}
\put(484.0,182.0){\rule[-0.400pt]{3.132pt}{0.800pt}}
\put(535,181.84){\rule{3.132pt}{0.800pt}}
\multiput(535.00,181.34)(6.500,1.000){2}{\rule{1.566pt}{0.800pt}}
\put(510.0,183.0){\rule[-0.400pt]{6.022pt}{0.800pt}}
\put(574,182.84){\rule{3.132pt}{0.800pt}}
\multiput(574.00,182.34)(6.500,1.000){2}{\rule{1.566pt}{0.800pt}}
\put(548.0,184.0){\rule[-0.400pt]{6.263pt}{0.800pt}}
\put(600,183.84){\rule{3.132pt}{0.800pt}}
\multiput(600.00,183.34)(6.500,1.000){2}{\rule{1.566pt}{0.800pt}}
\put(587.0,185.0){\rule[-0.400pt]{3.132pt}{0.800pt}}
\put(639,184.84){\rule{3.132pt}{0.800pt}}
\multiput(639.00,184.34)(6.500,1.000){2}{\rule{1.566pt}{0.800pt}}
\put(613.0,186.0){\rule[-0.400pt]{6.263pt}{0.800pt}}
\put(664,185.84){\rule{3.132pt}{0.800pt}}
\multiput(664.00,185.34)(6.500,1.000){2}{\rule{1.566pt}{0.800pt}}
\put(652.0,187.0){\rule[-0.400pt]{2.891pt}{0.800pt}}
\put(690,186.84){\rule{3.132pt}{0.800pt}}
\multiput(690.00,186.34)(6.500,1.000){2}{\rule{1.566pt}{0.800pt}}
\put(677.0,188.0){\rule[-0.400pt]{3.132pt}{0.800pt}}
\put(729,187.84){\rule{3.132pt}{0.800pt}}
\multiput(729.00,187.34)(6.500,1.000){2}{\rule{1.566pt}{0.800pt}}
\put(703.0,189.0){\rule[-0.400pt]{6.263pt}{0.800pt}}
\put(755,188.84){\rule{3.132pt}{0.800pt}}
\multiput(755.00,188.34)(6.500,1.000){2}{\rule{1.566pt}{0.800pt}}
\put(742.0,190.0){\rule[-0.400pt]{3.132pt}{0.800pt}}
\put(781,189.84){\rule{3.132pt}{0.800pt}}
\multiput(781.00,189.34)(6.500,1.000){2}{\rule{1.566pt}{0.800pt}}
\put(768.0,191.0){\rule[-0.400pt]{3.132pt}{0.800pt}}
\put(806,190.84){\rule{3.132pt}{0.800pt}}
\multiput(806.00,190.34)(6.500,1.000){2}{\rule{1.566pt}{0.800pt}}
\put(794.0,192.0){\rule[-0.400pt]{2.891pt}{0.800pt}}
\put(845,191.84){\rule{3.132pt}{0.800pt}}
\multiput(845.00,191.34)(6.500,1.000){2}{\rule{1.566pt}{0.800pt}}
\put(819.0,193.0){\rule[-0.400pt]{6.263pt}{0.800pt}}
\put(871,192.84){\rule{3.132pt}{0.800pt}}
\multiput(871.00,192.34)(6.500,1.000){2}{\rule{1.566pt}{0.800pt}}
\put(858.0,194.0){\rule[-0.400pt]{3.132pt}{0.800pt}}
\put(897,193.84){\rule{3.132pt}{0.800pt}}
\multiput(897.00,193.34)(6.500,1.000){2}{\rule{1.566pt}{0.800pt}}
\put(884.0,195.0){\rule[-0.400pt]{3.132pt}{0.800pt}}
\put(923,194.84){\rule{3.132pt}{0.800pt}}
\multiput(923.00,194.34)(6.500,1.000){2}{\rule{1.566pt}{0.800pt}}
\put(910.0,196.0){\rule[-0.400pt]{3.132pt}{0.800pt}}
\put(948,195.84){\rule{3.132pt}{0.800pt}}
\multiput(948.00,195.34)(6.500,1.000){2}{\rule{1.566pt}{0.800pt}}
\put(936.0,197.0){\rule[-0.400pt]{2.891pt}{0.800pt}}
\put(974,196.84){\rule{3.132pt}{0.800pt}}
\multiput(974.00,196.34)(6.500,1.000){2}{\rule{1.566pt}{0.800pt}}
\put(961.0,198.0){\rule[-0.400pt]{3.132pt}{0.800pt}}
\put(1000,197.84){\rule{3.132pt}{0.800pt}}
\multiput(1000.00,197.34)(6.500,1.000){2}{\rule{1.566pt}{0.800pt}}
\put(987.0,199.0){\rule[-0.400pt]{3.132pt}{0.800pt}}
\put(1026,198.84){\rule{3.132pt}{0.800pt}}
\multiput(1026.00,198.34)(6.500,1.000){2}{\rule{1.566pt}{0.800pt}}
\put(1013.0,200.0){\rule[-0.400pt]{3.132pt}{0.800pt}}
\put(1052,199.84){\rule{3.132pt}{0.800pt}}
\multiput(1052.00,199.34)(6.500,1.000){2}{\rule{1.566pt}{0.800pt}}
\put(1039.0,201.0){\rule[-0.400pt]{3.132pt}{0.800pt}}
\put(1078,200.84){\rule{2.891pt}{0.800pt}}
\multiput(1078.00,200.34)(6.000,1.000){2}{\rule{1.445pt}{0.800pt}}
\put(1065.0,202.0){\rule[-0.400pt]{3.132pt}{0.800pt}}
\put(1103,201.84){\rule{3.132pt}{0.800pt}}
\multiput(1103.00,201.34)(6.500,1.000){2}{\rule{1.566pt}{0.800pt}}
\put(1090.0,203.0){\rule[-0.400pt]{3.132pt}{0.800pt}}
\put(1129,202.84){\rule{3.132pt}{0.800pt}}
\multiput(1129.00,202.34)(6.500,1.000){2}{\rule{1.566pt}{0.800pt}}
\put(1142,203.84){\rule{3.132pt}{0.800pt}}
\multiput(1142.00,203.34)(6.500,1.000){2}{\rule{1.566pt}{0.800pt}}
\put(1116.0,204.0){\rule[-0.400pt]{3.132pt}{0.800pt}}
\put(1168,204.84){\rule{3.132pt}{0.800pt}}
\multiput(1168.00,204.34)(6.500,1.000){2}{\rule{1.566pt}{0.800pt}}
\put(1155.0,206.0){\rule[-0.400pt]{3.132pt}{0.800pt}}
\put(1194,205.84){\rule{3.132pt}{0.800pt}}
\multiput(1194.00,205.34)(6.500,1.000){2}{\rule{1.566pt}{0.800pt}}
\put(1181.0,207.0){\rule[-0.400pt]{3.132pt}{0.800pt}}
\put(1220,206.84){\rule{2.891pt}{0.800pt}}
\multiput(1220.00,206.34)(6.000,1.000){2}{\rule{1.445pt}{0.800pt}}
\put(1232,207.84){\rule{3.132pt}{0.800pt}}
\multiput(1232.00,207.34)(6.500,1.000){2}{\rule{1.566pt}{0.800pt}}
\put(1207.0,208.0){\rule[-0.400pt]{3.132pt}{0.800pt}}
\put(1258,208.84){\rule{3.132pt}{0.800pt}}
\multiput(1258.00,208.34)(6.500,1.000){2}{\rule{1.566pt}{0.800pt}}
\put(1245.0,210.0){\rule[-0.400pt]{3.132pt}{0.800pt}}
\put(1284,209.84){\rule{3.132pt}{0.800pt}}
\multiput(1284.00,209.34)(6.500,1.000){2}{\rule{1.566pt}{0.800pt}}
\put(1297,210.84){\rule{3.132pt}{0.800pt}}
\multiput(1297.00,210.34)(6.500,1.000){2}{\rule{1.566pt}{0.800pt}}
\put(1271.0,211.0){\rule[-0.400pt]{3.132pt}{0.800pt}}
\put(1323,211.84){\rule{3.132pt}{0.800pt}}
\multiput(1323.00,211.34)(6.500,1.000){2}{\rule{1.566pt}{0.800pt}}
\put(1310.0,213.0){\rule[-0.400pt]{3.132pt}{0.800pt}}
\put(1349,212.84){\rule{3.132pt}{0.800pt}}
\multiput(1349.00,212.34)(6.500,1.000){2}{\rule{1.566pt}{0.800pt}}
\put(1362,213.84){\rule{2.891pt}{0.800pt}}
\multiput(1362.00,213.34)(6.000,1.000){2}{\rule{1.445pt}{0.800pt}}
\put(1336.0,214.0){\rule[-0.400pt]{3.132pt}{0.800pt}}
\put(1387,214.84){\rule{3.132pt}{0.800pt}}
\multiput(1387.00,214.34)(6.500,1.000){2}{\rule{1.566pt}{0.800pt}}
\put(1400,215.84){\rule{3.132pt}{0.800pt}}
\multiput(1400.00,215.34)(6.500,1.000){2}{\rule{1.566pt}{0.800pt}}
\put(1374.0,216.0){\rule[-0.400pt]{3.132pt}{0.800pt}}
\put(1426,216.84){\rule{3.132pt}{0.800pt}}
\multiput(1426.00,216.34)(6.500,1.000){2}{\rule{1.566pt}{0.800pt}}
\put(1413.0,218.0){\rule[-0.400pt]{3.132pt}{0.800pt}}
\sbox{\plotpoint}{\rule[-0.500pt]{1.000pt}{1.000pt}}%
\put(161,333){\usebox{\plotpoint}}
\put(161.00,333.00){\usebox{\plotpoint}}
\put(181.76,333.00){\usebox{\plotpoint}}
\put(202.51,333.00){\usebox{\plotpoint}}
\put(223.27,333.00){\usebox{\plotpoint}}
\put(244.02,333.00){\usebox{\plotpoint}}
\put(264.78,333.00){\usebox{\plotpoint}}
\put(285.53,333.00){\usebox{\plotpoint}}
\put(306.29,333.00){\usebox{\plotpoint}}
\put(327.04,333.00){\usebox{\plotpoint}}
\put(347.80,333.00){\usebox{\plotpoint}}
\put(368.55,333.00){\usebox{\plotpoint}}
\put(389.31,333.00){\usebox{\plotpoint}}
\put(410.07,333.00){\usebox{\plotpoint}}
\put(430.82,333.00){\usebox{\plotpoint}}
\put(451.58,333.00){\usebox{\plotpoint}}
\put(472.33,333.00){\usebox{\plotpoint}}
\put(493.09,333.00){\usebox{\plotpoint}}
\put(513.84,333.00){\usebox{\plotpoint}}
\put(534.60,333.00){\usebox{\plotpoint}}
\put(555.35,333.00){\usebox{\plotpoint}}
\put(576.11,333.00){\usebox{\plotpoint}}
\put(596.87,333.00){\usebox{\plotpoint}}
\put(617.62,333.00){\usebox{\plotpoint}}
\put(638.38,333.00){\usebox{\plotpoint}}
\put(659.13,333.00){\usebox{\plotpoint}}
\put(679.89,333.00){\usebox{\plotpoint}}
\put(700.64,333.00){\usebox{\plotpoint}}
\put(721.40,333.00){\usebox{\plotpoint}}
\put(742.15,333.00){\usebox{\plotpoint}}
\put(762.91,333.00){\usebox{\plotpoint}}
\put(783.66,333.00){\usebox{\plotpoint}}
\put(804.42,333.00){\usebox{\plotpoint}}
\put(825.18,333.00){\usebox{\plotpoint}}
\put(845.93,333.00){\usebox{\plotpoint}}
\put(866.69,333.00){\usebox{\plotpoint}}
\put(887.44,333.00){\usebox{\plotpoint}}
\put(908.20,333.00){\usebox{\plotpoint}}
\put(928.95,333.00){\usebox{\plotpoint}}
\put(949.71,333.00){\usebox{\plotpoint}}
\put(970.46,333.00){\usebox{\plotpoint}}
\put(991.22,333.00){\usebox{\plotpoint}}
\put(1011.98,333.00){\usebox{\plotpoint}}
\put(1032.73,333.00){\usebox{\plotpoint}}
\put(1053.49,333.00){\usebox{\plotpoint}}
\put(1074.24,333.00){\usebox{\plotpoint}}
\put(1095.00,333.00){\usebox{\plotpoint}}
\put(1115.75,333.00){\usebox{\plotpoint}}
\put(1136.51,333.00){\usebox{\plotpoint}}
\put(1157.26,333.00){\usebox{\plotpoint}}
\put(1178.02,333.00){\usebox{\plotpoint}}
\put(1198.77,333.00){\usebox{\plotpoint}}
\put(1219.53,333.00){\usebox{\plotpoint}}
\put(1240.29,333.00){\usebox{\plotpoint}}
\put(1261.04,333.00){\usebox{\plotpoint}}
\put(1281.80,333.00){\usebox{\plotpoint}}
\put(1302.55,333.00){\usebox{\plotpoint}}
\put(1323.31,333.00){\usebox{\plotpoint}}
\put(1344.06,333.00){\usebox{\plotpoint}}
\put(1364.82,333.00){\usebox{\plotpoint}}
\put(1385.57,333.00){\usebox{\plotpoint}}
\put(1406.33,333.00){\usebox{\plotpoint}}
\put(1427.09,333.00){\usebox{\plotpoint}}
\put(1439,333){\usebox{\plotpoint}}
\put(161,228){\usebox{\plotpoint}}
\put(161.00,228.00){\usebox{\plotpoint}}
\put(181.76,228.00){\usebox{\plotpoint}}
\put(202.51,228.00){\usebox{\plotpoint}}
\put(223.27,228.00){\usebox{\plotpoint}}
\put(244.02,228.00){\usebox{\plotpoint}}
\put(264.78,228.00){\usebox{\plotpoint}}
\put(285.53,228.00){\usebox{\plotpoint}}
\put(306.29,228.00){\usebox{\plotpoint}}
\put(327.04,228.00){\usebox{\plotpoint}}
\put(347.80,228.00){\usebox{\plotpoint}}
\put(368.55,228.00){\usebox{\plotpoint}}
\put(389.31,228.00){\usebox{\plotpoint}}
\put(410.07,228.00){\usebox{\plotpoint}}
\put(430.82,228.00){\usebox{\plotpoint}}
\put(451.58,228.00){\usebox{\plotpoint}}
\put(472.33,228.00){\usebox{\plotpoint}}
\put(493.09,228.00){\usebox{\plotpoint}}
\put(513.84,228.00){\usebox{\plotpoint}}
\put(534.60,228.00){\usebox{\plotpoint}}
\put(555.35,228.00){\usebox{\plotpoint}}
\put(576.11,228.00){\usebox{\plotpoint}}
\put(596.87,228.00){\usebox{\plotpoint}}
\put(617.62,228.00){\usebox{\plotpoint}}
\put(638.38,228.00){\usebox{\plotpoint}}
\put(659.13,228.00){\usebox{\plotpoint}}
\put(679.89,228.00){\usebox{\plotpoint}}
\put(700.64,228.00){\usebox{\plotpoint}}
\put(721.40,228.00){\usebox{\plotpoint}}
\put(742.15,228.00){\usebox{\plotpoint}}
\put(762.91,228.00){\usebox{\plotpoint}}
\put(783.66,228.00){\usebox{\plotpoint}}
\put(804.42,228.00){\usebox{\plotpoint}}
\put(825.18,228.00){\usebox{\plotpoint}}
\put(845.93,228.00){\usebox{\plotpoint}}
\put(866.69,228.00){\usebox{\plotpoint}}
\put(887.44,228.00){\usebox{\plotpoint}}
\put(908.20,228.00){\usebox{\plotpoint}}
\put(928.95,228.00){\usebox{\plotpoint}}
\put(949.71,228.00){\usebox{\plotpoint}}
\put(970.46,228.00){\usebox{\plotpoint}}
\put(991.22,228.00){\usebox{\plotpoint}}
\put(1011.98,228.00){\usebox{\plotpoint}}
\put(1032.73,228.00){\usebox{\plotpoint}}
\put(1053.49,228.00){\usebox{\plotpoint}}
\put(1074.24,228.00){\usebox{\plotpoint}}
\put(1095.00,228.00){\usebox{\plotpoint}}
\put(1115.75,228.00){\usebox{\plotpoint}}
\put(1136.51,228.00){\usebox{\plotpoint}}
\put(1157.26,228.00){\usebox{\plotpoint}}
\put(1178.02,228.00){\usebox{\plotpoint}}
\put(1198.77,228.00){\usebox{\plotpoint}}
\put(1219.53,228.00){\usebox{\plotpoint}}
\put(1240.29,228.00){\usebox{\plotpoint}}
\put(1261.04,228.00){\usebox{\plotpoint}}
\put(1281.80,228.00){\usebox{\plotpoint}}
\put(1302.55,228.00){\usebox{\plotpoint}}
\put(1323.31,228.00){\usebox{\plotpoint}}
\put(1344.06,228.00){\usebox{\plotpoint}}
\put(1364.82,228.00){\usebox{\plotpoint}}
\put(1385.57,228.00){\usebox{\plotpoint}}
\put(1406.33,228.00){\usebox{\plotpoint}}
\put(1427.09,228.00){\usebox{\plotpoint}}
\put(1439,228){\usebox{\plotpoint}}
\end{picture}

%% file: spec-es.tex
\setlength{\unitlength}{0.240900pt}
\ifx\plotpoint\undefined\newsavebox{\plotpoint}\fi
\sbox{\plotpoint}{\rule[-0.200pt]{0.400pt}{0.400pt}}%
\begin{picture}(1500,900)(0,0)
\font\gnuplot=cmr10 at 10pt
\gnuplot
\sbox{\plotpoint}{\rule[-0.200pt]{0.400pt}{0.400pt}}%
\put(181.0,123.0){\rule[-0.200pt]{4.818pt}{0.400pt}}
\put(161,123){\makebox(0,0)[r]{0}}
\put(1419.0,123.0){\rule[-0.200pt]{4.818pt}{0.400pt}}
\put(181.0,197.0){\rule[-0.200pt]{4.818pt}{0.400pt}}
\put(161,197){\makebox(0,0)[r]{0.05}}
\put(1419.0,197.0){\rule[-0.200pt]{4.818pt}{0.400pt}}
\put(181.0,270.0){\rule[-0.200pt]{4.818pt}{0.400pt}}
\put(161,270){\makebox(0,0)[r]{0.1}}
\put(1419.0,270.0){\rule[-0.200pt]{4.818pt}{0.400pt}}
\put(181.0,344.0){\rule[-0.200pt]{4.818pt}{0.400pt}}
\put(161,344){\makebox(0,0)[r]{0.15}}
\put(1419.0,344.0){\rule[-0.200pt]{4.818pt}{0.400pt}}
\put(181.0,418.0){\rule[-0.200pt]{4.818pt}{0.400pt}}
\put(161,418){\makebox(0,0)[r]{0.2}}
\put(1419.0,418.0){\rule[-0.200pt]{4.818pt}{0.400pt}}
\put(181.0,492.0){\rule[-0.200pt]{4.818pt}{0.400pt}}
\put(161,492){\makebox(0,0)[r]{0.25}}
\put(1419.0,492.0){\rule[-0.200pt]{4.818pt}{0.400pt}}
\put(181.0,565.0){\rule[-0.200pt]{4.818pt}{0.400pt}}
\put(161,565){\makebox(0,0)[r]{0.3}}
\put(1419.0,565.0){\rule[-0.200pt]{4.818pt}{0.400pt}}
\put(181.0,639.0){\rule[-0.200pt]{4.818pt}{0.400pt}}
\put(161,639){\makebox(0,0)[r]{0.35}}
\put(1419.0,639.0){\rule[-0.200pt]{4.818pt}{0.400pt}}
\put(181.0,713.0){\rule[-0.200pt]{4.818pt}{0.400pt}}
\put(161,713){\makebox(0,0)[r]{0.4}}
\put(1419.0,713.0){\rule[-0.200pt]{4.818pt}{0.400pt}}
\put(181.0,786.0){\rule[-0.200pt]{4.818pt}{0.400pt}}
\put(161,786){\makebox(0,0)[r]{0.45}}
\put(1419.0,786.0){\rule[-0.200pt]{4.818pt}{0.400pt}}
\put(181.0,860.0){\rule[-0.200pt]{4.818pt}{0.400pt}}
\put(161,860){\makebox(0,0)[r]{0.5}}
\put(1419.0,860.0){\rule[-0.200pt]{4.818pt}{0.400pt}}
\put(181.0,123.0){\rule[-0.200pt]{0.400pt}{4.818pt}}
\put(181,82){\makebox(0,0){10}}
\put(181.0,840.0){\rule[-0.200pt]{0.400pt}{4.818pt}}
\put(391.0,123.0){\rule[-0.200pt]{0.400pt}{4.818pt}}
\put(391,82){\makebox(0,0){11}}
\put(391.0,840.0){\rule[-0.200pt]{0.400pt}{4.818pt}}
\put(600.0,123.0){\rule[-0.200pt]{0.400pt}{4.818pt}}
\put(600,82){\makebox(0,0){12}}
\put(600.0,840.0){\rule[-0.200pt]{0.400pt}{4.818pt}}
\put(810.0,123.0){\rule[-0.200pt]{0.400pt}{4.818pt}}
\put(810,82){\makebox(0,0){13}}
\put(810.0,840.0){\rule[-0.200pt]{0.400pt}{4.818pt}}
\put(1020.0,123.0){\rule[-0.200pt]{0.400pt}{4.818pt}}
\put(1020,82){\makebox(0,0){14}}
\put(1020.0,840.0){\rule[-0.200pt]{0.400pt}{4.818pt}}
\put(1229.0,123.0){\rule[-0.200pt]{0.400pt}{4.818pt}}
\put(1229,82){\makebox(0,0){15}}
\put(1229.0,840.0){\rule[-0.200pt]{0.400pt}{4.818pt}}
\put(1439.0,123.0){\rule[-0.200pt]{0.400pt}{4.818pt}}
\put(1439,82){\makebox(0,0){16}}
\put(1439.0,840.0){\rule[-0.200pt]{0.400pt}{4.818pt}}
\put(181.0,123.0){\rule[-0.200pt]{303.052pt}{0.400pt}}
\put(1439.0,123.0){\rule[-0.200pt]{0.400pt}{177.543pt}}
\put(181.0,860.0){\rule[-0.200pt]{303.052pt}{0.400pt}}
\put(10,461){\makebox(0,0){$m_{\tilde e}/M_3$}}
\put(810,21){\makebox(0,0){$\log_{10}(M_C)$ [GeV]}}
\put(1229,197){\makebox(0,0)[r]{$S=0$}}
\put(1229,492){\makebox(0,0)[r]{$S=-0.5M^2_3$}}
\put(181.0,123.0){\rule[-0.200pt]{0.400pt}{177.543pt}}
\sbox{\plotpoint}{\rule[-0.400pt]{0.800pt}{0.800pt}}%
\put(181,223){\usebox{\plotpoint}}
\put(194,221.84){\rule{2.891pt}{0.800pt}}
\multiput(194.00,221.34)(6.000,1.000){2}{\rule{1.445pt}{0.800pt}}
\put(206,222.84){\rule{3.132pt}{0.800pt}}
\multiput(206.00,222.34)(6.500,1.000){2}{\rule{1.566pt}{0.800pt}}
\put(181.0,223.0){\rule[-0.400pt]{3.132pt}{0.800pt}}
\put(232,223.84){\rule{3.132pt}{0.800pt}}
\multiput(232.00,223.34)(6.500,1.000){2}{\rule{1.566pt}{0.800pt}}
\put(245,224.84){\rule{2.891pt}{0.800pt}}
\multiput(245.00,224.34)(6.000,1.000){2}{\rule{1.445pt}{0.800pt}}
\put(219.0,225.0){\rule[-0.400pt]{3.132pt}{0.800pt}}
\put(270,225.84){\rule{3.132pt}{0.800pt}}
\multiput(270.00,225.34)(6.500,1.000){2}{\rule{1.566pt}{0.800pt}}
\put(283,226.84){\rule{2.891pt}{0.800pt}}
\multiput(283.00,226.34)(6.000,1.000){2}{\rule{1.445pt}{0.800pt}}
\put(295,227.84){\rule{3.132pt}{0.800pt}}
\multiput(295.00,227.34)(6.500,1.000){2}{\rule{1.566pt}{0.800pt}}
\put(257.0,227.0){\rule[-0.400pt]{3.132pt}{0.800pt}}
\put(321,228.84){\rule{2.891pt}{0.800pt}}
\multiput(321.00,228.34)(6.000,1.000){2}{\rule{1.445pt}{0.800pt}}
\put(333,229.84){\rule{3.132pt}{0.800pt}}
\multiput(333.00,229.34)(6.500,1.000){2}{\rule{1.566pt}{0.800pt}}
\put(308.0,230.0){\rule[-0.400pt]{3.132pt}{0.800pt}}
\put(359,230.84){\rule{3.132pt}{0.800pt}}
\multiput(359.00,230.34)(6.500,1.000){2}{\rule{1.566pt}{0.800pt}}
\put(372,231.84){\rule{2.891pt}{0.800pt}}
\multiput(372.00,231.34)(6.000,1.000){2}{\rule{1.445pt}{0.800pt}}
\put(384,232.84){\rule{3.132pt}{0.800pt}}
\multiput(384.00,232.34)(6.500,1.000){2}{\rule{1.566pt}{0.800pt}}
\put(346.0,232.0){\rule[-0.400pt]{3.132pt}{0.800pt}}
\put(410,233.84){\rule{2.891pt}{0.800pt}}
\multiput(410.00,233.34)(6.000,1.000){2}{\rule{1.445pt}{0.800pt}}
\put(422,234.84){\rule{3.132pt}{0.800pt}}
\multiput(422.00,234.34)(6.500,1.000){2}{\rule{1.566pt}{0.800pt}}
\put(397.0,235.0){\rule[-0.400pt]{3.132pt}{0.800pt}}
\put(448,235.84){\rule{3.132pt}{0.800pt}}
\multiput(448.00,235.34)(6.500,1.000){2}{\rule{1.566pt}{0.800pt}}
\put(461,236.84){\rule{2.891pt}{0.800pt}}
\multiput(461.00,236.34)(6.000,1.000){2}{\rule{1.445pt}{0.800pt}}
\put(473,237.84){\rule{3.132pt}{0.800pt}}
\multiput(473.00,237.34)(6.500,1.000){2}{\rule{1.566pt}{0.800pt}}
\put(435.0,237.0){\rule[-0.400pt]{3.132pt}{0.800pt}}
\put(499,238.84){\rule{2.891pt}{0.800pt}}
\multiput(499.00,238.34)(6.000,1.000){2}{\rule{1.445pt}{0.800pt}}
\put(511,239.84){\rule{3.132pt}{0.800pt}}
\multiput(511.00,239.34)(6.500,1.000){2}{\rule{1.566pt}{0.800pt}}
\put(524,240.84){\rule{3.132pt}{0.800pt}}
\multiput(524.00,240.34)(6.500,1.000){2}{\rule{1.566pt}{0.800pt}}
\put(537,241.84){\rule{3.132pt}{0.800pt}}
\multiput(537.00,241.34)(6.500,1.000){2}{\rule{1.566pt}{0.800pt}}
\put(486.0,240.0){\rule[-0.400pt]{3.132pt}{0.800pt}}
\put(562,242.84){\rule{3.132pt}{0.800pt}}
\multiput(562.00,242.34)(6.500,1.000){2}{\rule{1.566pt}{0.800pt}}
\put(575,243.84){\rule{3.132pt}{0.800pt}}
\multiput(575.00,243.34)(6.500,1.000){2}{\rule{1.566pt}{0.800pt}}
\put(588,244.84){\rule{2.891pt}{0.800pt}}
\multiput(588.00,244.34)(6.000,1.000){2}{\rule{1.445pt}{0.800pt}}
\put(600,245.84){\rule{3.132pt}{0.800pt}}
\multiput(600.00,245.34)(6.500,1.000){2}{\rule{1.566pt}{0.800pt}}
\put(550.0,244.0){\rule[-0.400pt]{2.891pt}{0.800pt}}
\put(626,246.84){\rule{2.891pt}{0.800pt}}
\multiput(626.00,246.34)(6.000,1.000){2}{\rule{1.445pt}{0.800pt}}
\put(638,247.84){\rule{3.132pt}{0.800pt}}
\multiput(638.00,247.34)(6.500,1.000){2}{\rule{1.566pt}{0.800pt}}
\put(651,248.84){\rule{3.132pt}{0.800pt}}
\multiput(651.00,248.34)(6.500,1.000){2}{\rule{1.566pt}{0.800pt}}
\put(664,249.84){\rule{3.132pt}{0.800pt}}
\multiput(664.00,249.34)(6.500,1.000){2}{\rule{1.566pt}{0.800pt}}
\put(613.0,248.0){\rule[-0.400pt]{3.132pt}{0.800pt}}
\put(689,250.84){\rule{3.132pt}{0.800pt}}
\multiput(689.00,250.34)(6.500,1.000){2}{\rule{1.566pt}{0.800pt}}
\put(702,251.84){\rule{3.132pt}{0.800pt}}
\multiput(702.00,251.34)(6.500,1.000){2}{\rule{1.566pt}{0.800pt}}
\put(715,252.84){\rule{2.891pt}{0.800pt}}
\multiput(715.00,252.34)(6.000,1.000){2}{\rule{1.445pt}{0.800pt}}
\put(727,253.84){\rule{3.132pt}{0.800pt}}
\multiput(727.00,253.34)(6.500,1.000){2}{\rule{1.566pt}{0.800pt}}
\put(740,254.84){\rule{3.132pt}{0.800pt}}
\multiput(740.00,254.34)(6.500,1.000){2}{\rule{1.566pt}{0.800pt}}
\put(753,255.84){\rule{3.132pt}{0.800pt}}
\multiput(753.00,255.34)(6.500,1.000){2}{\rule{1.566pt}{0.800pt}}
\put(677.0,252.0){\rule[-0.400pt]{2.891pt}{0.800pt}}
\put(778,256.84){\rule{3.132pt}{0.800pt}}
\multiput(778.00,256.34)(6.500,1.000){2}{\rule{1.566pt}{0.800pt}}
\put(791,257.84){\rule{3.132pt}{0.800pt}}
\multiput(791.00,257.34)(6.500,1.000){2}{\rule{1.566pt}{0.800pt}}
\put(804,258.84){\rule{2.891pt}{0.800pt}}
\multiput(804.00,258.34)(6.000,1.000){2}{\rule{1.445pt}{0.800pt}}
\put(816,259.84){\rule{3.132pt}{0.800pt}}
\multiput(816.00,259.34)(6.500,1.000){2}{\rule{1.566pt}{0.800pt}}
\put(829,260.84){\rule{3.132pt}{0.800pt}}
\multiput(829.00,260.34)(6.500,1.000){2}{\rule{1.566pt}{0.800pt}}
\put(842,261.84){\rule{2.891pt}{0.800pt}}
\multiput(842.00,261.34)(6.000,1.000){2}{\rule{1.445pt}{0.800pt}}
\put(854,262.84){\rule{3.132pt}{0.800pt}}
\multiput(854.00,262.34)(6.500,1.000){2}{\rule{1.566pt}{0.800pt}}
\put(867,263.84){\rule{3.132pt}{0.800pt}}
\multiput(867.00,263.34)(6.500,1.000){2}{\rule{1.566pt}{0.800pt}}
\put(880,264.84){\rule{3.132pt}{0.800pt}}
\multiput(880.00,264.34)(6.500,1.000){2}{\rule{1.566pt}{0.800pt}}
\put(893,265.84){\rule{2.891pt}{0.800pt}}
\multiput(893.00,265.34)(6.000,1.000){2}{\rule{1.445pt}{0.800pt}}
\put(905,266.84){\rule{3.132pt}{0.800pt}}
\multiput(905.00,266.34)(6.500,1.000){2}{\rule{1.566pt}{0.800pt}}
\put(766.0,258.0){\rule[-0.400pt]{2.891pt}{0.800pt}}
\put(931,267.84){\rule{2.891pt}{0.800pt}}
\multiput(931.00,267.34)(6.000,1.000){2}{\rule{1.445pt}{0.800pt}}
\put(943,268.84){\rule{3.132pt}{0.800pt}}
\multiput(943.00,268.34)(6.500,1.000){2}{\rule{1.566pt}{0.800pt}}
\put(956,269.84){\rule{3.132pt}{0.800pt}}
\multiput(956.00,269.34)(6.500,1.000){2}{\rule{1.566pt}{0.800pt}}
\put(969,270.84){\rule{3.132pt}{0.800pt}}
\multiput(969.00,270.34)(6.500,1.000){2}{\rule{1.566pt}{0.800pt}}
\put(982,271.84){\rule{2.891pt}{0.800pt}}
\multiput(982.00,271.34)(6.000,1.000){2}{\rule{1.445pt}{0.800pt}}
\put(994,272.84){\rule{3.132pt}{0.800pt}}
\multiput(994.00,272.34)(6.500,1.000){2}{\rule{1.566pt}{0.800pt}}
\put(1007,273.84){\rule{3.132pt}{0.800pt}}
\multiput(1007.00,273.34)(6.500,1.000){2}{\rule{1.566pt}{0.800pt}}
\put(1020,274.84){\rule{2.891pt}{0.800pt}}
\multiput(1020.00,274.34)(6.000,1.000){2}{\rule{1.445pt}{0.800pt}}
\put(1032,275.84){\rule{3.132pt}{0.800pt}}
\multiput(1032.00,275.34)(6.500,1.000){2}{\rule{1.566pt}{0.800pt}}
\put(1045,276.84){\rule{3.132pt}{0.800pt}}
\multiput(1045.00,276.34)(6.500,1.000){2}{\rule{1.566pt}{0.800pt}}
\put(1058,277.84){\rule{2.891pt}{0.800pt}}
\multiput(1058.00,277.34)(6.000,1.000){2}{\rule{1.445pt}{0.800pt}}
\put(1070,279.34){\rule{3.132pt}{0.800pt}}
\multiput(1070.00,278.34)(6.500,2.000){2}{\rule{1.566pt}{0.800pt}}
\put(1083,280.84){\rule{3.132pt}{0.800pt}}
\multiput(1083.00,280.34)(6.500,1.000){2}{\rule{1.566pt}{0.800pt}}
\put(1096,281.84){\rule{3.132pt}{0.800pt}}
\multiput(1096.00,281.34)(6.500,1.000){2}{\rule{1.566pt}{0.800pt}}
\put(1109,282.84){\rule{2.891pt}{0.800pt}}
\multiput(1109.00,282.34)(6.000,1.000){2}{\rule{1.445pt}{0.800pt}}
\put(1121,283.84){\rule{3.132pt}{0.800pt}}
\multiput(1121.00,283.34)(6.500,1.000){2}{\rule{1.566pt}{0.800pt}}
\put(1134,284.84){\rule{3.132pt}{0.800pt}}
\multiput(1134.00,284.34)(6.500,1.000){2}{\rule{1.566pt}{0.800pt}}
\put(1147,285.84){\rule{2.891pt}{0.800pt}}
\multiput(1147.00,285.34)(6.000,1.000){2}{\rule{1.445pt}{0.800pt}}
\put(1159,286.84){\rule{3.132pt}{0.800pt}}
\multiput(1159.00,286.34)(6.500,1.000){2}{\rule{1.566pt}{0.800pt}}
\put(1172,287.84){\rule{3.132pt}{0.800pt}}
\multiput(1172.00,287.34)(6.500,1.000){2}{\rule{1.566pt}{0.800pt}}
\put(1185,288.84){\rule{3.132pt}{0.800pt}}
\multiput(1185.00,288.34)(6.500,1.000){2}{\rule{1.566pt}{0.800pt}}
\put(1198,290.34){\rule{2.891pt}{0.800pt}}
\multiput(1198.00,289.34)(6.000,2.000){2}{\rule{1.445pt}{0.800pt}}
\put(1210,291.84){\rule{3.132pt}{0.800pt}}
\multiput(1210.00,291.34)(6.500,1.000){2}{\rule{1.566pt}{0.800pt}}
\put(1223,292.84){\rule{3.132pt}{0.800pt}}
\multiput(1223.00,292.34)(6.500,1.000){2}{\rule{1.566pt}{0.800pt}}
\put(1236,293.84){\rule{2.891pt}{0.800pt}}
\multiput(1236.00,293.34)(6.000,1.000){2}{\rule{1.445pt}{0.800pt}}
\put(1248,294.84){\rule{3.132pt}{0.800pt}}
\multiput(1248.00,294.34)(6.500,1.000){2}{\rule{1.566pt}{0.800pt}}
\put(1261,295.84){\rule{3.132pt}{0.800pt}}
\multiput(1261.00,295.34)(6.500,1.000){2}{\rule{1.566pt}{0.800pt}}
\put(1274,297.34){\rule{3.132pt}{0.800pt}}
\multiput(1274.00,296.34)(6.500,2.000){2}{\rule{1.566pt}{0.800pt}}
\put(1287,298.84){\rule{2.891pt}{0.800pt}}
\multiput(1287.00,298.34)(6.000,1.000){2}{\rule{1.445pt}{0.800pt}}
\put(1299,299.84){\rule{3.132pt}{0.800pt}}
\multiput(1299.00,299.34)(6.500,1.000){2}{\rule{1.566pt}{0.800pt}}
\put(1312,300.84){\rule{3.132pt}{0.800pt}}
\multiput(1312.00,300.34)(6.500,1.000){2}{\rule{1.566pt}{0.800pt}}
\put(1325,302.34){\rule{2.891pt}{0.800pt}}
\multiput(1325.00,301.34)(6.000,2.000){2}{\rule{1.445pt}{0.800pt}}
\put(1337,303.84){\rule{3.132pt}{0.800pt}}
\multiput(1337.00,303.34)(6.500,1.000){2}{\rule{1.566pt}{0.800pt}}
\put(1350,304.84){\rule{3.132pt}{0.800pt}}
\multiput(1350.00,304.34)(6.500,1.000){2}{\rule{1.566pt}{0.800pt}}
\put(1363,306.34){\rule{2.891pt}{0.800pt}}
\multiput(1363.00,305.34)(6.000,2.000){2}{\rule{1.445pt}{0.800pt}}
\put(1375,307.84){\rule{3.132pt}{0.800pt}}
\multiput(1375.00,307.34)(6.500,1.000){2}{\rule{1.566pt}{0.800pt}}
\put(1388,308.84){\rule{3.132pt}{0.800pt}}
\multiput(1388.00,308.34)(6.500,1.000){2}{\rule{1.566pt}{0.800pt}}
\put(1401,310.34){\rule{3.132pt}{0.800pt}}
\multiput(1401.00,309.34)(6.500,2.000){2}{\rule{1.566pt}{0.800pt}}
\put(1414,311.84){\rule{2.891pt}{0.800pt}}
\multiput(1414.00,311.34)(6.000,1.000){2}{\rule{1.445pt}{0.800pt}}
\put(1426,312.84){\rule{3.132pt}{0.800pt}}
\multiput(1426.00,312.34)(6.500,1.000){2}{\rule{1.566pt}{0.800pt}}
\put(918.0,269.0){\rule[-0.400pt]{3.132pt}{0.800pt}}
\put(181,329){\usebox{\plotpoint}}
\put(181,327.84){\rule{3.132pt}{0.800pt}}
\multiput(181.00,327.34)(6.500,1.000){2}{\rule{1.566pt}{0.800pt}}
\put(194,328.84){\rule{2.891pt}{0.800pt}}
\multiput(194.00,328.34)(6.000,1.000){2}{\rule{1.445pt}{0.800pt}}
\put(206,329.84){\rule{3.132pt}{0.800pt}}
\multiput(206.00,329.34)(6.500,1.000){2}{\rule{1.566pt}{0.800pt}}
\put(219,330.84){\rule{3.132pt}{0.800pt}}
\multiput(219.00,330.34)(6.500,1.000){2}{\rule{1.566pt}{0.800pt}}
\put(232,331.84){\rule{3.132pt}{0.800pt}}
\multiput(232.00,331.34)(6.500,1.000){2}{\rule{1.566pt}{0.800pt}}
\put(245,332.84){\rule{2.891pt}{0.800pt}}
\multiput(245.00,332.34)(6.000,1.000){2}{\rule{1.445pt}{0.800pt}}
\put(257,333.84){\rule{3.132pt}{0.800pt}}
\multiput(257.00,333.34)(6.500,1.000){2}{\rule{1.566pt}{0.800pt}}
\put(270,334.84){\rule{3.132pt}{0.800pt}}
\multiput(270.00,334.34)(6.500,1.000){2}{\rule{1.566pt}{0.800pt}}
\put(295,335.84){\rule{3.132pt}{0.800pt}}
\multiput(295.00,335.34)(6.500,1.000){2}{\rule{1.566pt}{0.800pt}}
\put(308,336.84){\rule{3.132pt}{0.800pt}}
\multiput(308.00,336.34)(6.500,1.000){2}{\rule{1.566pt}{0.800pt}}
\put(321,337.84){\rule{2.891pt}{0.800pt}}
\multiput(321.00,337.34)(6.000,1.000){2}{\rule{1.445pt}{0.800pt}}
\put(333,338.84){\rule{3.132pt}{0.800pt}}
\multiput(333.00,338.34)(6.500,1.000){2}{\rule{1.566pt}{0.800pt}}
\put(346,339.84){\rule{3.132pt}{0.800pt}}
\multiput(346.00,339.34)(6.500,1.000){2}{\rule{1.566pt}{0.800pt}}
\put(359,340.84){\rule{3.132pt}{0.800pt}}
\multiput(359.00,340.34)(6.500,1.000){2}{\rule{1.566pt}{0.800pt}}
\put(372,341.84){\rule{2.891pt}{0.800pt}}
\multiput(372.00,341.34)(6.000,1.000){2}{\rule{1.445pt}{0.800pt}}
\put(384,342.84){\rule{3.132pt}{0.800pt}}
\multiput(384.00,342.34)(6.500,1.000){2}{\rule{1.566pt}{0.800pt}}
\put(397,343.84){\rule{3.132pt}{0.800pt}}
\multiput(397.00,343.34)(6.500,1.000){2}{\rule{1.566pt}{0.800pt}}
\put(410,344.84){\rule{2.891pt}{0.800pt}}
\multiput(410.00,344.34)(6.000,1.000){2}{\rule{1.445pt}{0.800pt}}
\put(422,345.84){\rule{3.132pt}{0.800pt}}
\multiput(422.00,345.34)(6.500,1.000){2}{\rule{1.566pt}{0.800pt}}
\put(283.0,337.0){\rule[-0.400pt]{2.891pt}{0.800pt}}
\put(448,346.84){\rule{3.132pt}{0.800pt}}
\multiput(448.00,346.34)(6.500,1.000){2}{\rule{1.566pt}{0.800pt}}
\put(461,347.84){\rule{2.891pt}{0.800pt}}
\multiput(461.00,347.34)(6.000,1.000){2}{\rule{1.445pt}{0.800pt}}
\put(473,348.84){\rule{3.132pt}{0.800pt}}
\multiput(473.00,348.34)(6.500,1.000){2}{\rule{1.566pt}{0.800pt}}
\put(486,349.84){\rule{3.132pt}{0.800pt}}
\multiput(486.00,349.34)(6.500,1.000){2}{\rule{1.566pt}{0.800pt}}
\put(499,350.84){\rule{2.891pt}{0.800pt}}
\multiput(499.00,350.34)(6.000,1.000){2}{\rule{1.445pt}{0.800pt}}
\put(511,351.84){\rule{3.132pt}{0.800pt}}
\multiput(511.00,351.34)(6.500,1.000){2}{\rule{1.566pt}{0.800pt}}
\put(524,352.84){\rule{3.132pt}{0.800pt}}
\multiput(524.00,352.34)(6.500,1.000){2}{\rule{1.566pt}{0.800pt}}
\put(537,353.84){\rule{3.132pt}{0.800pt}}
\multiput(537.00,353.34)(6.500,1.000){2}{\rule{1.566pt}{0.800pt}}
\put(550,354.84){\rule{2.891pt}{0.800pt}}
\multiput(550.00,354.34)(6.000,1.000){2}{\rule{1.445pt}{0.800pt}}
\put(562,355.84){\rule{3.132pt}{0.800pt}}
\multiput(562.00,355.34)(6.500,1.000){2}{\rule{1.566pt}{0.800pt}}
\put(575,356.84){\rule{3.132pt}{0.800pt}}
\multiput(575.00,356.34)(6.500,1.000){2}{\rule{1.566pt}{0.800pt}}
\put(588,357.84){\rule{2.891pt}{0.800pt}}
\multiput(588.00,357.34)(6.000,1.000){2}{\rule{1.445pt}{0.800pt}}
\put(600,358.84){\rule{3.132pt}{0.800pt}}
\multiput(600.00,358.34)(6.500,1.000){2}{\rule{1.566pt}{0.800pt}}
\put(613,359.84){\rule{3.132pt}{0.800pt}}
\multiput(613.00,359.34)(6.500,1.000){2}{\rule{1.566pt}{0.800pt}}
\put(435.0,348.0){\rule[-0.400pt]{3.132pt}{0.800pt}}
\put(638,360.84){\rule{3.132pt}{0.800pt}}
\multiput(638.00,360.34)(6.500,1.000){2}{\rule{1.566pt}{0.800pt}}
\put(651,361.84){\rule{3.132pt}{0.800pt}}
\multiput(651.00,361.34)(6.500,1.000){2}{\rule{1.566pt}{0.800pt}}
\put(664,362.84){\rule{3.132pt}{0.800pt}}
\multiput(664.00,362.34)(6.500,1.000){2}{\rule{1.566pt}{0.800pt}}
\put(677,363.84){\rule{2.891pt}{0.800pt}}
\multiput(677.00,363.34)(6.000,1.000){2}{\rule{1.445pt}{0.800pt}}
\put(689,364.84){\rule{3.132pt}{0.800pt}}
\multiput(689.00,364.34)(6.500,1.000){2}{\rule{1.566pt}{0.800pt}}
\put(702,365.84){\rule{3.132pt}{0.800pt}}
\multiput(702.00,365.34)(6.500,1.000){2}{\rule{1.566pt}{0.800pt}}
\put(715,366.84){\rule{2.891pt}{0.800pt}}
\multiput(715.00,366.34)(6.000,1.000){2}{\rule{1.445pt}{0.800pt}}
\put(727,367.84){\rule{3.132pt}{0.800pt}}
\multiput(727.00,367.34)(6.500,1.000){2}{\rule{1.566pt}{0.800pt}}
\put(740,368.84){\rule{3.132pt}{0.800pt}}
\multiput(740.00,368.34)(6.500,1.000){2}{\rule{1.566pt}{0.800pt}}
\put(753,369.84){\rule{3.132pt}{0.800pt}}
\multiput(753.00,369.34)(6.500,1.000){2}{\rule{1.566pt}{0.800pt}}
\put(766,370.84){\rule{2.891pt}{0.800pt}}
\multiput(766.00,370.34)(6.000,1.000){2}{\rule{1.445pt}{0.800pt}}
\put(778,371.84){\rule{3.132pt}{0.800pt}}
\multiput(778.00,371.34)(6.500,1.000){2}{\rule{1.566pt}{0.800pt}}
\put(791,372.84){\rule{3.132pt}{0.800pt}}
\multiput(791.00,372.34)(6.500,1.000){2}{\rule{1.566pt}{0.800pt}}
\put(804,373.84){\rule{2.891pt}{0.800pt}}
\multiput(804.00,373.34)(6.000,1.000){2}{\rule{1.445pt}{0.800pt}}
\put(816,374.84){\rule{3.132pt}{0.800pt}}
\multiput(816.00,374.34)(6.500,1.000){2}{\rule{1.566pt}{0.800pt}}
\put(829,375.84){\rule{3.132pt}{0.800pt}}
\multiput(829.00,375.34)(6.500,1.000){2}{\rule{1.566pt}{0.800pt}}
\put(842,376.84){\rule{2.891pt}{0.800pt}}
\multiput(842.00,376.34)(6.000,1.000){2}{\rule{1.445pt}{0.800pt}}
\put(854,377.84){\rule{3.132pt}{0.800pt}}
\multiput(854.00,377.34)(6.500,1.000){2}{\rule{1.566pt}{0.800pt}}
\put(867,378.84){\rule{3.132pt}{0.800pt}}
\multiput(867.00,378.34)(6.500,1.000){2}{\rule{1.566pt}{0.800pt}}
\put(880,379.84){\rule{3.132pt}{0.800pt}}
\multiput(880.00,379.34)(6.500,1.000){2}{\rule{1.566pt}{0.800pt}}
\put(893,380.84){\rule{2.891pt}{0.800pt}}
\multiput(893.00,380.34)(6.000,1.000){2}{\rule{1.445pt}{0.800pt}}
\put(905,381.84){\rule{3.132pt}{0.800pt}}
\multiput(905.00,381.34)(6.500,1.000){2}{\rule{1.566pt}{0.800pt}}
\put(918,382.84){\rule{3.132pt}{0.800pt}}
\multiput(918.00,382.34)(6.500,1.000){2}{\rule{1.566pt}{0.800pt}}
\put(931,383.84){\rule{2.891pt}{0.800pt}}
\multiput(931.00,383.34)(6.000,1.000){2}{\rule{1.445pt}{0.800pt}}
\put(943,384.84){\rule{3.132pt}{0.800pt}}
\multiput(943.00,384.34)(6.500,1.000){2}{\rule{1.566pt}{0.800pt}}
\put(956,385.84){\rule{3.132pt}{0.800pt}}
\multiput(956.00,385.34)(6.500,1.000){2}{\rule{1.566pt}{0.800pt}}
\put(969,386.84){\rule{3.132pt}{0.800pt}}
\multiput(969.00,386.34)(6.500,1.000){2}{\rule{1.566pt}{0.800pt}}
\put(982,387.84){\rule{2.891pt}{0.800pt}}
\multiput(982.00,387.34)(6.000,1.000){2}{\rule{1.445pt}{0.800pt}}
\put(994,388.84){\rule{3.132pt}{0.800pt}}
\multiput(994.00,388.34)(6.500,1.000){2}{\rule{1.566pt}{0.800pt}}
\put(1007,389.84){\rule{3.132pt}{0.800pt}}
\multiput(1007.00,389.34)(6.500,1.000){2}{\rule{1.566pt}{0.800pt}}
\put(1020,390.84){\rule{2.891pt}{0.800pt}}
\multiput(1020.00,390.34)(6.000,1.000){2}{\rule{1.445pt}{0.800pt}}
\put(1032,391.84){\rule{3.132pt}{0.800pt}}
\multiput(1032.00,391.34)(6.500,1.000){2}{\rule{1.566pt}{0.800pt}}
\put(1045,392.84){\rule{3.132pt}{0.800pt}}
\multiput(1045.00,392.34)(6.500,1.000){2}{\rule{1.566pt}{0.800pt}}
\put(1058,393.84){\rule{2.891pt}{0.800pt}}
\multiput(1058.00,393.34)(6.000,1.000){2}{\rule{1.445pt}{0.800pt}}
\put(1070,394.84){\rule{3.132pt}{0.800pt}}
\multiput(1070.00,394.34)(6.500,1.000){2}{\rule{1.566pt}{0.800pt}}
\put(1083,395.84){\rule{3.132pt}{0.800pt}}
\multiput(1083.00,395.34)(6.500,1.000){2}{\rule{1.566pt}{0.800pt}}
\put(1096,396.84){\rule{3.132pt}{0.800pt}}
\multiput(1096.00,396.34)(6.500,1.000){2}{\rule{1.566pt}{0.800pt}}
\put(1109,397.84){\rule{2.891pt}{0.800pt}}
\multiput(1109.00,397.34)(6.000,1.000){2}{\rule{1.445pt}{0.800pt}}
\put(1121,398.84){\rule{3.132pt}{0.800pt}}
\multiput(1121.00,398.34)(6.500,1.000){2}{\rule{1.566pt}{0.800pt}}
\put(1134,400.34){\rule{3.132pt}{0.800pt}}
\multiput(1134.00,399.34)(6.500,2.000){2}{\rule{1.566pt}{0.800pt}}
\put(1147,401.84){\rule{2.891pt}{0.800pt}}
\multiput(1147.00,401.34)(6.000,1.000){2}{\rule{1.445pt}{0.800pt}}
\put(1159,402.84){\rule{3.132pt}{0.800pt}}
\multiput(1159.00,402.34)(6.500,1.000){2}{\rule{1.566pt}{0.800pt}}
\put(1172,403.84){\rule{3.132pt}{0.800pt}}
\multiput(1172.00,403.34)(6.500,1.000){2}{\rule{1.566pt}{0.800pt}}
\put(1185,404.84){\rule{3.132pt}{0.800pt}}
\multiput(1185.00,404.34)(6.500,1.000){2}{\rule{1.566pt}{0.800pt}}
\put(1198,405.84){\rule{2.891pt}{0.800pt}}
\multiput(1198.00,405.34)(6.000,1.000){2}{\rule{1.445pt}{0.800pt}}
\put(1210,406.84){\rule{3.132pt}{0.800pt}}
\multiput(1210.00,406.34)(6.500,1.000){2}{\rule{1.566pt}{0.800pt}}
\put(1223,407.84){\rule{3.132pt}{0.800pt}}
\multiput(1223.00,407.34)(6.500,1.000){2}{\rule{1.566pt}{0.800pt}}
\put(1236,408.84){\rule{2.891pt}{0.800pt}}
\multiput(1236.00,408.34)(6.000,1.000){2}{\rule{1.445pt}{0.800pt}}
\put(1248,410.34){\rule{3.132pt}{0.800pt}}
\multiput(1248.00,409.34)(6.500,2.000){2}{\rule{1.566pt}{0.800pt}}
\put(1261,411.84){\rule{3.132pt}{0.800pt}}
\multiput(1261.00,411.34)(6.500,1.000){2}{\rule{1.566pt}{0.800pt}}
\put(1274,412.84){\rule{3.132pt}{0.800pt}}
\multiput(1274.00,412.34)(6.500,1.000){2}{\rule{1.566pt}{0.800pt}}
\put(1287,413.84){\rule{2.891pt}{0.800pt}}
\multiput(1287.00,413.34)(6.000,1.000){2}{\rule{1.445pt}{0.800pt}}
\put(1299,414.84){\rule{3.132pt}{0.800pt}}
\multiput(1299.00,414.34)(6.500,1.000){2}{\rule{1.566pt}{0.800pt}}
\put(1312,415.84){\rule{3.132pt}{0.800pt}}
\multiput(1312.00,415.34)(6.500,1.000){2}{\rule{1.566pt}{0.800pt}}
\put(1325,417.34){\rule{2.891pt}{0.800pt}}
\multiput(1325.00,416.34)(6.000,2.000){2}{\rule{1.445pt}{0.800pt}}
\put(1337,418.84){\rule{3.132pt}{0.800pt}}
\multiput(1337.00,418.34)(6.500,1.000){2}{\rule{1.566pt}{0.800pt}}
\put(1350,419.84){\rule{3.132pt}{0.800pt}}
\multiput(1350.00,419.34)(6.500,1.000){2}{\rule{1.566pt}{0.800pt}}
\put(1363,420.84){\rule{2.891pt}{0.800pt}}
\multiput(1363.00,420.34)(6.000,1.000){2}{\rule{1.445pt}{0.800pt}}
\put(1375,421.84){\rule{3.132pt}{0.800pt}}
\multiput(1375.00,421.34)(6.500,1.000){2}{\rule{1.566pt}{0.800pt}}
\put(1388,423.34){\rule{3.132pt}{0.800pt}}
\multiput(1388.00,422.34)(6.500,2.000){2}{\rule{1.566pt}{0.800pt}}
\put(1401,424.84){\rule{3.132pt}{0.800pt}}
\multiput(1401.00,424.34)(6.500,1.000){2}{\rule{1.566pt}{0.800pt}}
\put(1414,425.84){\rule{2.891pt}{0.800pt}}
\multiput(1414.00,425.34)(6.000,1.000){2}{\rule{1.445pt}{0.800pt}}
\put(1426,426.84){\rule{3.132pt}{0.800pt}}
\multiput(1426.00,426.34)(6.500,1.000){2}{\rule{1.566pt}{0.800pt}}
\put(626.0,362.0){\rule[-0.400pt]{2.891pt}{0.800pt}}
\sbox{\plotpoint}{\rule[-0.500pt]{1.000pt}{1.000pt}}%
\put(181,334){\usebox{\plotpoint}}
\put(181.00,334.00){\usebox{\plotpoint}}
\put(201.76,334.00){\usebox{\plotpoint}}
\put(222.51,334.00){\usebox{\plotpoint}}
\put(243.27,334.00){\usebox{\plotpoint}}
\put(264.02,334.00){\usebox{\plotpoint}}
\put(284.78,334.00){\usebox{\plotpoint}}
\put(305.53,334.00){\usebox{\plotpoint}}
\put(326.29,334.00){\usebox{\plotpoint}}
\put(347.04,334.00){\usebox{\plotpoint}}
\put(367.80,334.00){\usebox{\plotpoint}}
\put(388.55,334.00){\usebox{\plotpoint}}
\put(409.31,334.00){\usebox{\plotpoint}}
\put(430.07,334.00){\usebox{\plotpoint}}
\put(450.82,334.00){\usebox{\plotpoint}}
\put(471.58,334.00){\usebox{\plotpoint}}
\put(492.33,334.00){\usebox{\plotpoint}}
\put(513.09,334.00){\usebox{\plotpoint}}
\put(533.84,334.00){\usebox{\plotpoint}}
\put(554.60,334.00){\usebox{\plotpoint}}
\put(575.35,334.00){\usebox{\plotpoint}}
\put(596.11,334.00){\usebox{\plotpoint}}
\put(616.87,334.00){\usebox{\plotpoint}}
\put(637.62,334.00){\usebox{\plotpoint}}
\put(658.38,334.00){\usebox{\plotpoint}}
\put(679.13,334.00){\usebox{\plotpoint}}
\put(699.89,334.00){\usebox{\plotpoint}}
\put(720.64,334.00){\usebox{\plotpoint}}
\put(741.40,334.00){\usebox{\plotpoint}}
\put(762.15,334.00){\usebox{\plotpoint}}
\put(782.91,334.00){\usebox{\plotpoint}}
\put(803.66,334.00){\usebox{\plotpoint}}
\put(824.42,334.00){\usebox{\plotpoint}}
\put(845.18,334.00){\usebox{\plotpoint}}
\put(865.93,334.00){\usebox{\plotpoint}}
\put(886.69,334.00){\usebox{\plotpoint}}
\put(907.44,334.00){\usebox{\plotpoint}}
\put(928.20,334.00){\usebox{\plotpoint}}
\put(948.95,334.00){\usebox{\plotpoint}}
\put(969.71,334.00){\usebox{\plotpoint}}
\put(990.46,334.00){\usebox{\plotpoint}}
\put(1011.22,334.00){\usebox{\plotpoint}}
\put(1031.98,334.00){\usebox{\plotpoint}}
\put(1052.73,334.00){\usebox{\plotpoint}}
\put(1073.49,334.00){\usebox{\plotpoint}}
\put(1094.24,334.00){\usebox{\plotpoint}}
\put(1115.00,334.00){\usebox{\plotpoint}}
\put(1135.75,334.00){\usebox{\plotpoint}}
\put(1156.51,334.00){\usebox{\plotpoint}}
\put(1177.26,334.00){\usebox{\plotpoint}}
\put(1198.02,334.00){\usebox{\plotpoint}}
\put(1218.77,334.00){\usebox{\plotpoint}}
\put(1239.53,334.00){\usebox{\plotpoint}}
\put(1260.29,334.00){\usebox{\plotpoint}}
\put(1281.04,334.00){\usebox{\plotpoint}}
\put(1301.80,334.00){\usebox{\plotpoint}}
\put(1322.55,334.00){\usebox{\plotpoint}}
\put(1343.31,334.00){\usebox{\plotpoint}}
\put(1364.06,334.00){\usebox{\plotpoint}}
\put(1384.82,334.00){\usebox{\plotpoint}}
\put(1405.57,334.00){\usebox{\plotpoint}}
\put(1426.33,334.00){\usebox{\plotpoint}}
\put(1439,334){\usebox{\plotpoint}}
\end{picture}

%% file: Delta-q.tex
\setlength{\unitlength}{0.240900pt}
\ifx\plotpoint\undefined\newsavebox{\plotpoint}\fi
\sbox{\plotpoint}{\rule[-0.200pt]{0.400pt}{0.400pt}}%
\begin{picture}(1500,900)(0,0)
\font\gnuplot=cmr10 at 10pt
\gnuplot
\sbox{\plotpoint}{\rule[-0.200pt]{0.400pt}{0.400pt}}%
\put(201.0,123.0){\rule[-0.200pt]{4.818pt}{0.400pt}}
\put(181,123){\makebox(0,0)[r]{0.004}}
\put(1419.0,123.0){\rule[-0.200pt]{4.818pt}{0.400pt}}
\put(201.0,215.0){\rule[-0.200pt]{4.818pt}{0.400pt}}
\put(181,215){\makebox(0,0)[r]{0.006}}
\put(1419.0,215.0){\rule[-0.200pt]{4.818pt}{0.400pt}}
\put(201.0,307.0){\rule[-0.200pt]{4.818pt}{0.400pt}}
\put(181,307){\makebox(0,0)[r]{0.008}}
\put(1419.0,307.0){\rule[-0.200pt]{4.818pt}{0.400pt}}
\put(201.0,399.0){\rule[-0.200pt]{4.818pt}{0.400pt}}
\put(181,399){\makebox(0,0)[r]{0.01}}
\put(1419.0,399.0){\rule[-0.200pt]{4.818pt}{0.400pt}}
\put(201.0,492.0){\rule[-0.200pt]{4.818pt}{0.400pt}}
\put(181,492){\makebox(0,0)[r]{0.012}}
\put(1419.0,492.0){\rule[-0.200pt]{4.818pt}{0.400pt}}
\put(201.0,584.0){\rule[-0.200pt]{4.818pt}{0.400pt}}
\put(181,584){\makebox(0,0)[r]{0.014}}
\put(1419.0,584.0){\rule[-0.200pt]{4.818pt}{0.400pt}}
\put(201.0,676.0){\rule[-0.200pt]{4.818pt}{0.400pt}}
\put(181,676){\makebox(0,0)[r]{0.016}}
\put(1419.0,676.0){\rule[-0.200pt]{4.818pt}{0.400pt}}
\put(201.0,768.0){\rule[-0.200pt]{4.818pt}{0.400pt}}
\put(181,768){\makebox(0,0)[r]{0.018}}
\put(1419.0,768.0){\rule[-0.200pt]{4.818pt}{0.400pt}}
\put(201.0,860.0){\rule[-0.200pt]{4.818pt}{0.400pt}}
\put(181,860){\makebox(0,0)[r]{0.02}}
\put(1419.0,860.0){\rule[-0.200pt]{4.818pt}{0.400pt}}
\put(201.0,123.0){\rule[-0.200pt]{0.400pt}{4.818pt}}
\put(201,82){\makebox(0,0){10}}
\put(201.0,840.0){\rule[-0.200pt]{0.400pt}{4.818pt}}
\put(407.0,123.0){\rule[-0.200pt]{0.400pt}{4.818pt}}
\put(407,82){\makebox(0,0){11}}
\put(407.0,840.0){\rule[-0.200pt]{0.400pt}{4.818pt}}
\put(614.0,123.0){\rule[-0.200pt]{0.400pt}{4.818pt}}
\put(614,82){\makebox(0,0){12}}
\put(614.0,840.0){\rule[-0.200pt]{0.400pt}{4.818pt}}
\put(820.0,123.0){\rule[-0.200pt]{0.400pt}{4.818pt}}
\put(820,82){\makebox(0,0){13}}
\put(820.0,840.0){\rule[-0.200pt]{0.400pt}{4.818pt}}
\put(1026.0,123.0){\rule[-0.200pt]{0.400pt}{4.818pt}}
\put(1026,82){\makebox(0,0){14}}
\put(1026.0,840.0){\rule[-0.200pt]{0.400pt}{4.818pt}}
\put(1233.0,123.0){\rule[-0.200pt]{0.400pt}{4.818pt}}
\put(1233,82){\makebox(0,0){15}}
\put(1233.0,840.0){\rule[-0.200pt]{0.400pt}{4.818pt}}
\put(1439.0,123.0){\rule[-0.200pt]{0.400pt}{4.818pt}}
\put(1439,82){\makebox(0,0){16}}
\put(1439.0,840.0){\rule[-0.200pt]{0.400pt}{4.818pt}}
\put(201.0,123.0){\rule[-0.200pt]{298.234pt}{0.400pt}}
\put(1439.0,123.0){\rule[-0.200pt]{0.400pt}{177.543pt}}
\put(201.0,860.0){\rule[-0.200pt]{298.234pt}{0.400pt}}
\put(40,491){\makebox(0,0){$\Delta_{\tilde f}$}}
\put(820,21){\makebox(0,0){$\log_{10} (M_C)$ [GeV]}}
\put(1026,492){\makebox(0,0)[r]{$\tilde Q$}}
\put(1026,215){\makebox(0,0)[r]{$\tilde d$}}
\put(201.0,123.0){\rule[-0.200pt]{0.400pt}{177.543pt}}
\sbox{\plotpoint}{\rule[-0.400pt]{0.800pt}{0.800pt}}%
\put(201,822){\usebox{\plotpoint}}
\multiput(201.00,820.08)(0.589,-0.512){15}{\rule{1.145pt}{0.123pt}}
\multiput(201.00,820.34)(10.623,-11.000){2}{\rule{0.573pt}{0.800pt}}
\multiput(214.00,809.08)(0.539,-0.512){15}{\rule{1.073pt}{0.123pt}}
\multiput(214.00,809.34)(9.774,-11.000){2}{\rule{0.536pt}{0.800pt}}
\multiput(226.00,798.08)(0.654,-0.514){13}{\rule{1.240pt}{0.124pt}}
\multiput(226.00,798.34)(10.426,-10.000){2}{\rule{0.620pt}{0.800pt}}
\multiput(239.00,788.08)(0.599,-0.514){13}{\rule{1.160pt}{0.124pt}}
\multiput(239.00,788.34)(9.592,-10.000){2}{\rule{0.580pt}{0.800pt}}
\multiput(251.00,778.08)(0.654,-0.514){13}{\rule{1.240pt}{0.124pt}}
\multiput(251.00,778.34)(10.426,-10.000){2}{\rule{0.620pt}{0.800pt}}
\multiput(264.00,768.08)(0.599,-0.514){13}{\rule{1.160pt}{0.124pt}}
\multiput(264.00,768.34)(9.592,-10.000){2}{\rule{0.580pt}{0.800pt}}
\multiput(276.00,758.08)(0.654,-0.514){13}{\rule{1.240pt}{0.124pt}}
\multiput(276.00,758.34)(10.426,-10.000){2}{\rule{0.620pt}{0.800pt}}
\multiput(289.00,748.08)(0.674,-0.516){11}{\rule{1.267pt}{0.124pt}}
\multiput(289.00,748.34)(9.371,-9.000){2}{\rule{0.633pt}{0.800pt}}
\multiput(301.00,739.08)(0.737,-0.516){11}{\rule{1.356pt}{0.124pt}}
\multiput(301.00,739.34)(10.186,-9.000){2}{\rule{0.678pt}{0.800pt}}
\multiput(314.00,730.08)(0.674,-0.516){11}{\rule{1.267pt}{0.124pt}}
\multiput(314.00,730.34)(9.371,-9.000){2}{\rule{0.633pt}{0.800pt}}
\multiput(326.00,721.08)(0.737,-0.516){11}{\rule{1.356pt}{0.124pt}}
\multiput(326.00,721.34)(10.186,-9.000){2}{\rule{0.678pt}{0.800pt}}
\multiput(339.00,712.08)(0.674,-0.516){11}{\rule{1.267pt}{0.124pt}}
\multiput(339.00,712.34)(9.371,-9.000){2}{\rule{0.633pt}{0.800pt}}
\multiput(351.00,703.08)(0.847,-0.520){9}{\rule{1.500pt}{0.125pt}}
\multiput(351.00,703.34)(9.887,-8.000){2}{\rule{0.750pt}{0.800pt}}
\multiput(364.00,695.08)(0.674,-0.516){11}{\rule{1.267pt}{0.124pt}}
\multiput(364.00,695.34)(9.371,-9.000){2}{\rule{0.633pt}{0.800pt}}
\multiput(376.00,686.08)(0.847,-0.520){9}{\rule{1.500pt}{0.125pt}}
\multiput(376.00,686.34)(9.887,-8.000){2}{\rule{0.750pt}{0.800pt}}
\multiput(389.00,678.08)(0.774,-0.520){9}{\rule{1.400pt}{0.125pt}}
\multiput(389.00,678.34)(9.094,-8.000){2}{\rule{0.700pt}{0.800pt}}
\multiput(401.00,670.08)(0.847,-0.520){9}{\rule{1.500pt}{0.125pt}}
\multiput(401.00,670.34)(9.887,-8.000){2}{\rule{0.750pt}{0.800pt}}
\multiput(414.00,662.08)(0.774,-0.520){9}{\rule{1.400pt}{0.125pt}}
\multiput(414.00,662.34)(9.094,-8.000){2}{\rule{0.700pt}{0.800pt}}
\multiput(426.00,654.08)(0.847,-0.520){9}{\rule{1.500pt}{0.125pt}}
\multiput(426.00,654.34)(9.887,-8.000){2}{\rule{0.750pt}{0.800pt}}
\multiput(439.00,646.08)(0.913,-0.526){7}{\rule{1.571pt}{0.127pt}}
\multiput(439.00,646.34)(8.738,-7.000){2}{\rule{0.786pt}{0.800pt}}
\multiput(451.00,639.08)(1.000,-0.526){7}{\rule{1.686pt}{0.127pt}}
\multiput(451.00,639.34)(9.501,-7.000){2}{\rule{0.843pt}{0.800pt}}
\multiput(464.00,632.08)(0.774,-0.520){9}{\rule{1.400pt}{0.125pt}}
\multiput(464.00,632.34)(9.094,-8.000){2}{\rule{0.700pt}{0.800pt}}
\multiput(476.00,624.08)(1.000,-0.526){7}{\rule{1.686pt}{0.127pt}}
\multiput(476.00,624.34)(9.501,-7.000){2}{\rule{0.843pt}{0.800pt}}
\multiput(489.00,617.08)(0.913,-0.526){7}{\rule{1.571pt}{0.127pt}}
\multiput(489.00,617.34)(8.738,-7.000){2}{\rule{0.786pt}{0.800pt}}
\multiput(501.00,610.08)(1.000,-0.526){7}{\rule{1.686pt}{0.127pt}}
\multiput(501.00,610.34)(9.501,-7.000){2}{\rule{0.843pt}{0.800pt}}
\multiput(514.00,603.07)(1.132,-0.536){5}{\rule{1.800pt}{0.129pt}}
\multiput(514.00,603.34)(8.264,-6.000){2}{\rule{0.900pt}{0.800pt}}
\multiput(526.00,597.08)(1.000,-0.526){7}{\rule{1.686pt}{0.127pt}}
\multiput(526.00,597.34)(9.501,-7.000){2}{\rule{0.843pt}{0.800pt}}
\multiput(539.00,590.08)(0.913,-0.526){7}{\rule{1.571pt}{0.127pt}}
\multiput(539.00,590.34)(8.738,-7.000){2}{\rule{0.786pt}{0.800pt}}
\multiput(551.00,583.07)(1.244,-0.536){5}{\rule{1.933pt}{0.129pt}}
\multiput(551.00,583.34)(8.987,-6.000){2}{\rule{0.967pt}{0.800pt}}
\multiput(564.00,577.07)(1.132,-0.536){5}{\rule{1.800pt}{0.129pt}}
\multiput(564.00,577.34)(8.264,-6.000){2}{\rule{0.900pt}{0.800pt}}
\multiput(576.00,571.08)(1.000,-0.526){7}{\rule{1.686pt}{0.127pt}}
\multiput(576.00,571.34)(9.501,-7.000){2}{\rule{0.843pt}{0.800pt}}
\multiput(589.00,564.07)(1.132,-0.536){5}{\rule{1.800pt}{0.129pt}}
\multiput(589.00,564.34)(8.264,-6.000){2}{\rule{0.900pt}{0.800pt}}
\multiput(601.00,558.07)(1.244,-0.536){5}{\rule{1.933pt}{0.129pt}}
\multiput(601.00,558.34)(8.987,-6.000){2}{\rule{0.967pt}{0.800pt}}
\multiput(614.00,552.06)(1.600,-0.560){3}{\rule{2.120pt}{0.135pt}}
\multiput(614.00,552.34)(7.600,-5.000){2}{\rule{1.060pt}{0.800pt}}
\multiput(626.00,547.07)(1.244,-0.536){5}{\rule{1.933pt}{0.129pt}}
\multiput(626.00,547.34)(8.987,-6.000){2}{\rule{0.967pt}{0.800pt}}
\multiput(639.00,541.07)(1.132,-0.536){5}{\rule{1.800pt}{0.129pt}}
\multiput(639.00,541.34)(8.264,-6.000){2}{\rule{0.900pt}{0.800pt}}
\multiput(651.00,535.06)(1.768,-0.560){3}{\rule{2.280pt}{0.135pt}}
\multiput(651.00,535.34)(8.268,-5.000){2}{\rule{1.140pt}{0.800pt}}
\multiput(664.00,530.07)(1.132,-0.536){5}{\rule{1.800pt}{0.129pt}}
\multiput(664.00,530.34)(8.264,-6.000){2}{\rule{0.900pt}{0.800pt}}
\multiput(676.00,524.06)(1.768,-0.560){3}{\rule{2.280pt}{0.135pt}}
\multiput(676.00,524.34)(8.268,-5.000){2}{\rule{1.140pt}{0.800pt}}
\multiput(689.00,519.07)(1.132,-0.536){5}{\rule{1.800pt}{0.129pt}}
\multiput(689.00,519.34)(8.264,-6.000){2}{\rule{0.900pt}{0.800pt}}
\multiput(701.00,513.06)(1.768,-0.560){3}{\rule{2.280pt}{0.135pt}}
\multiput(701.00,513.34)(8.268,-5.000){2}{\rule{1.140pt}{0.800pt}}
\multiput(714.00,508.06)(1.600,-0.560){3}{\rule{2.120pt}{0.135pt}}
\multiput(714.00,508.34)(7.600,-5.000){2}{\rule{1.060pt}{0.800pt}}
\multiput(726.00,503.06)(1.768,-0.560){3}{\rule{2.280pt}{0.135pt}}
\multiput(726.00,503.34)(8.268,-5.000){2}{\rule{1.140pt}{0.800pt}}
\multiput(739.00,498.06)(1.600,-0.560){3}{\rule{2.120pt}{0.135pt}}
\multiput(739.00,498.34)(7.600,-5.000){2}{\rule{1.060pt}{0.800pt}}
\multiput(751.00,493.06)(1.768,-0.560){3}{\rule{2.280pt}{0.135pt}}
\multiput(751.00,493.34)(8.268,-5.000){2}{\rule{1.140pt}{0.800pt}}
\multiput(764.00,488.06)(1.600,-0.560){3}{\rule{2.120pt}{0.135pt}}
\multiput(764.00,488.34)(7.600,-5.000){2}{\rule{1.060pt}{0.800pt}}
\put(776,481.34){\rule{2.800pt}{0.800pt}}
\multiput(776.00,483.34)(7.188,-4.000){2}{\rule{1.400pt}{0.800pt}}
\multiput(789.00,479.06)(1.600,-0.560){3}{\rule{2.120pt}{0.135pt}}
\multiput(789.00,479.34)(7.600,-5.000){2}{\rule{1.060pt}{0.800pt}}
\multiput(801.00,474.06)(1.768,-0.560){3}{\rule{2.280pt}{0.135pt}}
\multiput(801.00,474.34)(8.268,-5.000){2}{\rule{1.140pt}{0.800pt}}
\put(814,467.34){\rule{2.600pt}{0.800pt}}
\multiput(814.00,469.34)(6.604,-4.000){2}{\rule{1.300pt}{0.800pt}}
\multiput(826.00,465.06)(1.768,-0.560){3}{\rule{2.280pt}{0.135pt}}
\multiput(826.00,465.34)(8.268,-5.000){2}{\rule{1.140pt}{0.800pt}}
\put(839,458.34){\rule{2.600pt}{0.800pt}}
\multiput(839.00,460.34)(6.604,-4.000){2}{\rule{1.300pt}{0.800pt}}
\put(851,454.34){\rule{2.800pt}{0.800pt}}
\multiput(851.00,456.34)(7.188,-4.000){2}{\rule{1.400pt}{0.800pt}}
\put(864,450.34){\rule{2.600pt}{0.800pt}}
\multiput(864.00,452.34)(6.604,-4.000){2}{\rule{1.300pt}{0.800pt}}
\multiput(876.00,448.06)(1.768,-0.560){3}{\rule{2.280pt}{0.135pt}}
\multiput(876.00,448.34)(8.268,-5.000){2}{\rule{1.140pt}{0.800pt}}
\put(889,441.34){\rule{2.600pt}{0.800pt}}
\multiput(889.00,443.34)(6.604,-4.000){2}{\rule{1.300pt}{0.800pt}}
\put(901,437.34){\rule{2.800pt}{0.800pt}}
\multiput(901.00,439.34)(7.188,-4.000){2}{\rule{1.400pt}{0.800pt}}
\put(914,433.34){\rule{2.600pt}{0.800pt}}
\multiput(914.00,435.34)(6.604,-4.000){2}{\rule{1.300pt}{0.800pt}}
\put(926,429.84){\rule{3.132pt}{0.800pt}}
\multiput(926.00,431.34)(6.500,-3.000){2}{\rule{1.566pt}{0.800pt}}
\put(939,426.34){\rule{2.600pt}{0.800pt}}
\multiput(939.00,428.34)(6.604,-4.000){2}{\rule{1.300pt}{0.800pt}}
\put(951,422.34){\rule{2.800pt}{0.800pt}}
\multiput(951.00,424.34)(7.188,-4.000){2}{\rule{1.400pt}{0.800pt}}
\put(964,418.34){\rule{2.600pt}{0.800pt}}
\multiput(964.00,420.34)(6.604,-4.000){2}{\rule{1.300pt}{0.800pt}}
\put(976,414.84){\rule{3.132pt}{0.800pt}}
\multiput(976.00,416.34)(6.500,-3.000){2}{\rule{1.566pt}{0.800pt}}
\put(989,411.34){\rule{2.600pt}{0.800pt}}
\multiput(989.00,413.34)(6.604,-4.000){2}{\rule{1.300pt}{0.800pt}}
\put(1001,407.34){\rule{2.800pt}{0.800pt}}
\multiput(1001.00,409.34)(7.188,-4.000){2}{\rule{1.400pt}{0.800pt}}
\put(1014,403.84){\rule{2.891pt}{0.800pt}}
\multiput(1014.00,405.34)(6.000,-3.000){2}{\rule{1.445pt}{0.800pt}}
\put(1026,400.84){\rule{3.132pt}{0.800pt}}
\multiput(1026.00,402.34)(6.500,-3.000){2}{\rule{1.566pt}{0.800pt}}
\put(1039,397.34){\rule{2.600pt}{0.800pt}}
\multiput(1039.00,399.34)(6.604,-4.000){2}{\rule{1.300pt}{0.800pt}}
\put(1051,393.84){\rule{3.132pt}{0.800pt}}
\multiput(1051.00,395.34)(6.500,-3.000){2}{\rule{1.566pt}{0.800pt}}
\put(1064,390.84){\rule{2.891pt}{0.800pt}}
\multiput(1064.00,392.34)(6.000,-3.000){2}{\rule{1.445pt}{0.800pt}}
\put(1076,387.34){\rule{2.800pt}{0.800pt}}
\multiput(1076.00,389.34)(7.188,-4.000){2}{\rule{1.400pt}{0.800pt}}
\put(1089,383.84){\rule{2.891pt}{0.800pt}}
\multiput(1089.00,385.34)(6.000,-3.000){2}{\rule{1.445pt}{0.800pt}}
\put(1101,380.84){\rule{3.132pt}{0.800pt}}
\multiput(1101.00,382.34)(6.500,-3.000){2}{\rule{1.566pt}{0.800pt}}
\put(1114,377.84){\rule{2.891pt}{0.800pt}}
\multiput(1114.00,379.34)(6.000,-3.000){2}{\rule{1.445pt}{0.800pt}}
\put(1126,374.84){\rule{3.132pt}{0.800pt}}
\multiput(1126.00,376.34)(6.500,-3.000){2}{\rule{1.566pt}{0.800pt}}
\put(1139,371.84){\rule{2.891pt}{0.800pt}}
\multiput(1139.00,373.34)(6.000,-3.000){2}{\rule{1.445pt}{0.800pt}}
\put(1151,368.84){\rule{3.132pt}{0.800pt}}
\multiput(1151.00,370.34)(6.500,-3.000){2}{\rule{1.566pt}{0.800pt}}
\put(1164,365.84){\rule{2.891pt}{0.800pt}}
\multiput(1164.00,367.34)(6.000,-3.000){2}{\rule{1.445pt}{0.800pt}}
\put(1176,362.84){\rule{3.132pt}{0.800pt}}
\multiput(1176.00,364.34)(6.500,-3.000){2}{\rule{1.566pt}{0.800pt}}
\put(1189,360.34){\rule{2.891pt}{0.800pt}}
\multiput(1189.00,361.34)(6.000,-2.000){2}{\rule{1.445pt}{0.800pt}}
\put(1201,357.84){\rule{3.132pt}{0.800pt}}
\multiput(1201.00,359.34)(6.500,-3.000){2}{\rule{1.566pt}{0.800pt}}
\put(1214,354.84){\rule{2.891pt}{0.800pt}}
\multiput(1214.00,356.34)(6.000,-3.000){2}{\rule{1.445pt}{0.800pt}}
\put(1226,351.84){\rule{3.132pt}{0.800pt}}
\multiput(1226.00,353.34)(6.500,-3.000){2}{\rule{1.566pt}{0.800pt}}
\put(1239,349.34){\rule{2.891pt}{0.800pt}}
\multiput(1239.00,350.34)(6.000,-2.000){2}{\rule{1.445pt}{0.800pt}}
\put(1251,346.84){\rule{3.132pt}{0.800pt}}
\multiput(1251.00,348.34)(6.500,-3.000){2}{\rule{1.566pt}{0.800pt}}
\put(1264,344.34){\rule{2.891pt}{0.800pt}}
\multiput(1264.00,345.34)(6.000,-2.000){2}{\rule{1.445pt}{0.800pt}}
\put(1276,341.84){\rule{3.132pt}{0.800pt}}
\multiput(1276.00,343.34)(6.500,-3.000){2}{\rule{1.566pt}{0.800pt}}
\put(1289,339.34){\rule{2.891pt}{0.800pt}}
\multiput(1289.00,340.34)(6.000,-2.000){2}{\rule{1.445pt}{0.800pt}}
\put(1301,336.84){\rule{3.132pt}{0.800pt}}
\multiput(1301.00,338.34)(6.500,-3.000){2}{\rule{1.566pt}{0.800pt}}
\put(1314,334.34){\rule{2.891pt}{0.800pt}}
\multiput(1314.00,335.34)(6.000,-2.000){2}{\rule{1.445pt}{0.800pt}}
\put(1326,331.84){\rule{3.132pt}{0.800pt}}
\multiput(1326.00,333.34)(6.500,-3.000){2}{\rule{1.566pt}{0.800pt}}
\put(1339,329.34){\rule{2.891pt}{0.800pt}}
\multiput(1339.00,330.34)(6.000,-2.000){2}{\rule{1.445pt}{0.800pt}}
\put(1351,327.34){\rule{3.132pt}{0.800pt}}
\multiput(1351.00,328.34)(6.500,-2.000){2}{\rule{1.566pt}{0.800pt}}
\put(1364,324.84){\rule{2.891pt}{0.800pt}}
\multiput(1364.00,326.34)(6.000,-3.000){2}{\rule{1.445pt}{0.800pt}}
\put(1376,322.34){\rule{3.132pt}{0.800pt}}
\multiput(1376.00,323.34)(6.500,-2.000){2}{\rule{1.566pt}{0.800pt}}
\put(1389,320.34){\rule{2.891pt}{0.800pt}}
\multiput(1389.00,321.34)(6.000,-2.000){2}{\rule{1.445pt}{0.800pt}}
\put(1401,318.34){\rule{3.132pt}{0.800pt}}
\multiput(1401.00,319.34)(6.500,-2.000){2}{\rule{1.566pt}{0.800pt}}
\put(1414,316.34){\rule{2.891pt}{0.800pt}}
\multiput(1414.00,317.34)(6.000,-2.000){2}{\rule{1.445pt}{0.800pt}}
\put(1426,314.34){\rule{3.132pt}{0.800pt}}
\multiput(1426.00,315.34)(6.500,-2.000){2}{\rule{1.566pt}{0.800pt}}
\put(201,739){\usebox{\plotpoint}}
\multiput(201.00,737.08)(0.589,-0.512){15}{\rule{1.145pt}{0.123pt}}
\multiput(201.00,737.34)(10.623,-11.000){2}{\rule{0.573pt}{0.800pt}}
\multiput(214.00,726.08)(0.539,-0.512){15}{\rule{1.073pt}{0.123pt}}
\multiput(214.00,726.34)(9.774,-11.000){2}{\rule{0.536pt}{0.800pt}}
\multiput(226.00,715.08)(0.589,-0.512){15}{\rule{1.145pt}{0.123pt}}
\multiput(226.00,715.34)(10.623,-11.000){2}{\rule{0.573pt}{0.800pt}}
\multiput(239.00,704.08)(0.599,-0.514){13}{\rule{1.160pt}{0.124pt}}
\multiput(239.00,704.34)(9.592,-10.000){2}{\rule{0.580pt}{0.800pt}}
\multiput(251.00,694.08)(0.654,-0.514){13}{\rule{1.240pt}{0.124pt}}
\multiput(251.00,694.34)(10.426,-10.000){2}{\rule{0.620pt}{0.800pt}}
\multiput(264.00,684.08)(0.599,-0.514){13}{\rule{1.160pt}{0.124pt}}
\multiput(264.00,684.34)(9.592,-10.000){2}{\rule{0.580pt}{0.800pt}}
\multiput(276.00,674.08)(0.654,-0.514){13}{\rule{1.240pt}{0.124pt}}
\multiput(276.00,674.34)(10.426,-10.000){2}{\rule{0.620pt}{0.800pt}}
\multiput(289.00,664.08)(0.599,-0.514){13}{\rule{1.160pt}{0.124pt}}
\multiput(289.00,664.34)(9.592,-10.000){2}{\rule{0.580pt}{0.800pt}}
\multiput(301.00,654.08)(0.737,-0.516){11}{\rule{1.356pt}{0.124pt}}
\multiput(301.00,654.34)(10.186,-9.000){2}{\rule{0.678pt}{0.800pt}}
\multiput(314.00,645.08)(0.674,-0.516){11}{\rule{1.267pt}{0.124pt}}
\multiput(314.00,645.34)(9.371,-9.000){2}{\rule{0.633pt}{0.800pt}}
\multiput(326.00,636.08)(0.737,-0.516){11}{\rule{1.356pt}{0.124pt}}
\multiput(326.00,636.34)(10.186,-9.000){2}{\rule{0.678pt}{0.800pt}}
\multiput(339.00,627.08)(0.674,-0.516){11}{\rule{1.267pt}{0.124pt}}
\multiput(339.00,627.34)(9.371,-9.000){2}{\rule{0.633pt}{0.800pt}}
\multiput(351.00,618.08)(0.737,-0.516){11}{\rule{1.356pt}{0.124pt}}
\multiput(351.00,618.34)(10.186,-9.000){2}{\rule{0.678pt}{0.800pt}}
\multiput(364.00,609.08)(0.674,-0.516){11}{\rule{1.267pt}{0.124pt}}
\multiput(364.00,609.34)(9.371,-9.000){2}{\rule{0.633pt}{0.800pt}}
\multiput(376.00,600.08)(0.847,-0.520){9}{\rule{1.500pt}{0.125pt}}
\multiput(376.00,600.34)(9.887,-8.000){2}{\rule{0.750pt}{0.800pt}}
\multiput(389.00,592.08)(0.674,-0.516){11}{\rule{1.267pt}{0.124pt}}
\multiput(389.00,592.34)(9.371,-9.000){2}{\rule{0.633pt}{0.800pt}}
\multiput(401.00,583.08)(0.847,-0.520){9}{\rule{1.500pt}{0.125pt}}
\multiput(401.00,583.34)(9.887,-8.000){2}{\rule{0.750pt}{0.800pt}}
\multiput(414.00,575.08)(0.774,-0.520){9}{\rule{1.400pt}{0.125pt}}
\multiput(414.00,575.34)(9.094,-8.000){2}{\rule{0.700pt}{0.800pt}}
\multiput(426.00,567.08)(0.847,-0.520){9}{\rule{1.500pt}{0.125pt}}
\multiput(426.00,567.34)(9.887,-8.000){2}{\rule{0.750pt}{0.800pt}}
\multiput(439.00,559.08)(0.913,-0.526){7}{\rule{1.571pt}{0.127pt}}
\multiput(439.00,559.34)(8.738,-7.000){2}{\rule{0.786pt}{0.800pt}}
\multiput(451.00,552.08)(0.847,-0.520){9}{\rule{1.500pt}{0.125pt}}
\multiput(451.00,552.34)(9.887,-8.000){2}{\rule{0.750pt}{0.800pt}}
\multiput(464.00,544.08)(0.913,-0.526){7}{\rule{1.571pt}{0.127pt}}
\multiput(464.00,544.34)(8.738,-7.000){2}{\rule{0.786pt}{0.800pt}}
\multiput(476.00,537.08)(0.847,-0.520){9}{\rule{1.500pt}{0.125pt}}
\multiput(476.00,537.34)(9.887,-8.000){2}{\rule{0.750pt}{0.800pt}}
\multiput(489.00,529.08)(0.913,-0.526){7}{\rule{1.571pt}{0.127pt}}
\multiput(489.00,529.34)(8.738,-7.000){2}{\rule{0.786pt}{0.800pt}}
\multiput(501.00,522.08)(1.000,-0.526){7}{\rule{1.686pt}{0.127pt}}
\multiput(501.00,522.34)(9.501,-7.000){2}{\rule{0.843pt}{0.800pt}}
\multiput(514.00,515.08)(0.913,-0.526){7}{\rule{1.571pt}{0.127pt}}
\multiput(514.00,515.34)(8.738,-7.000){2}{\rule{0.786pt}{0.800pt}}
\multiput(526.00,508.08)(1.000,-0.526){7}{\rule{1.686pt}{0.127pt}}
\multiput(526.00,508.34)(9.501,-7.000){2}{\rule{0.843pt}{0.800pt}}
\multiput(539.00,501.07)(1.132,-0.536){5}{\rule{1.800pt}{0.129pt}}
\multiput(539.00,501.34)(8.264,-6.000){2}{\rule{0.900pt}{0.800pt}}
\multiput(551.00,495.08)(1.000,-0.526){7}{\rule{1.686pt}{0.127pt}}
\multiput(551.00,495.34)(9.501,-7.000){2}{\rule{0.843pt}{0.800pt}}
\multiput(564.00,488.08)(0.913,-0.526){7}{\rule{1.571pt}{0.127pt}}
\multiput(564.00,488.34)(8.738,-7.000){2}{\rule{0.786pt}{0.800pt}}
\multiput(576.00,481.07)(1.244,-0.536){5}{\rule{1.933pt}{0.129pt}}
\multiput(576.00,481.34)(8.987,-6.000){2}{\rule{0.967pt}{0.800pt}}
\multiput(589.00,475.07)(1.132,-0.536){5}{\rule{1.800pt}{0.129pt}}
\multiput(589.00,475.34)(8.264,-6.000){2}{\rule{0.900pt}{0.800pt}}
\multiput(601.00,469.07)(1.244,-0.536){5}{\rule{1.933pt}{0.129pt}}
\multiput(601.00,469.34)(8.987,-6.000){2}{\rule{0.967pt}{0.800pt}}
\multiput(614.00,463.08)(0.913,-0.526){7}{\rule{1.571pt}{0.127pt}}
\multiput(614.00,463.34)(8.738,-7.000){2}{\rule{0.786pt}{0.800pt}}
\multiput(626.00,456.06)(1.768,-0.560){3}{\rule{2.280pt}{0.135pt}}
\multiput(626.00,456.34)(8.268,-5.000){2}{\rule{1.140pt}{0.800pt}}
\multiput(639.00,451.07)(1.132,-0.536){5}{\rule{1.800pt}{0.129pt}}
\multiput(639.00,451.34)(8.264,-6.000){2}{\rule{0.900pt}{0.800pt}}
\multiput(651.00,445.07)(1.244,-0.536){5}{\rule{1.933pt}{0.129pt}}
\multiput(651.00,445.34)(8.987,-6.000){2}{\rule{0.967pt}{0.800pt}}
\multiput(664.00,439.07)(1.132,-0.536){5}{\rule{1.800pt}{0.129pt}}
\multiput(664.00,439.34)(8.264,-6.000){2}{\rule{0.900pt}{0.800pt}}
\multiput(676.00,433.06)(1.768,-0.560){3}{\rule{2.280pt}{0.135pt}}
\multiput(676.00,433.34)(8.268,-5.000){2}{\rule{1.140pt}{0.800pt}}
\multiput(689.00,428.07)(1.132,-0.536){5}{\rule{1.800pt}{0.129pt}}
\multiput(689.00,428.34)(8.264,-6.000){2}{\rule{0.900pt}{0.800pt}}
\multiput(701.00,422.06)(1.768,-0.560){3}{\rule{2.280pt}{0.135pt}}
\multiput(701.00,422.34)(8.268,-5.000){2}{\rule{1.140pt}{0.800pt}}
\multiput(714.00,417.07)(1.132,-0.536){5}{\rule{1.800pt}{0.129pt}}
\multiput(714.00,417.34)(8.264,-6.000){2}{\rule{0.900pt}{0.800pt}}
\multiput(726.00,411.06)(1.768,-0.560){3}{\rule{2.280pt}{0.135pt}}
\multiput(726.00,411.34)(8.268,-5.000){2}{\rule{1.140pt}{0.800pt}}
\multiput(739.00,406.06)(1.600,-0.560){3}{\rule{2.120pt}{0.135pt}}
\multiput(739.00,406.34)(7.600,-5.000){2}{\rule{1.060pt}{0.800pt}}
\multiput(751.00,401.06)(1.768,-0.560){3}{\rule{2.280pt}{0.135pt}}
\multiput(751.00,401.34)(8.268,-5.000){2}{\rule{1.140pt}{0.800pt}}
\multiput(764.00,396.06)(1.600,-0.560){3}{\rule{2.120pt}{0.135pt}}
\multiput(764.00,396.34)(7.600,-5.000){2}{\rule{1.060pt}{0.800pt}}
\multiput(776.00,391.06)(1.768,-0.560){3}{\rule{2.280pt}{0.135pt}}
\multiput(776.00,391.34)(8.268,-5.000){2}{\rule{1.140pt}{0.800pt}}
\multiput(789.00,386.06)(1.600,-0.560){3}{\rule{2.120pt}{0.135pt}}
\multiput(789.00,386.34)(7.600,-5.000){2}{\rule{1.060pt}{0.800pt}}
\multiput(801.00,381.06)(1.768,-0.560){3}{\rule{2.280pt}{0.135pt}}
\multiput(801.00,381.34)(8.268,-5.000){2}{\rule{1.140pt}{0.800pt}}
\put(814,374.34){\rule{2.600pt}{0.800pt}}
\multiput(814.00,376.34)(6.604,-4.000){2}{\rule{1.300pt}{0.800pt}}
\multiput(826.00,372.06)(1.768,-0.560){3}{\rule{2.280pt}{0.135pt}}
\multiput(826.00,372.34)(8.268,-5.000){2}{\rule{1.140pt}{0.800pt}}
\multiput(839.00,367.06)(1.600,-0.560){3}{\rule{2.120pt}{0.135pt}}
\multiput(839.00,367.34)(7.600,-5.000){2}{\rule{1.060pt}{0.800pt}}
\put(851,360.34){\rule{2.800pt}{0.800pt}}
\multiput(851.00,362.34)(7.188,-4.000){2}{\rule{1.400pt}{0.800pt}}
\put(864,356.34){\rule{2.600pt}{0.800pt}}
\multiput(864.00,358.34)(6.604,-4.000){2}{\rule{1.300pt}{0.800pt}}
\multiput(876.00,354.06)(1.768,-0.560){3}{\rule{2.280pt}{0.135pt}}
\multiput(876.00,354.34)(8.268,-5.000){2}{\rule{1.140pt}{0.800pt}}
\put(889,347.34){\rule{2.600pt}{0.800pt}}
\multiput(889.00,349.34)(6.604,-4.000){2}{\rule{1.300pt}{0.800pt}}
\put(901,343.34){\rule{2.800pt}{0.800pt}}
\multiput(901.00,345.34)(7.188,-4.000){2}{\rule{1.400pt}{0.800pt}}
\put(914,339.34){\rule{2.600pt}{0.800pt}}
\multiput(914.00,341.34)(6.604,-4.000){2}{\rule{1.300pt}{0.800pt}}
\multiput(926.00,337.06)(1.768,-0.560){3}{\rule{2.280pt}{0.135pt}}
\multiput(926.00,337.34)(8.268,-5.000){2}{\rule{1.140pt}{0.800pt}}
\put(939,330.34){\rule{2.600pt}{0.800pt}}
\multiput(939.00,332.34)(6.604,-4.000){2}{\rule{1.300pt}{0.800pt}}
\put(951,326.34){\rule{2.800pt}{0.800pt}}
\multiput(951.00,328.34)(7.188,-4.000){2}{\rule{1.400pt}{0.800pt}}
\put(964,322.84){\rule{2.891pt}{0.800pt}}
\multiput(964.00,324.34)(6.000,-3.000){2}{\rule{1.445pt}{0.800pt}}
\put(976,319.34){\rule{2.800pt}{0.800pt}}
\multiput(976.00,321.34)(7.188,-4.000){2}{\rule{1.400pt}{0.800pt}}
\put(989,315.34){\rule{2.600pt}{0.800pt}}
\multiput(989.00,317.34)(6.604,-4.000){2}{\rule{1.300pt}{0.800pt}}
\put(1001,311.34){\rule{2.800pt}{0.800pt}}
\multiput(1001.00,313.34)(7.188,-4.000){2}{\rule{1.400pt}{0.800pt}}
\put(1014,307.84){\rule{2.891pt}{0.800pt}}
\multiput(1014.00,309.34)(6.000,-3.000){2}{\rule{1.445pt}{0.800pt}}
\put(1026,304.34){\rule{2.800pt}{0.800pt}}
\multiput(1026.00,306.34)(7.188,-4.000){2}{\rule{1.400pt}{0.800pt}}
\put(1039,300.34){\rule{2.600pt}{0.800pt}}
\multiput(1039.00,302.34)(6.604,-4.000){2}{\rule{1.300pt}{0.800pt}}
\put(1051,296.84){\rule{3.132pt}{0.800pt}}
\multiput(1051.00,298.34)(6.500,-3.000){2}{\rule{1.566pt}{0.800pt}}
\put(1064,293.34){\rule{2.600pt}{0.800pt}}
\multiput(1064.00,295.34)(6.604,-4.000){2}{\rule{1.300pt}{0.800pt}}
\put(1076,289.84){\rule{3.132pt}{0.800pt}}
\multiput(1076.00,291.34)(6.500,-3.000){2}{\rule{1.566pt}{0.800pt}}
\put(1089,286.84){\rule{2.891pt}{0.800pt}}
\multiput(1089.00,288.34)(6.000,-3.000){2}{\rule{1.445pt}{0.800pt}}
\put(1101,283.34){\rule{2.800pt}{0.800pt}}
\multiput(1101.00,285.34)(7.188,-4.000){2}{\rule{1.400pt}{0.800pt}}
\put(1114,279.84){\rule{2.891pt}{0.800pt}}
\multiput(1114.00,281.34)(6.000,-3.000){2}{\rule{1.445pt}{0.800pt}}
\put(1126,276.84){\rule{3.132pt}{0.800pt}}
\multiput(1126.00,278.34)(6.500,-3.000){2}{\rule{1.566pt}{0.800pt}}
\put(1139,273.84){\rule{2.891pt}{0.800pt}}
\multiput(1139.00,275.34)(6.000,-3.000){2}{\rule{1.445pt}{0.800pt}}
\put(1151,270.84){\rule{3.132pt}{0.800pt}}
\multiput(1151.00,272.34)(6.500,-3.000){2}{\rule{1.566pt}{0.800pt}}
\put(1164,267.34){\rule{2.600pt}{0.800pt}}
\multiput(1164.00,269.34)(6.604,-4.000){2}{\rule{1.300pt}{0.800pt}}
\put(1176,263.84){\rule{3.132pt}{0.800pt}}
\multiput(1176.00,265.34)(6.500,-3.000){2}{\rule{1.566pt}{0.800pt}}
\put(1189,260.84){\rule{2.891pt}{0.800pt}}
\multiput(1189.00,262.34)(6.000,-3.000){2}{\rule{1.445pt}{0.800pt}}
\put(1201,258.34){\rule{3.132pt}{0.800pt}}
\multiput(1201.00,259.34)(6.500,-2.000){2}{\rule{1.566pt}{0.800pt}}
\put(1214,255.84){\rule{2.891pt}{0.800pt}}
\multiput(1214.00,257.34)(6.000,-3.000){2}{\rule{1.445pt}{0.800pt}}
\put(1226,252.84){\rule{3.132pt}{0.800pt}}
\multiput(1226.00,254.34)(6.500,-3.000){2}{\rule{1.566pt}{0.800pt}}
\put(1239,249.84){\rule{2.891pt}{0.800pt}}
\multiput(1239.00,251.34)(6.000,-3.000){2}{\rule{1.445pt}{0.800pt}}
\put(1251,246.84){\rule{3.132pt}{0.800pt}}
\multiput(1251.00,248.34)(6.500,-3.000){2}{\rule{1.566pt}{0.800pt}}
\put(1264,243.84){\rule{2.891pt}{0.800pt}}
\multiput(1264.00,245.34)(6.000,-3.000){2}{\rule{1.445pt}{0.800pt}}
\put(1276,241.34){\rule{3.132pt}{0.800pt}}
\multiput(1276.00,242.34)(6.500,-2.000){2}{\rule{1.566pt}{0.800pt}}
\put(1289,238.84){\rule{2.891pt}{0.800pt}}
\multiput(1289.00,240.34)(6.000,-3.000){2}{\rule{1.445pt}{0.800pt}}
\put(1301,236.34){\rule{3.132pt}{0.800pt}}
\multiput(1301.00,237.34)(6.500,-2.000){2}{\rule{1.566pt}{0.800pt}}
\put(1314,233.84){\rule{2.891pt}{0.800pt}}
\multiput(1314.00,235.34)(6.000,-3.000){2}{\rule{1.445pt}{0.800pt}}
\put(1326,230.84){\rule{3.132pt}{0.800pt}}
\multiput(1326.00,232.34)(6.500,-3.000){2}{\rule{1.566pt}{0.800pt}}
\put(1339,228.34){\rule{2.891pt}{0.800pt}}
\multiput(1339.00,229.34)(6.000,-2.000){2}{\rule{1.445pt}{0.800pt}}
\put(1351,225.84){\rule{3.132pt}{0.800pt}}
\multiput(1351.00,227.34)(6.500,-3.000){2}{\rule{1.566pt}{0.800pt}}
\put(1364,223.34){\rule{2.891pt}{0.800pt}}
\multiput(1364.00,224.34)(6.000,-2.000){2}{\rule{1.445pt}{0.800pt}}
\put(1376,221.34){\rule{3.132pt}{0.800pt}}
\multiput(1376.00,222.34)(6.500,-2.000){2}{\rule{1.566pt}{0.800pt}}
\put(1389,218.84){\rule{2.891pt}{0.800pt}}
\multiput(1389.00,220.34)(6.000,-3.000){2}{\rule{1.445pt}{0.800pt}}
\put(1401,216.34){\rule{3.132pt}{0.800pt}}
\multiput(1401.00,217.34)(6.500,-2.000){2}{\rule{1.566pt}{0.800pt}}
\put(1414,214.34){\rule{2.891pt}{0.800pt}}
\multiput(1414.00,215.34)(6.000,-2.000){2}{\rule{1.445pt}{0.800pt}}
\put(1426,212.34){\rule{3.132pt}{0.800pt}}
\multiput(1426.00,213.34)(6.500,-2.000){2}{\rule{1.566pt}{0.800pt}}
\end{picture}

%% file: Delta-l.tex
\setlength{\unitlength}{0.240900pt}
\ifx\plotpoint\undefined\newsavebox{\plotpoint}\fi
\sbox{\plotpoint}{\rule[-0.200pt]{0.400pt}{0.400pt}}%
\begin{picture}(1500,900)(0,0)
\font\gnuplot=cmr10 at 10pt
\gnuplot
\sbox{\plotpoint}{\rule[-0.200pt]{0.400pt}{0.400pt}}%
\put(181.0,123.0){\rule[-0.200pt]{4.818pt}{0.400pt}}
\put(161,123){\makebox(0,0)[r]{0.04}}
\put(1419.0,123.0){\rule[-0.200pt]{4.818pt}{0.400pt}}
\put(181.0,228.0){\rule[-0.200pt]{4.818pt}{0.400pt}}
\put(161,228){\makebox(0,0)[r]{0.05}}
\put(1419.0,228.0){\rule[-0.200pt]{4.818pt}{0.400pt}}
\put(181.0,334.0){\rule[-0.200pt]{4.818pt}{0.400pt}}
\put(161,334){\makebox(0,0)[r]{0.06}}
\put(1419.0,334.0){\rule[-0.200pt]{4.818pt}{0.400pt}}
\put(181.0,439.0){\rule[-0.200pt]{4.818pt}{0.400pt}}
\put(161,439){\makebox(0,0)[r]{0.07}}
\put(1419.0,439.0){\rule[-0.200pt]{4.818pt}{0.400pt}}
\put(181.0,544.0){\rule[-0.200pt]{4.818pt}{0.400pt}}
\put(161,544){\makebox(0,0)[r]{0.08}}
\put(1419.0,544.0){\rule[-0.200pt]{4.818pt}{0.400pt}}
\put(181.0,649.0){\rule[-0.200pt]{4.818pt}{0.400pt}}
\put(161,649){\makebox(0,0)[r]{0.09}}
\put(1419.0,649.0){\rule[-0.200pt]{4.818pt}{0.400pt}}
\put(181.0,755.0){\rule[-0.200pt]{4.818pt}{0.400pt}}
\put(161,755){\makebox(0,0)[r]{0.1}}
\put(1419.0,755.0){\rule[-0.200pt]{4.818pt}{0.400pt}}
\put(181.0,860.0){\rule[-0.200pt]{4.818pt}{0.400pt}}
\put(161,860){\makebox(0,0)[r]{0.11}}
\put(1419.0,860.0){\rule[-0.200pt]{4.818pt}{0.400pt}}
\put(181.0,123.0){\rule[-0.200pt]{0.400pt}{4.818pt}}
\put(181,82){\makebox(0,0){10}}
\put(181.0,840.0){\rule[-0.200pt]{0.400pt}{4.818pt}}
\put(391.0,123.0){\rule[-0.200pt]{0.400pt}{4.818pt}}
\put(391,82){\makebox(0,0){11}}
\put(391.0,840.0){\rule[-0.200pt]{0.400pt}{4.818pt}}
\put(600.0,123.0){\rule[-0.200pt]{0.400pt}{4.818pt}}
\put(600,82){\makebox(0,0){12}}
\put(600.0,840.0){\rule[-0.200pt]{0.400pt}{4.818pt}}
\put(810.0,123.0){\rule[-0.200pt]{0.400pt}{4.818pt}}
\put(810,82){\makebox(0,0){13}}
\put(810.0,840.0){\rule[-0.200pt]{0.400pt}{4.818pt}}
\put(1020.0,123.0){\rule[-0.200pt]{0.400pt}{4.818pt}}
\put(1020,82){\makebox(0,0){14}}
\put(1020.0,840.0){\rule[-0.200pt]{0.400pt}{4.818pt}}
\put(1229.0,123.0){\rule[-0.200pt]{0.400pt}{4.818pt}}
\put(1229,82){\makebox(0,0){15}}
\put(1229.0,840.0){\rule[-0.200pt]{0.400pt}{4.818pt}}
\put(1439.0,123.0){\rule[-0.200pt]{0.400pt}{4.818pt}}
\put(1439,82){\makebox(0,0){16}}
\put(1439.0,840.0){\rule[-0.200pt]{0.400pt}{4.818pt}}
\put(181.0,123.0){\rule[-0.200pt]{303.052pt}{0.400pt}}
\put(1439.0,123.0){\rule[-0.200pt]{0.400pt}{177.543pt}}
\put(181.0,860.0){\rule[-0.200pt]{303.052pt}{0.400pt}}
\put(40,491){\makebox(0,0){$\Delta_{\tilde f}$}}
\put(810,21){\makebox(0,0){$\log_{10} (M_C)$ [GeV]}}
\put(1020,334){\makebox(0,0)[r]{$\tilde L$}}
\put(1020,649){\makebox(0,0)[r]{$\tilde e$}}
\put(181.0,123.0){\rule[-0.200pt]{0.400pt}{177.543pt}}
\sbox{\plotpoint}{\rule[-0.400pt]{0.800pt}{0.800pt}}%
\put(181,380){\usebox{\plotpoint}}
\put(181,376.34){\rule{2.800pt}{0.800pt}}
\multiput(181.00,378.34)(7.188,-4.000){2}{\rule{1.400pt}{0.800pt}}
\put(194,372.34){\rule{2.600pt}{0.800pt}}
\multiput(194.00,374.34)(6.604,-4.000){2}{\rule{1.300pt}{0.800pt}}
\put(206,368.34){\rule{2.800pt}{0.800pt}}
\multiput(206.00,370.34)(7.188,-4.000){2}{\rule{1.400pt}{0.800pt}}
\put(219,364.34){\rule{2.800pt}{0.800pt}}
\multiput(219.00,366.34)(7.188,-4.000){2}{\rule{1.400pt}{0.800pt}}
\put(232,360.84){\rule{3.132pt}{0.800pt}}
\multiput(232.00,362.34)(6.500,-3.000){2}{\rule{1.566pt}{0.800pt}}
\put(245,357.34){\rule{2.600pt}{0.800pt}}
\multiput(245.00,359.34)(6.604,-4.000){2}{\rule{1.300pt}{0.800pt}}
\put(257,353.34){\rule{2.800pt}{0.800pt}}
\multiput(257.00,355.34)(7.188,-4.000){2}{\rule{1.400pt}{0.800pt}}
\put(270,349.84){\rule{3.132pt}{0.800pt}}
\multiput(270.00,351.34)(6.500,-3.000){2}{\rule{1.566pt}{0.800pt}}
\put(283,346.34){\rule{2.600pt}{0.800pt}}
\multiput(283.00,348.34)(6.604,-4.000){2}{\rule{1.300pt}{0.800pt}}
\put(295,342.84){\rule{3.132pt}{0.800pt}}
\multiput(295.00,344.34)(6.500,-3.000){2}{\rule{1.566pt}{0.800pt}}
\put(308,339.34){\rule{2.800pt}{0.800pt}}
\multiput(308.00,341.34)(7.188,-4.000){2}{\rule{1.400pt}{0.800pt}}
\put(321,335.84){\rule{2.891pt}{0.800pt}}
\multiput(321.00,337.34)(6.000,-3.000){2}{\rule{1.445pt}{0.800pt}}
\put(333,332.34){\rule{2.800pt}{0.800pt}}
\multiput(333.00,334.34)(7.188,-4.000){2}{\rule{1.400pt}{0.800pt}}
\put(346,328.84){\rule{3.132pt}{0.800pt}}
\multiput(346.00,330.34)(6.500,-3.000){2}{\rule{1.566pt}{0.800pt}}
\put(359,325.84){\rule{3.132pt}{0.800pt}}
\multiput(359.00,327.34)(6.500,-3.000){2}{\rule{1.566pt}{0.800pt}}
\put(372,322.84){\rule{2.891pt}{0.800pt}}
\multiput(372.00,324.34)(6.000,-3.000){2}{\rule{1.445pt}{0.800pt}}
\put(384,319.84){\rule{3.132pt}{0.800pt}}
\multiput(384.00,321.34)(6.500,-3.000){2}{\rule{1.566pt}{0.800pt}}
\put(397,316.84){\rule{3.132pt}{0.800pt}}
\multiput(397.00,318.34)(6.500,-3.000){2}{\rule{1.566pt}{0.800pt}}
\put(410,313.84){\rule{2.891pt}{0.800pt}}
\multiput(410.00,315.34)(6.000,-3.000){2}{\rule{1.445pt}{0.800pt}}
\put(422,310.84){\rule{3.132pt}{0.800pt}}
\multiput(422.00,312.34)(6.500,-3.000){2}{\rule{1.566pt}{0.800pt}}
\put(435,307.84){\rule{3.132pt}{0.800pt}}
\multiput(435.00,309.34)(6.500,-3.000){2}{\rule{1.566pt}{0.800pt}}
\put(448,304.84){\rule{3.132pt}{0.800pt}}
\multiput(448.00,306.34)(6.500,-3.000){2}{\rule{1.566pt}{0.800pt}}
\put(461,301.84){\rule{2.891pt}{0.800pt}}
\multiput(461.00,303.34)(6.000,-3.000){2}{\rule{1.445pt}{0.800pt}}
\put(473,298.84){\rule{3.132pt}{0.800pt}}
\multiput(473.00,300.34)(6.500,-3.000){2}{\rule{1.566pt}{0.800pt}}
\put(486,295.84){\rule{3.132pt}{0.800pt}}
\multiput(486.00,297.34)(6.500,-3.000){2}{\rule{1.566pt}{0.800pt}}
\put(499,293.34){\rule{2.891pt}{0.800pt}}
\multiput(499.00,294.34)(6.000,-2.000){2}{\rule{1.445pt}{0.800pt}}
\put(511,290.84){\rule{3.132pt}{0.800pt}}
\multiput(511.00,292.34)(6.500,-3.000){2}{\rule{1.566pt}{0.800pt}}
\put(524,287.84){\rule{3.132pt}{0.800pt}}
\multiput(524.00,289.34)(6.500,-3.000){2}{\rule{1.566pt}{0.800pt}}
\put(537,285.34){\rule{3.132pt}{0.800pt}}
\multiput(537.00,286.34)(6.500,-2.000){2}{\rule{1.566pt}{0.800pt}}
\put(550,282.84){\rule{2.891pt}{0.800pt}}
\multiput(550.00,284.34)(6.000,-3.000){2}{\rule{1.445pt}{0.800pt}}
\put(562,279.84){\rule{3.132pt}{0.800pt}}
\multiput(562.00,281.34)(6.500,-3.000){2}{\rule{1.566pt}{0.800pt}}
\put(575,277.34){\rule{3.132pt}{0.800pt}}
\multiput(575.00,278.34)(6.500,-2.000){2}{\rule{1.566pt}{0.800pt}}
\put(588,275.34){\rule{2.891pt}{0.800pt}}
\multiput(588.00,276.34)(6.000,-2.000){2}{\rule{1.445pt}{0.800pt}}
\put(600,272.84){\rule{3.132pt}{0.800pt}}
\multiput(600.00,274.34)(6.500,-3.000){2}{\rule{1.566pt}{0.800pt}}
\put(613,270.34){\rule{3.132pt}{0.800pt}}
\multiput(613.00,271.34)(6.500,-2.000){2}{\rule{1.566pt}{0.800pt}}
\put(626,267.84){\rule{2.891pt}{0.800pt}}
\multiput(626.00,269.34)(6.000,-3.000){2}{\rule{1.445pt}{0.800pt}}
\put(638,265.34){\rule{3.132pt}{0.800pt}}
\multiput(638.00,266.34)(6.500,-2.000){2}{\rule{1.566pt}{0.800pt}}
\put(651,263.34){\rule{3.132pt}{0.800pt}}
\multiput(651.00,264.34)(6.500,-2.000){2}{\rule{1.566pt}{0.800pt}}
\put(664,261.34){\rule{3.132pt}{0.800pt}}
\multiput(664.00,262.34)(6.500,-2.000){2}{\rule{1.566pt}{0.800pt}}
\put(677,258.84){\rule{2.891pt}{0.800pt}}
\multiput(677.00,260.34)(6.000,-3.000){2}{\rule{1.445pt}{0.800pt}}
\put(689,256.34){\rule{3.132pt}{0.800pt}}
\multiput(689.00,257.34)(6.500,-2.000){2}{\rule{1.566pt}{0.800pt}}
\put(702,254.34){\rule{3.132pt}{0.800pt}}
\multiput(702.00,255.34)(6.500,-2.000){2}{\rule{1.566pt}{0.800pt}}
\put(715,252.34){\rule{2.891pt}{0.800pt}}
\multiput(715.00,253.34)(6.000,-2.000){2}{\rule{1.445pt}{0.800pt}}
\put(727,250.34){\rule{3.132pt}{0.800pt}}
\multiput(727.00,251.34)(6.500,-2.000){2}{\rule{1.566pt}{0.800pt}}
\put(740,248.34){\rule{3.132pt}{0.800pt}}
\multiput(740.00,249.34)(6.500,-2.000){2}{\rule{1.566pt}{0.800pt}}
\put(753,246.34){\rule{3.132pt}{0.800pt}}
\multiput(753.00,247.34)(6.500,-2.000){2}{\rule{1.566pt}{0.800pt}}
\put(766,244.34){\rule{2.891pt}{0.800pt}}
\multiput(766.00,245.34)(6.000,-2.000){2}{\rule{1.445pt}{0.800pt}}
\put(778,242.34){\rule{3.132pt}{0.800pt}}
\multiput(778.00,243.34)(6.500,-2.000){2}{\rule{1.566pt}{0.800pt}}
\put(791,240.34){\rule{3.132pt}{0.800pt}}
\multiput(791.00,241.34)(6.500,-2.000){2}{\rule{1.566pt}{0.800pt}}
\put(804,238.34){\rule{2.891pt}{0.800pt}}
\multiput(804.00,239.34)(6.000,-2.000){2}{\rule{1.445pt}{0.800pt}}
\put(816,236.34){\rule{3.132pt}{0.800pt}}
\multiput(816.00,237.34)(6.500,-2.000){2}{\rule{1.566pt}{0.800pt}}
\put(829,234.34){\rule{3.132pt}{0.800pt}}
\multiput(829.00,235.34)(6.500,-2.000){2}{\rule{1.566pt}{0.800pt}}
\put(842,232.34){\rule{2.891pt}{0.800pt}}
\multiput(842.00,233.34)(6.000,-2.000){2}{\rule{1.445pt}{0.800pt}}
\put(854,230.84){\rule{3.132pt}{0.800pt}}
\multiput(854.00,231.34)(6.500,-1.000){2}{\rule{1.566pt}{0.800pt}}
\put(867,229.34){\rule{3.132pt}{0.800pt}}
\multiput(867.00,230.34)(6.500,-2.000){2}{\rule{1.566pt}{0.800pt}}
\put(880,227.34){\rule{3.132pt}{0.800pt}}
\multiput(880.00,228.34)(6.500,-2.000){2}{\rule{1.566pt}{0.800pt}}
\put(893,225.84){\rule{2.891pt}{0.800pt}}
\multiput(893.00,226.34)(6.000,-1.000){2}{\rule{1.445pt}{0.800pt}}
\put(905,224.34){\rule{3.132pt}{0.800pt}}
\multiput(905.00,225.34)(6.500,-2.000){2}{\rule{1.566pt}{0.800pt}}
\put(918,222.34){\rule{3.132pt}{0.800pt}}
\multiput(918.00,223.34)(6.500,-2.000){2}{\rule{1.566pt}{0.800pt}}
\put(931,220.84){\rule{2.891pt}{0.800pt}}
\multiput(931.00,221.34)(6.000,-1.000){2}{\rule{1.445pt}{0.800pt}}
\put(943,219.34){\rule{3.132pt}{0.800pt}}
\multiput(943.00,220.34)(6.500,-2.000){2}{\rule{1.566pt}{0.800pt}}
\put(956,217.34){\rule{3.132pt}{0.800pt}}
\multiput(956.00,218.34)(6.500,-2.000){2}{\rule{1.566pt}{0.800pt}}
\put(969,215.84){\rule{3.132pt}{0.800pt}}
\multiput(969.00,216.34)(6.500,-1.000){2}{\rule{1.566pt}{0.800pt}}
\put(982,214.34){\rule{2.891pt}{0.800pt}}
\multiput(982.00,215.34)(6.000,-2.000){2}{\rule{1.445pt}{0.800pt}}
\put(994,212.84){\rule{3.132pt}{0.800pt}}
\multiput(994.00,213.34)(6.500,-1.000){2}{\rule{1.566pt}{0.800pt}}
\put(1007,211.34){\rule{3.132pt}{0.800pt}}
\multiput(1007.00,212.34)(6.500,-2.000){2}{\rule{1.566pt}{0.800pt}}
\put(1020,209.84){\rule{2.891pt}{0.800pt}}
\multiput(1020.00,210.34)(6.000,-1.000){2}{\rule{1.445pt}{0.800pt}}
\put(1032,208.84){\rule{3.132pt}{0.800pt}}
\multiput(1032.00,209.34)(6.500,-1.000){2}{\rule{1.566pt}{0.800pt}}
\put(1045,207.34){\rule{3.132pt}{0.800pt}}
\multiput(1045.00,208.34)(6.500,-2.000){2}{\rule{1.566pt}{0.800pt}}
\put(1058,205.84){\rule{2.891pt}{0.800pt}}
\multiput(1058.00,206.34)(6.000,-1.000){2}{\rule{1.445pt}{0.800pt}}
\put(1070,204.84){\rule{3.132pt}{0.800pt}}
\multiput(1070.00,205.34)(6.500,-1.000){2}{\rule{1.566pt}{0.800pt}}
\put(1083,203.34){\rule{3.132pt}{0.800pt}}
\multiput(1083.00,204.34)(6.500,-2.000){2}{\rule{1.566pt}{0.800pt}}
\put(1096,201.84){\rule{3.132pt}{0.800pt}}
\multiput(1096.00,202.34)(6.500,-1.000){2}{\rule{1.566pt}{0.800pt}}
\put(1109,200.84){\rule{2.891pt}{0.800pt}}
\multiput(1109.00,201.34)(6.000,-1.000){2}{\rule{1.445pt}{0.800pt}}
\put(1121,199.84){\rule{3.132pt}{0.800pt}}
\multiput(1121.00,200.34)(6.500,-1.000){2}{\rule{1.566pt}{0.800pt}}
\put(1134,198.34){\rule{3.132pt}{0.800pt}}
\multiput(1134.00,199.34)(6.500,-2.000){2}{\rule{1.566pt}{0.800pt}}
\put(1147,196.84){\rule{2.891pt}{0.800pt}}
\multiput(1147.00,197.34)(6.000,-1.000){2}{\rule{1.445pt}{0.800pt}}
\put(1159,195.84){\rule{3.132pt}{0.800pt}}
\multiput(1159.00,196.34)(6.500,-1.000){2}{\rule{1.566pt}{0.800pt}}
\put(1172,194.84){\rule{3.132pt}{0.800pt}}
\multiput(1172.00,195.34)(6.500,-1.000){2}{\rule{1.566pt}{0.800pt}}
\put(1185,193.84){\rule{3.132pt}{0.800pt}}
\multiput(1185.00,194.34)(6.500,-1.000){2}{\rule{1.566pt}{0.800pt}}
\put(1198,192.84){\rule{2.891pt}{0.800pt}}
\multiput(1198.00,193.34)(6.000,-1.000){2}{\rule{1.445pt}{0.800pt}}
\put(1210,191.84){\rule{3.132pt}{0.800pt}}
\multiput(1210.00,192.34)(6.500,-1.000){2}{\rule{1.566pt}{0.800pt}}
\put(1223,190.84){\rule{3.132pt}{0.800pt}}
\multiput(1223.00,191.34)(6.500,-1.000){2}{\rule{1.566pt}{0.800pt}}
\put(1236,189.84){\rule{2.891pt}{0.800pt}}
\multiput(1236.00,190.34)(6.000,-1.000){2}{\rule{1.445pt}{0.800pt}}
\put(1248,188.84){\rule{3.132pt}{0.800pt}}
\multiput(1248.00,189.34)(6.500,-1.000){2}{\rule{1.566pt}{0.800pt}}
\put(1261,187.84){\rule{3.132pt}{0.800pt}}
\multiput(1261.00,188.34)(6.500,-1.000){2}{\rule{1.566pt}{0.800pt}}
\put(1274,186.84){\rule{3.132pt}{0.800pt}}
\multiput(1274.00,187.34)(6.500,-1.000){2}{\rule{1.566pt}{0.800pt}}
\put(1287,185.84){\rule{2.891pt}{0.800pt}}
\multiput(1287.00,186.34)(6.000,-1.000){2}{\rule{1.445pt}{0.800pt}}
\put(1299,184.84){\rule{3.132pt}{0.800pt}}
\multiput(1299.00,185.34)(6.500,-1.000){2}{\rule{1.566pt}{0.800pt}}
\put(1312,183.84){\rule{3.132pt}{0.800pt}}
\multiput(1312.00,184.34)(6.500,-1.000){2}{\rule{1.566pt}{0.800pt}}
\put(1325,182.84){\rule{2.891pt}{0.800pt}}
\multiput(1325.00,183.34)(6.000,-1.000){2}{\rule{1.445pt}{0.800pt}}
\put(1350,181.84){\rule{3.132pt}{0.800pt}}
\multiput(1350.00,182.34)(6.500,-1.000){2}{\rule{1.566pt}{0.800pt}}
\put(1363,180.84){\rule{2.891pt}{0.800pt}}
\multiput(1363.00,181.34)(6.000,-1.000){2}{\rule{1.445pt}{0.800pt}}
\put(1375,179.84){\rule{3.132pt}{0.800pt}}
\multiput(1375.00,180.34)(6.500,-1.000){2}{\rule{1.566pt}{0.800pt}}
\put(1337.0,184.0){\rule[-0.400pt]{3.132pt}{0.800pt}}
\put(1401,178.84){\rule{3.132pt}{0.800pt}}
\multiput(1401.00,179.34)(6.500,-1.000){2}{\rule{1.566pt}{0.800pt}}
\put(1414,177.84){\rule{2.891pt}{0.800pt}}
\multiput(1414.00,178.34)(6.000,-1.000){2}{\rule{1.445pt}{0.800pt}}
\put(1388.0,181.0){\rule[-0.400pt]{3.132pt}{0.800pt}}
\put(1426.0,179.0){\rule[-0.400pt]{3.132pt}{0.800pt}}
\put(181,718){\usebox{\plotpoint}}
\put(181,715.34){\rule{3.132pt}{0.800pt}}
\multiput(181.00,716.34)(6.500,-2.000){2}{\rule{1.566pt}{0.800pt}}
\put(194,713.34){\rule{2.891pt}{0.800pt}}
\multiput(194.00,714.34)(6.000,-2.000){2}{\rule{1.445pt}{0.800pt}}
\put(206,711.34){\rule{3.132pt}{0.800pt}}
\multiput(206.00,712.34)(6.500,-2.000){2}{\rule{1.566pt}{0.800pt}}
\put(219,708.84){\rule{3.132pt}{0.800pt}}
\multiput(219.00,710.34)(6.500,-3.000){2}{\rule{1.566pt}{0.800pt}}
\put(232,706.34){\rule{3.132pt}{0.800pt}}
\multiput(232.00,707.34)(6.500,-2.000){2}{\rule{1.566pt}{0.800pt}}
\put(245,704.34){\rule{2.891pt}{0.800pt}}
\multiput(245.00,705.34)(6.000,-2.000){2}{\rule{1.445pt}{0.800pt}}
\put(257,702.84){\rule{3.132pt}{0.800pt}}
\multiput(257.00,703.34)(6.500,-1.000){2}{\rule{1.566pt}{0.800pt}}
\put(270,701.34){\rule{3.132pt}{0.800pt}}
\multiput(270.00,702.34)(6.500,-2.000){2}{\rule{1.566pt}{0.800pt}}
\put(283,699.34){\rule{2.891pt}{0.800pt}}
\multiput(283.00,700.34)(6.000,-2.000){2}{\rule{1.445pt}{0.800pt}}
\put(295,697.34){\rule{3.132pt}{0.800pt}}
\multiput(295.00,698.34)(6.500,-2.000){2}{\rule{1.566pt}{0.800pt}}
\put(308,695.84){\rule{3.132pt}{0.800pt}}
\multiput(308.00,696.34)(6.500,-1.000){2}{\rule{1.566pt}{0.800pt}}
\put(321,694.34){\rule{2.891pt}{0.800pt}}
\multiput(321.00,695.34)(6.000,-2.000){2}{\rule{1.445pt}{0.800pt}}
\put(333,692.84){\rule{3.132pt}{0.800pt}}
\multiput(333.00,693.34)(6.500,-1.000){2}{\rule{1.566pt}{0.800pt}}
\put(346,691.34){\rule{3.132pt}{0.800pt}}
\multiput(346.00,692.34)(6.500,-2.000){2}{\rule{1.566pt}{0.800pt}}
\put(359,689.84){\rule{3.132pt}{0.800pt}}
\multiput(359.00,690.34)(6.500,-1.000){2}{\rule{1.566pt}{0.800pt}}
\put(372,688.84){\rule{2.891pt}{0.800pt}}
\multiput(372.00,689.34)(6.000,-1.000){2}{\rule{1.445pt}{0.800pt}}
\put(384,687.34){\rule{3.132pt}{0.800pt}}
\multiput(384.00,688.34)(6.500,-2.000){2}{\rule{1.566pt}{0.800pt}}
\put(397,685.84){\rule{3.132pt}{0.800pt}}
\multiput(397.00,686.34)(6.500,-1.000){2}{\rule{1.566pt}{0.800pt}}
\put(410,684.84){\rule{2.891pt}{0.800pt}}
\multiput(410.00,685.34)(6.000,-1.000){2}{\rule{1.445pt}{0.800pt}}
\put(422,683.84){\rule{3.132pt}{0.800pt}}
\multiput(422.00,684.34)(6.500,-1.000){2}{\rule{1.566pt}{0.800pt}}
\put(435,682.84){\rule{3.132pt}{0.800pt}}
\multiput(435.00,683.34)(6.500,-1.000){2}{\rule{1.566pt}{0.800pt}}
\put(448,681.84){\rule{3.132pt}{0.800pt}}
\multiput(448.00,682.34)(6.500,-1.000){2}{\rule{1.566pt}{0.800pt}}
\put(473,680.84){\rule{3.132pt}{0.800pt}}
\multiput(473.00,681.34)(6.500,-1.000){2}{\rule{1.566pt}{0.800pt}}
\put(486,679.84){\rule{3.132pt}{0.800pt}}
\multiput(486.00,680.34)(6.500,-1.000){2}{\rule{1.566pt}{0.800pt}}
\put(499,678.84){\rule{2.891pt}{0.800pt}}
\multiput(499.00,679.34)(6.000,-1.000){2}{\rule{1.445pt}{0.800pt}}
\put(461.0,683.0){\rule[-0.400pt]{2.891pt}{0.800pt}}
\put(524,677.84){\rule{3.132pt}{0.800pt}}
\multiput(524.00,678.34)(6.500,-1.000){2}{\rule{1.566pt}{0.800pt}}
\put(511.0,680.0){\rule[-0.400pt]{3.132pt}{0.800pt}}
\put(550,676.84){\rule{2.891pt}{0.800pt}}
\multiput(550.00,677.34)(6.000,-1.000){2}{\rule{1.445pt}{0.800pt}}
\put(537.0,679.0){\rule[-0.400pt]{3.132pt}{0.800pt}}
\put(588,675.84){\rule{2.891pt}{0.800pt}}
\multiput(588.00,676.34)(6.000,-1.000){2}{\rule{1.445pt}{0.800pt}}
\put(562.0,678.0){\rule[-0.400pt]{6.263pt}{0.800pt}}
\put(715,675.84){\rule{2.891pt}{0.800pt}}
\multiput(715.00,675.34)(6.000,1.000){2}{\rule{1.445pt}{0.800pt}}
\put(600.0,677.0){\rule[-0.400pt]{27.703pt}{0.800pt}}
\put(753,676.84){\rule{3.132pt}{0.800pt}}
\multiput(753.00,676.34)(6.500,1.000){2}{\rule{1.566pt}{0.800pt}}
\put(727.0,678.0){\rule[-0.400pt]{6.263pt}{0.800pt}}
\put(778,677.84){\rule{3.132pt}{0.800pt}}
\multiput(778.00,677.34)(6.500,1.000){2}{\rule{1.566pt}{0.800pt}}
\put(791,678.84){\rule{3.132pt}{0.800pt}}
\multiput(791.00,678.34)(6.500,1.000){2}{\rule{1.566pt}{0.800pt}}
\put(766.0,679.0){\rule[-0.400pt]{2.891pt}{0.800pt}}
\put(816,679.84){\rule{3.132pt}{0.800pt}}
\multiput(816.00,679.34)(6.500,1.000){2}{\rule{1.566pt}{0.800pt}}
\put(829,680.84){\rule{3.132pt}{0.800pt}}
\multiput(829.00,680.34)(6.500,1.000){2}{\rule{1.566pt}{0.800pt}}
\put(804.0,681.0){\rule[-0.400pt]{2.891pt}{0.800pt}}
\put(854,681.84){\rule{3.132pt}{0.800pt}}
\multiput(854.00,681.34)(6.500,1.000){2}{\rule{1.566pt}{0.800pt}}
\put(867,682.84){\rule{3.132pt}{0.800pt}}
\multiput(867.00,682.34)(6.500,1.000){2}{\rule{1.566pt}{0.800pt}}
\put(880,683.84){\rule{3.132pt}{0.800pt}}
\multiput(880.00,683.34)(6.500,1.000){2}{\rule{1.566pt}{0.800pt}}
\put(893,684.84){\rule{2.891pt}{0.800pt}}
\multiput(893.00,684.34)(6.000,1.000){2}{\rule{1.445pt}{0.800pt}}
\put(905,685.84){\rule{3.132pt}{0.800pt}}
\multiput(905.00,685.34)(6.500,1.000){2}{\rule{1.566pt}{0.800pt}}
\put(918,686.84){\rule{3.132pt}{0.800pt}}
\multiput(918.00,686.34)(6.500,1.000){2}{\rule{1.566pt}{0.800pt}}
\put(931,688.34){\rule{2.891pt}{0.800pt}}
\multiput(931.00,687.34)(6.000,2.000){2}{\rule{1.445pt}{0.800pt}}
\put(943,689.84){\rule{3.132pt}{0.800pt}}
\multiput(943.00,689.34)(6.500,1.000){2}{\rule{1.566pt}{0.800pt}}
\put(956,690.84){\rule{3.132pt}{0.800pt}}
\multiput(956.00,690.34)(6.500,1.000){2}{\rule{1.566pt}{0.800pt}}
\put(969,691.84){\rule{3.132pt}{0.800pt}}
\multiput(969.00,691.34)(6.500,1.000){2}{\rule{1.566pt}{0.800pt}}
\put(982,693.34){\rule{2.891pt}{0.800pt}}
\multiput(982.00,692.34)(6.000,2.000){2}{\rule{1.445pt}{0.800pt}}
\put(994,694.84){\rule{3.132pt}{0.800pt}}
\multiput(994.00,694.34)(6.500,1.000){2}{\rule{1.566pt}{0.800pt}}
\put(1007,696.34){\rule{3.132pt}{0.800pt}}
\multiput(1007.00,695.34)(6.500,2.000){2}{\rule{1.566pt}{0.800pt}}
\put(1020,697.84){\rule{2.891pt}{0.800pt}}
\multiput(1020.00,697.34)(6.000,1.000){2}{\rule{1.445pt}{0.800pt}}
\put(1032,699.34){\rule{3.132pt}{0.800pt}}
\multiput(1032.00,698.34)(6.500,2.000){2}{\rule{1.566pt}{0.800pt}}
\put(1045,701.34){\rule{3.132pt}{0.800pt}}
\multiput(1045.00,700.34)(6.500,2.000){2}{\rule{1.566pt}{0.800pt}}
\put(1058,702.84){\rule{2.891pt}{0.800pt}}
\multiput(1058.00,702.34)(6.000,1.000){2}{\rule{1.445pt}{0.800pt}}
\put(1070,704.34){\rule{3.132pt}{0.800pt}}
\multiput(1070.00,703.34)(6.500,2.000){2}{\rule{1.566pt}{0.800pt}}
\put(1083,706.34){\rule{3.132pt}{0.800pt}}
\multiput(1083.00,705.34)(6.500,2.000){2}{\rule{1.566pt}{0.800pt}}
\put(1096,708.34){\rule{3.132pt}{0.800pt}}
\multiput(1096.00,707.34)(6.500,2.000){2}{\rule{1.566pt}{0.800pt}}
\put(1109,710.34){\rule{2.891pt}{0.800pt}}
\multiput(1109.00,709.34)(6.000,2.000){2}{\rule{1.445pt}{0.800pt}}
\put(1121,712.34){\rule{3.132pt}{0.800pt}}
\multiput(1121.00,711.34)(6.500,2.000){2}{\rule{1.566pt}{0.800pt}}
\put(1134,714.34){\rule{3.132pt}{0.800pt}}
\multiput(1134.00,713.34)(6.500,2.000){2}{\rule{1.566pt}{0.800pt}}
\put(1147,716.34){\rule{2.891pt}{0.800pt}}
\multiput(1147.00,715.34)(6.000,2.000){2}{\rule{1.445pt}{0.800pt}}
\put(1159,718.34){\rule{3.132pt}{0.800pt}}
\multiput(1159.00,717.34)(6.500,2.000){2}{\rule{1.566pt}{0.800pt}}
\put(1172,720.34){\rule{3.132pt}{0.800pt}}
\multiput(1172.00,719.34)(6.500,2.000){2}{\rule{1.566pt}{0.800pt}}
\put(1185,722.84){\rule{3.132pt}{0.800pt}}
\multiput(1185.00,721.34)(6.500,3.000){2}{\rule{1.566pt}{0.800pt}}
\put(1198,725.34){\rule{2.891pt}{0.800pt}}
\multiput(1198.00,724.34)(6.000,2.000){2}{\rule{1.445pt}{0.800pt}}
\put(1210,727.34){\rule{3.132pt}{0.800pt}}
\multiput(1210.00,726.34)(6.500,2.000){2}{\rule{1.566pt}{0.800pt}}
\put(1223,729.84){\rule{3.132pt}{0.800pt}}
\multiput(1223.00,728.34)(6.500,3.000){2}{\rule{1.566pt}{0.800pt}}
\put(1236,732.34){\rule{2.891pt}{0.800pt}}
\multiput(1236.00,731.34)(6.000,2.000){2}{\rule{1.445pt}{0.800pt}}
\put(1248,734.84){\rule{3.132pt}{0.800pt}}
\multiput(1248.00,733.34)(6.500,3.000){2}{\rule{1.566pt}{0.800pt}}
\put(1261,737.84){\rule{3.132pt}{0.800pt}}
\multiput(1261.00,736.34)(6.500,3.000){2}{\rule{1.566pt}{0.800pt}}
\put(1274,740.34){\rule{3.132pt}{0.800pt}}
\multiput(1274.00,739.34)(6.500,2.000){2}{\rule{1.566pt}{0.800pt}}
\put(1287,742.84){\rule{2.891pt}{0.800pt}}
\multiput(1287.00,741.34)(6.000,3.000){2}{\rule{1.445pt}{0.800pt}}
\put(1299,745.84){\rule{3.132pt}{0.800pt}}
\multiput(1299.00,744.34)(6.500,3.000){2}{\rule{1.566pt}{0.800pt}}
\put(1312,748.84){\rule{3.132pt}{0.800pt}}
\multiput(1312.00,747.34)(6.500,3.000){2}{\rule{1.566pt}{0.800pt}}
\put(1325,751.84){\rule{2.891pt}{0.800pt}}
\multiput(1325.00,750.34)(6.000,3.000){2}{\rule{1.445pt}{0.800pt}}
\put(1337,754.84){\rule{3.132pt}{0.800pt}}
\multiput(1337.00,753.34)(6.500,3.000){2}{\rule{1.566pt}{0.800pt}}
\put(1350,757.84){\rule{3.132pt}{0.800pt}}
\multiput(1350.00,756.34)(6.500,3.000){2}{\rule{1.566pt}{0.800pt}}
\put(1363,760.84){\rule{2.891pt}{0.800pt}}
\multiput(1363.00,759.34)(6.000,3.000){2}{\rule{1.445pt}{0.800pt}}
\put(1375,763.84){\rule{3.132pt}{0.800pt}}
\multiput(1375.00,762.34)(6.500,3.000){2}{\rule{1.566pt}{0.800pt}}
\put(1388,767.34){\rule{2.800pt}{0.800pt}}
\multiput(1388.00,765.34)(7.188,4.000){2}{\rule{1.400pt}{0.800pt}}
\put(1401,770.84){\rule{3.132pt}{0.800pt}}
\multiput(1401.00,769.34)(6.500,3.000){2}{\rule{1.566pt}{0.800pt}}
\put(1414,774.34){\rule{2.600pt}{0.800pt}}
\multiput(1414.00,772.34)(6.604,4.000){2}{\rule{1.300pt}{0.800pt}}
\put(1426,777.84){\rule{3.132pt}{0.800pt}}
\multiput(1426.00,776.34)(6.500,3.000){2}{\rule{1.566pt}{0.800pt}}
\put(842.0,683.0){\rule[-0.400pt]{2.891pt}{0.800pt}}
\end{picture}

%% file: SI01rep.bbl
\begin{thebibliography}{99}
\bibitem{gg}
L.~Girardello and M.~T.~Grisaru, Nucl.~Phys. {\bf B194} (1982) 65.

\bibitem{yamada}
Y.~Yamada, Phys.~Rev. {\bf D 50} (1994) 3537.

\bibitem{jjmvy}
I.~Jack, D.~R.~T.~Jones, S.~P.~Martin, M.~T.~Vaughn and Y.~Yamada,
Phys.~Rev. {\bf D50} (1994) 5481.

\bibitem{hs}
J.~Hisano and M.~Shifman, Phys.~Rev. {\bf D56} (1997) 5475.

\bibitem{jj1}
I.~Jack and D.~R.~T.~Jones, Phys.~Lett. {\bf B415} (1997) 383.

\bibitem{kazakov}
L.~V.~Avdeev, D.~I.~Kazakov and I.~N.~Kondrashuk,
Nucl.~Phys. {\bf B 510} (1998) 289.

\bibitem{jjp1}
I.~Jack, D.~R.~T.~Jones and A.~Pickering, Phys.~Lett. {\bf B 426} (1998) 73;
Phys.~Lett. {\bf B 432} (1998) 114.

\bibitem{kkz}
T.~Kobayashi, J.~Kubo and G.~Zoupanos, Phys.~Lett. {\bf B427} (1998) 291.

\bibitem{kkk}
Y.~Kawamura, T.~Kobayashi and  J.~Kubo, Phys.~Lett. {\bf B432} (1998) 108.

\bibitem{jjp2}
I.~Jack, D.~R.~T.~Jones and A.~Pickering, Phys.~Lett. {\bf B432} (1998) 114.

\bibitem{jjp3} 
I.~Jack, D.~R.~T.~Jones and A.~Pickering, Phys.~Lett. {\bf B435} (1998) 61.

\bibitem{gr}
G.~F.~Giudice and R.~Rattazi, Nucl.~Phys. {\bf B511} (1998) 25.

\bibitem{aglr}
N.~Arkani-Hamed, G.~F.~Giudice, M.~A.~Luty and R.~Rattazzi,
Phys.~Rev. {\bf D58} (1998) 115005.

\bibitem{kv}
D.~I.~Kazakov and V.~N.~Velizhanin, Phys. Lett {\bf B426} (2000) 393.

\bibitem{feynman}
M.~T.~Grisaru, W.~Siegel and M.~Ro\v{c}ek, Nucl~Phys. {\bf B159} (1979) 429;
S.~J.~Gates, Jr., M.~T.~Grisaru, M.~Ro\v{c}ek and W.~Siegel,
{\it SUPERSPACE or One Thousand and One}, Benjamin, (1983).

\bibitem{seiberg1}
N.~Seiberg. Phys.~Lett. {\bf B318} (1993) 469.

\bibitem{weinberg}
S.~Weinberg, Phys.~Rev.~Lett. {\bf 80} (1998) 3702.

\bibitem{fl}
K.~Fujikawa and W.~Lang, Nucl.~Phys. {\bf B88} (1975) 61.

\bibitem{fiterm}
W.Fischler, H.~P.~Nilles, J.~Polchinski, S.~Raby and L.~Susskind,
Phys.~Rev.~Lett. {\bf 47} (1981) 757.

\bibitem{nsvz} 
V.~Novikov, M.~Shifman, A.~Vainstein and V.~Zakharov, 
Nucl. Phys. {\bf B229} (1983) 381; Phys. Lett. {\bf B166} (1986) 329;
M.~Shifman, Int.~J.~Mod.~Phys. {\bf  A11} (1996) 5761 and references therein. 

\bibitem{am}
N.~Arkani-Hamed and H.~Murayama, Phys.~Rev. {\bf D57} (1998) 6638;
JHEP {\bf 0006} (2000) 030.

\bibitem{konishi}
K.~Konishi, Phys.~Lett. {\bf B135} 1984 439.

\bibitem{jj2} 
I.~Jack and D.~R.~T.~Jones, Phys.~Lett. {\bf B465} (1999) 148.

\bibitem{anomaly}
L.~Randall and R.~Sundrum, Nucl. Phys. B557 (1999) 79;
G.~F.~Giudice, M.~A.~Luty, H.~Murayama and R.~Rattazi, JHEP {\bf 9812} (1998) 27;
A.~Pomarol and R.~Rattazi, JHEP 9905 (1999) 013.

\bibitem{ns1}
A.E.~Nelson and M.J.~Strassler, JHEP {\bf 0009} (2000) 030.

\bibitem{kt}
T.~Kobayashi and H.~Terao, Phys.~Rev. {\bf D64} (2001) 075003.

\bibitem{ns2}
A.E.~Nelson and M.J.~Strassler, hep-ph/0104051.

\bibitem{fn}
C.~D.~Froggatt and H.~B.~Nielsen, Nucl. Phys. {\bf B147} (1979) 277;
L.~Ib\'{a}\~{n}ez and G.~G.~Ross, Phys.~Lett. {\bf 332} (1994) 100.

\bibitem{fpt}
T.~Banks and A.~Zaks, Nucl.~Phys. {\bf B196} (1982) 189;
N.~Seiberg, Nucl.~Phys. {\bf B435} (1995) 129;
K.~Intrilligator and N.~Seiberg, 
Nucl.~Phys.~Proc.~Suppl. {\bf 45BC} (1996) 1.

\bibitem{fcnc}
F.~Gabbiani, E.~Gabrielli, A. Masiero and L.~Silvestrini,
Nucl.~Phys. {\bf B477} (1996) 321;
See for a review, {\it e.g.} M.~Misiak, S.~Pokorski and J.~Rosiek, hep-ph/9703442; 
J.~L.~Feng, hep-ph/0101122 and references therein.

\bibitem{degenerate}
M.~Dine, R.~Leigh and A.~Kagan, Phys.~Rev. {\bf D48} (1993) 4269.

\bibitem{alignment}
Y.~Nir and M.~Seiberg. Phys.~Lett. {\bf B309} (1993) 337;
M.~Leurer, Y.~Nir and N.~Seiberg, Nucl.~Phys. {\bf B398} (1993) 319.

\bibitem{decoupling}
S. Dimopoulos and G.~F.~Giudice, Phys.~Lett. {\bf B357} (1995) 573.

\bibitem{anomalousU1}
P.~Bin\'etruy and E.~Dudas, Nucl.~Phys. {\bf B442} (1995) 21;
G.~Dvali and A.~Pomarol, Phys.~Rev.~Lett. {\bf 77} (1996) 3728.

\bibitem{decoupling2}
N.~Arkani-Hamed and H.~Murayama, Phys.~Rev. {\bf D56} (1997) R6733.

\bibitem{kkkz} 
A.~Karch, T.~Kobayashi, J.~Kubo and G.~Zoupanos, 
Phys.~Lett. {\bf B 441} (1998) 235.

\bibitem{lr}
M.~A.~Luty and R.~Rattazi, JHEP 9911 (1999) 001.

\bibitem{knt}
T.~Kobayashi, H.~Nakano and H.~Terao, to be published in Phys.~Rev. {\bf D};
T.~Kobayashi, H.~Nakano and H.~Terao, in preparation.

\bibitem{knnt}
T.~Kobayashi, H.~Nakano, T.~Noguchi and H.~Terao, in preparation.

\end{thebibliography}
